\documentclass[11pt, a4paper]{article}
\usepackage[affil-it]{authblk}

\usepackage{fancyhdr}
\fancypagestyle{firstpage}{
  \fancyhf{}  
   
 \fancyhead[R]{TTI-MATHPHYS-44}
}
\fancypagestyle{nonfirstpage}{
  \fancyhf{}  
   
 \fancyfoot[C]{\hyperlink{contents}{\thepage}} 
}

\usepackage[utf8]{inputenc}

\usepackage[top=20truemm,bottom=20truemm,left=16truemm,right=16truemm]{geometry}
\usepackage{graphicx}
\usepackage{amsmath, latexsym, amssymb, mathrsfs}
\usepackage{mathtools} 
\usepackage{nccmath}
\usepackage{multirow,here}
\usepackage{anyfontsize}
\usepackage{braket,bm,tensor,comment,cite}
\usepackage{physics}
\usepackage{xcolor}
\usepackage[Euler]{upgreek}
\usepackage{enumitem}
\usepackage{comment}
\usepackage{tikz}
\usetikzlibrary{arrows.meta,positioning,fit,backgrounds}
\newcommand{\nn}{\nonumber \\}
\newcommand{\e}[1]{\mathrm{e}^{#1}}
\newcommand{\ep}{\epsilon}

\newcommand{\rhov}{\varrho}
\newcommand{\phiv}{\varphi}

\newcommand{\del}{\partial}
\newcommand{\ar}{\Rightarrow}

\newcommand{\mr}[1]{\mathrm{#1}}
\newcommand{\mc}[1]{\mathcal{#1}}

\newcommand{\mb}[1]{\mathbb{#1}}
\newcommand{\msf}[1]{\mathsf{#1}}

\newcommand{\rD}{\reflectbox{D}}

\usepackage[colorlinks=true,urlcolor=blue,anchorcolor=black,citecolor=blue,linkcolor=black,filecolor=black,menucolor=black,linktocpage=true,pdfproducer=medialab,pdfa=true]{hyperref}
\usepackage{cleveref}
\crefname{equation}{Eq.}{Eqs.}
\Crefname{equation}{Eq.}{Eqs.}
\usepackage{setspace}

 \makeatletter
 
 \@addtoreset{equation}{section}
 \makeatother

\begin{document}
\newgeometry{top=30truemm,bottom=30truemm,left=20truemm,right=20truemm}

\title{\bfseries \fontsize{20pt}{24pt}\selectfont
Generalized Black Holes with\\ Fully Non-aligned Electromagnetic Fields
\vspace{1em}
}

\date{}

\renewcommand\Authfont{\normalsize\normalfont\raggedright\bfseries}
\renewcommand\Affilfont{\normalfont\itshape\raggedright\fontsize{10.4pt}{12}\selectfont}
\renewcommand\Authand{ and }
\renewcommand\Authands{, and }
\newcommand{\email}[1]{\thanks{\href{mailto:#1}{\texttt{#1}}}}

\author[1]{Hideo Furugori\email{hideo@toyota-ti.ac.jp}}
\affil[1]{Mathematical Physics Laboratory, Toyota Technological Institute\vspace{2mm}\\Hisakata 2-12-1, Tempaku-ku, Nagoya, Japan 468-8511}

\author[1]{Shinya Tomizawa\email{tomizawa@toyota-ti.ac.jp}}

\maketitle
\thispagestyle{firstpage} 


\noindent
{\fontsize{11pt}{14pt}\selectfont
{\bfseries Abstract:}
We develop a Pleba\'nski--Demia\'nski-adapted parametrization of the general Ovcharenko--Podolsk\'y solution and construct new families of generalized black holes in four-dimensional Einstein--Maxwell theory.
These Petrov type~D spacetimes possess non-null, fully non-aligned electromagnetic fields. A restriction imposed in the previous parametrization is not required by the field equations in shifted coordinates.
Removing it retains an independent charge parameter $q$, which may be real or purely imaginary.
For $q^2\geq0$, both the metric and electromagnetic field admit a smooth aligned limit to the charged Pleba\'nski--Demia\'nski solution with $\Lambda=0$ and $e^2+g^2=q^2$. 
We impose angular conditions in the Griffiths--Podolsk\'y form allowing Killing horizon cross sections with spherical topology.
The resulting eight-parameter class extends the seven-parameter class of Ovcharenko and Podolsk\'y~\cite{Ovcharenko:2025cpm}.
The remaining normalization condition is generically quartic, defining four algebraic branches.
We give explicit metrics and electromagnetic fields for the non-twisting and $l=0$ twisting families and analyze their black hole and acceleration horizons, extremality, conicity, and further limits.
We determine the effective-NUT-free branches of the latter family and their degenerate cases.
Using the genuine Griffiths--Podolsk\'y form, we establish an explicit correspondence between the non-accelerating effective-NUT-free subclass and the Kerr--Newman--Bertotti--Robinson family and express its electromagnetic flux charges in our parameters.
The $q^2<0$ sector includes the genuine uncharged Kerr--Bertotti--Robinson solution recently identified by Ovcharenko and Podolsk\'y ~\cite{Ovcharenko:2026tos}.

}

\restoregeometry

\newpage

\pagestyle{nonfirstpage} 

\noindent\rule{\linewidth}{0.4pt}
\hypertarget{contents}{\tableofcontents}
\noindent\rule{\linewidth}{0.4pt}

\setstretch{1.2} 

\section{Introduction}
\label{Sec. Intro}

Black holes have long provided a fertile arena for theoretical investigations in high energy physics, including string theory, the holographic principle, black hole thermodynamics, and the information-loss problem~\cite{Strominger:1996sh,tHooft:1993dmi,Susskind:1994vu,Bekenstein:1973ur,Bardeen:1973gs,Hawking:1975vcx,Hawking:1976ra,Strominger:2017zoo}.
More directly, the astrophysical reality of black holes is now strongly supported by gravitational wave observations of binary black hole mergers~\cite{LIGOScientific:2016aoc,LIGOScientific:2026wfs} and by horizon scale imaging with the Event Horizon Telescope~\cite{EventHorizonTelescope:2019dse,EventHorizonTelescope:2025vum}.
Black holes are therefore becoming an important testing ground for the validity of physical theories.

The black hole uniqueness theorem states that, under suitable global and regularity assumptions, the domain of outer communication of a four-dimensional, asymptotically flat, stationary electro-vacuum black hole with a connected event horizon is described by the Kerr--Newman solution~\cite{Carter:1971zc,Robinson:1974nf,Mazur:1982db,Amsel:2009et}.
Its metric and electromagnetic field are characterized by the mass parameter $m$, the spin parameter $a$, and the electric and magnetic charge parameters $e$ and $g$.
The Kerr--Newman spacetime is of Petrov type D and has two repeated principal null directions $\bm{k}$ and $\bm{\ell}$ of the Weyl tensor, which are also principal null directions of the Faraday tensor.
Its electromagnetic field is thus doubly aligned.
Moreover, for $e^2+g^2\neq0$, the electromagnetic invariants $\mc{F}=F_{ab}F^{ab}$ and $\mc{G}=F_{ab}(\ast F)^{ab}$ satisfy $\mc{F}+i\mc{G}\neq0$; the field is therefore non-null.

The general local Petrov type~D Einstein--Maxwell solutions with a possibly nonzero cosmological constant and non-null, doubly aligned electromagnetic fields are known~\cite{VandenBergh:2016ooq}.
The Kundt--Tr\"umper theorem~\cite{Kundt:2016iky} underlies their classification into the following three families:
\begin{enumerate}[label=(\roman*)]
  \item class $\mc{D}$ solutions of Debever--Kamran--McLenaghan~\cite{classD81,Debever:1983pi,Debever:1984yxe}, namely local solutions satisfying the generalized Goldberg--Sachs conditions $\kappa_s=0=\nu_s\,,\,\sigma_s=0=\lambda_s$\footnote{
  We use the Newman--Penrose formalism~\cite{Newman:1961qr}, with conventions given in Appendix~\ref{App. NP}.};
  \item Pleba\'nski--Hacyan solutions~\cite{Plebanski-Hacyan79}, which satisfy $3\Psi_2=2\Phi_{11}$;
  \item Garc\'ia--Pleba\'nski solutions~\cite{Garca1982AnET}, which satisfy $3\Psi_2=-2\Phi_{11}$.
\end{enumerate}
The class $\mc D$ solutions in item~(i) are encompassed by the Pleba\'nski--Demia\'nski family when appropriate limiting and degenerate transformation procedures are included~\cite{Plebanski:1976gy,GarciaPlebanski82rel,Griffiths:2005qp,Griffiths:2009dfa}.

In recent years, our understanding of the Pleba\'nski--Demia\'nski solution has advanced substantially, in particular with respect to its global structure and the transitions between solutions induced by degenerate coordinate transformations~\cite{Podolsky:2018dpr,Podolsky:2021zwr,Podolsky:2022xxd,Ovcharenko:2024yyu,Ovcharenko:2025fxg}.
By contrast, exact type~D Einstein--Maxwell solutions with non-aligned, rather than doubly aligned, electromagnetic fields remain incompletely understood.

Petrov type~D Einstein--Maxwell solutions with null electromagnetic fields have been completely classified and belong to the Robinson--Trautman class~\cite{Debever1989,Bergh1989}.
For non-null fields, the remaining possibilities are partially non-aligned electromagnetic fields\footnote{
These fields are often called partially aligned.
We use ``partially non-aligned'' to emphasize the contrast with doubly aligned fields.}
and fully non-aligned electromagnetic fields.
In a null tetrad adapted to the repeated principal null directions of the Weyl tensor, the former satisfy $\phi_0=0$ and $\phi_2\neq0$, up to an interchange of $\bm{k}$ and $\bm{\ell}$, and several examples are known~\cite{Leroy:1979rss,Kowalczynski:1977wg}.
The latter satisfy $\phi_0\phi_2\neq0$.
An important development was the construction by Van den Bergh and Carminati~\cite{VandenBergh:2020lvf} of the general solution under the conditions\footnote{See also the exact solution constructed by Alekseev and Garc\'ia in 1996~\cite{Alekseev:1996fq}, whose Petrov type~D character was clarified only much later~\cite{Ortaggio:2018ikt}.}
\[
\kappa_s=\nu_s=\sigma_s=\lambda_s=0,\qquad
\Im\rho_s=\Im\mu_s=0,\qquad
\rho_s\mu_s\neq0,\qquad
\tau_s+\bar{\pi}_s=0.
\]
Their construction was motivated by Van den Bergh's result that a type~D Einstein--Maxwell solution with a non-null, fully non-aligned electromagnetic field must have $\Lambda=0$ if it satisfies the generalized Goldberg--Sachs conditions~\cite{VandenBergh:2016ooq}.

Building on these developments, Ovcharenko and Podolsk\'y constructed a broad class of Petrov type~D Einstein--Maxwell solutions with non-null, fully non-aligned electromagnetic fields~\cite{Ovcharenko:2025cpm,Ovcharenko:2026byw}.
We refer to this class as the Ovcharenko--Podolsk\'y solution.
This class includes the Kerr--Bertotti--Robinson spacetime~\cite{Podolsky:2025tle}, which describes a rotating black hole immersed in an external Bertotti--Robinson electromagnetic field~\cite{Bertotti:1959pf,Robinson:1959ev}.
These solutions provide an exact setting for studying the interaction between black holes and external electromagnetic fields.
This spacetime and related exact solutions have already been studied both for their astrophysical implications~\cite{Zeng:2025olq,Wang:2025vsx,Zeng:2025tji,Vachher:2025jsq,Zhang:2025ole,Wang:2026czl,Mirkhaydarov:2026fyn,Wan:2026lca} and from a theoretical perspective~\cite{Hu:2026slp,Siahaan:2025ngu,Gray:2025lwy,Ovcharenko:2026pow,DiPinto:2026rvp}.

The local uniqueness result established by Ovcharenko and Podolsk\'y~\cite{Ovcharenko:2026pow} is particularly important for the classification of type~D Einstein--Maxwell solutions.
They showed that a four-dimensional spacetime satisfying the following five conditions
\begin{enumerate}[label=(\arabic*)]
  \item it is stationary and axisymmetric and possesses non-null Killing vectors;
  \item it is of Petrov type~D;
  \item it satisfies the generalized Goldberg--Sachs conditions;
  \item both repeated principal null directions $\bm{k},\bm{\ell}$ of the Weyl tensor are orthogonal to the polar direction; 
  \item the one-form $\bm{\theta}\coloneqq \mu_s \bm{k}-\rho_s \bm{\ell}-\pi_s \bm{m}+\tau_s\bar{\bm{m}}$ is closed,
\end{enumerate}
can locally be written in the conformal-to-Carter form, Eq.~\eqref{CtoC metric}.\footnote{
While this manuscript was being finalized, Nakajima, Guo, and Lin~\cite{Nakajima:2026lkr} showed that the conformal-to-Carter form can be derived without assumptions~(4) and~(5), in the generic type~D sector considered in their analysis.
}
Within this metric ansatz, the general Einstein--Maxwell solution with a non-null, fully non-aligned electromagnetic field is the Ovcharenko--Podolsk\'y family.
Thus, within the scope of this local uniqueness result, any such solution outside the Ovcharenko--Podolsk\'y family must violate at least one of the stated assumptions.

The Pleba\'nski--Demia\'nski solution with a non-null, doubly aligned electromagnetic field also satisfies these five conditions~\cite{Ovcharenko:2026pow}.
To clarify the physical meaning of the Ovcharenko--Podolsk\'y parameters, Ovcharenko and Podolsk\'y related them to those of the Pleba\'nski--Demia\'nski solution~\cite{Ovcharenko:2025cpm}.
Using this parametrization, they also expressed the metric and electromagnetic field in the Griffiths--Podolsk\'y form, which is convenient for analyzing black hole spacetimes, and imposed angular conditions allowing Killing horizon cross sections with $\mb{S}^2$ topology.
Their construction follows the procedure previously used to obtain generalized black holes in the Pleba\'nski--Demia\'nski family~\cite{Griffiths:2005se,Griffiths:2005qp,Griffiths:2009dfa}.
For fixed values of the remaining parameters, these conditions generically determine the auxiliary parameters $k$, $\epsilon$, and $n$ uniquely in the Pleba\'nski--Demia\'nski case, whereas they yield up to five algebraic branches in the Ovcharenko--Podolsk\'y parametrization, indicating a richer algebraic structure.

Despite these advances, the Pleba\'nski--Demia\'nski-adapted parametrization used in Ref.~\cite{Ovcharenko:2025cpm} does not cover the full parameter space of the Ovcharenko--Podolsk\'y solution.
In the original conformal-to-Carter form of the metric, the Einstein--Maxwell equations require the coefficients of the conformal factor to satisfy
\[
\check c_{20}=-\check c_{02}.
\]
After the two non-Killing coordinates are shifted so that $\hat c_{10}=\hat c_{01}=0$, with hats denoting the transformed coefficients, this relation takes the form of Eq.~\eqref{p0q0 constraint}, which does not require $\hat c_{20}=-\hat c_{02}$.
Ovcharenko and Podolsk\'y subsequently specialized to the latter condition when introducing their Pleba\'nski--Demia\'nski-type parameters.
This additional restriction selects a consistent subclass but removes one independent degree of freedom from the parametrization.

The two main results of this paper are as follows.
First, we construct a Pleba\'nski--Demia\'nski-adapted parametrization that preserves the generality of the Ovcharenko--Podolsk\'y solution.
The additional degree of freedom is represented by an independent charge parameter $\mr{q}$, which can be real or purely imaginary.
For $\mr{q}^2\geq0$, we exhibit a smooth aligned limit in which the coordinate shifts vanish and both the metric and electromagnetic field reduce to those of the charged Pleba\'nski--Demia\'nski solution with $\Lambda=0$ and $\mr{q}^2=e^2+g^2$.
This limit clarifies the interpretation of $\mr{q}$ and provides a nontrivial consistency check of the parametrization.

Second, using this parametrization, we construct an eight-parameter class of generalized black holes, extending the previously studied seven-parameter class by one independent parameter.
We express the solution in the Griffiths--Podolsk\'y form and impose the angular conditions allowing Killing horizon cross sections with $\mb{S}^2$ topology.
We use the term ``generalized black holes'' in this local sense, following Griffiths and Podolsk\'y.
In the generic twisting sector, the remaining normalization condition is quartic in $k$, yielding four algebraic branches in our parametrization, in contrast to the fifth-order condition in the earlier Ovcharenko--Podolsk\'y parametrization.
For the standard real aligned limit, one branch generically approaches finite Pleba\'nski--Demia\'nski parameter values, whereas the other three algebraic roots become unbounded with the remaining parameters held fixed.
We obtain explicit metrics and electromagnetic fields for the non-twisting family and the twisting family with vanishing NUT-like parameter $l$, and analyze their black hole and acceleration horizons, extremality conditions, conicity, and further limiting spacetimes.
In particular, within the latter family, we determine the two radial-shift branches with vanishing effective NUT-like parameter and the conditions under which they coincide.

The genuine Griffiths--Podolsk\'y representations further clarify the physical interpretation of these families.
We introduce apparent mass and charge parameters through the radial metric functions and, for the non-accelerating effective-NUT-free twisting subclass, give an explicit coordinate and parameter correspondence with the Kerr--Newman--Bertotti--Robinson family recently reported in Ref.~\cite{Ovcharenko:2026tos}.
Using the flux charges obtained in that work, we express the physical electric and magnetic charges in our parameters.
The resulting relations show that neither $\mr{q}=0$ nor a vanishing apparent charge parameter $\mr{Q}$ implies vanishing electromagnetic flux charges in this rotating subclass.
Conversely, the sector $\mr{q}^2<0$ contains the genuine uncharged Kerr--Bertotti--Robinson solution identified in the same reference.
Thus, allowing $\mr{q}$ to be purely imaginary includes real black hole solutions rather than merely a formal complex extension.
The explicit forms and parameter correspondence are given in Appendix~\ref{App. genuine GP}.
A complete determination of the admissible parameter and global coordinate domains is left for future work.

\begin{figure}[p]
\centering
\begin{tikzpicture}[
  solution/.style={draw, rounded corners=2pt, align=center, text width=36mm,
    minimum height=10mm, fill=blue!7, font=\small},
  general/.style={solution, text width=145mm, minimum height=14mm},
  pdform/.style={solution, text width=145mm, minimum height=14mm},
  sector/.style={solution, text width=62mm},
  nontwisting/.style={solution, text width=52mm},
  family/.style={solution, text width=62mm},
  pdsolution/.style={solution, text width=18mm, inner xsep=2mm,
    fill=green!8},
  gplimit/.style={solution, text width=48mm, fill=green!8,
    font=\footnotesize},
  extendedbox/.style={draw=red!65!black, very thick, rounded corners=3pt,
    fill=yellow!15},
  familytitle/.style={align=center, text width=62mm, font=\small},
  known/.style={solution, text width=50mm, minimum height=14mm,
    inner xsep=4mm, inner ysep=3mm, fill=orange!10,
    font=\footnotesize\linespread{1.18}\selectfont},
  limitpanel/.style={draw=black!65, rounded corners=2pt, fill=gray!8},
  limititems/.style={align=left, text width=62mm, font=\footnotesize},
  nonut/.style={solution, text width=62mm, minimum height=8mm,
    fill=cyan!10, draw=cyan!60!black, very thick, font=\footnotesize},
  groupbox/.style={draw=blue!60!black, dashed, rounded corners=4pt,
    inner xsep=4mm, inner ysep=5mm},
  grouptitle/.style={font=\small\bfseries, text=blue!50!black,
    fill=white, inner sep=1.5pt, align=center},
  arrow/.style={-{Stealth[length=3mm,width=2.4mm]}, thick},
  label/.style={font=\scriptsize, fill=white, inner sep=1pt, align=center}
]
\node[general] (general) at (0,10.9)
  {General Ovcharenko--Podolsk\'y spacetime in conformal-to-Carter form\\
   $(\eta,q,p,\sigma):\ \check P(p),\ \check Q(q),\
   \check\Omega^2(q,p),\ \check\rho^2(q,p)$};
\node[pdform] (pdform) at (0,8.6)
  {Ovcharenko--Podolsk\'y spacetime in Pleba\'nski--Demia\'nski form\\
   $(\eta,q,p,\sigma):\ \hat P(p),\ \hat Q(q),\
   \hat\Omega^2(q,p),\ \hat\rho^2(q,p)$};
\node[sector] (twist) at (0,6.7)
  {$\mr{OP}_{\alpha\omega}$ twisting sector ($\omega\neq0$)\\
   $(\tau,\phi,r,\xi):\ P(\xi),Q(r),\Omega^2(r,\xi),\rho^2(r,\xi)$};
\node[nontwisting] (nontwist) at (2.7,4.55)
  {$\mr{OP}_{\alpha}^{0}$ non-twisting sector\\
   $(\tau,\phi,r,\xi):\ P_0(\xi),Q_0(r),\Omega_0^2(r,\xi)$};
\node[pdsolution] (pdtwist) at (-7.0,6.7)
  {$\mr{PD}_{\alpha\omega}$\\$\Lambda=0$};
\node[pdsolution] (pdnontwist) at (7.25,4.55)
  {$\mr{PD}_{\alpha}^{0}$\\$\Lambda=0$};

\node[family] (bhtwist) at (-3.8,4.55)
  {Generalized black hole family\\in the twisting sector\\
   Four algebraic branches (generically)\\
   $(t,\phiv,r,\theta):\
   \mc P(\theta),\mc Q(r),\Omega^2(r,\theta),\rhov^2(r,\theta)$};

\node[familytitle] (ntfamilytitle) at (4.9,1.75)
  {New generalized black hole family\\in the non-twisting sector\\
   Two branches: $\mr{OP}_{\alpha\pm}^{0}$\\with charge parameter $\mr q$\\
   $(t,\phiv,r,\theta):\
   \mc P_0(\theta),\mc Q_0(r),\Omega_0^2(r,\theta)$};
\node[known] (opoldnontwist) at (4.9,-0.25)
  {$\mr q=0$ subclass obtained by\\Ovcharenko and Podolsk\'y};

\node[grouptitle, text width=35mm] (lzerotitle) at (-1.6,2.65)
  {Detailed subclasses\\in this paper: $l=0$};
\node[familytitle] (twfamilytitle) at (-3.8,0.85)
  {New generalized black hole family\\in the $l=0$ twisting sector\\with charge parameter $\mr q$};
\node[known] (opoldtwist) at (-3.8,-0.95)
  {$\mr q=0$ subclass obtained by\\Ovcharenko and Podolsk\'y};
\node[limititems, font=\footnotesize\bfseries] (limitstitle) at (-3.8,-3.12)
  {Further subclasses};
\node[nonut] (nonutlimit) at (-3.8,-3.85)
  {\textbf{No NUT:} $\ell_{\mathrm{eff}}=\omega\xi_0=0$};
\node[limititems] (otherlimits) at (-3.8,-4.95)
  {$\bullet$ No spin parameter: $a=0$\\
   $\bullet$ No acceleration parameter: $\upalpha=0$\\
   $\bullet$ No mass parameter: $m=0$};

\node[gplimit] (gpnontwist) at (4.9,-2.80)
  {Pleba\'nski--Demia\'nski\\generalized black holes\\
   non-twisting sector, $\Lambda=0$};
\node[gplimit] (gptwist) at (4.9,-4.65)
  {Pleba\'nski--Demia\'nski\\generalized black holes\\
   twisting sector, $\Lambda=0$};

\coordinate (lzerotop) at ([yshift=3mm]lzerotitle.north);
\begin{scope}[on background layer]
  \node[extendedbox, fit=(ntfamilytitle)(opoldnontwist), inner sep=3mm]
    (ntfamily) {};
  \node[extendedbox, fit=(twfamilytitle)(opoldtwist), inner sep=3mm]
    (twfamily) {};
  \node[limitpanel, fit=(limitstitle)(nonutlimit)(otherlimits), inner sep=2mm]
    (twistlimits) {};
  \node[groupbox, fit=(lzerotop)(lzerotitle)(twfamily)(twistlimits),
    inner xsep=4mm, inner ysep=1mm]
    (lzerogroup) {};
\end{scope}

\draw[arrow] (general) -- node[label, right=2mm] {Pleba\'nski--Demia\'nski gauge and\\new parametrization with charge parameter $\mr{q}$} (pdform);
\draw[arrow] (pdform) -- node[label, right=2mm] {$\omega\neq0$, $\alpha\neq0$} (twist);
\draw[arrow] (twist.south) -- node[label, pos=.52, right=8mm, yshift=2mm] {$\omega\to0$ with\\$n_0= n/\omega$ fixed} (nontwist.north);
\draw[arrow] (twist.west) -- node[label, above=3mm] {$b\to0$} (pdtwist.east);
\draw[arrow] (nontwist.east) -- node[label, above=3mm] {$b\to0$} (pdnontwist.west);
\draw[arrow] (twist.south) -- node[label, pos=.55, left=4mm,
  xshift=-4mm, yshift=2mm] {black hole candidate conditions} (bhtwist.north);
\draw[arrow] (nontwist.south) -- node[label, pos=.48, right=4mm] {black hole candidate conditions} (ntfamily.north);
\draw[arrow] (bhtwist.south) -- node[label, pos=.55, left=1mm] {$l\to0$} (twfamily.north);
\draw[arrow] (twfamily.south) -- (twistlimits.north);
\draw[arrow] (ntfamily.south) -- node[label, right=2mm] {$B\to0$} (gpnontwist.north);
\draw[arrow, preaction={draw=white, line width=4pt, -}]
  (nonutlimit.east) .. controls (0.40,-3.25) and (0.30,-0.20) ..
  node[label, pos=.10, right=1mm] {$a\to0$} (ntfamily.west);
\path (twfamily.south) --
  node[label, right=1mm] {conditions and limits}
  (twistlimits.north);
\coordinate (twistBturn) at (1.05,-4.65);
\draw[thick] (bhtwist.south east) -| (twistBturn);
\node[label, anchor=west] at (1.20,-1.75) {$b\to0$\\finite branch};
\draw[arrow] (twistBturn) -- (gptwist.west);
\end{tikzpicture}
\caption{
Relations among the metric forms, sectors, and subclasses considered in this paper.
The diagram also indicates the coordinates and metric functions used in the different forms.
The arrows indicate coordinate and parameter transformations, restrictions to subclasses, or limits, as labeled.
The parameters $b$ or $B$ denote the strength of the non-aligned component of the electromagnetic field.
The red-bordered boxes highlight the non-twisting and $l=0$ twisting generalized black hole families with the additional charge parameter $q$, while the nested orange boxes represent the corresponding $q=0$ subclasses previously studied by Ovcharenko and Podolsk\'y.
The green boxes denote the standard aligned Pleba\'nski--Demia\'nski limits with $\Lambda=0$ and $\mr{q}^2\geq0$, in which the coordinate shifts vanish.
The lower two represent generalized black holes in the Griffiths--Podolsk\'y form.
For the generic twisting generalized black hole family, the aligned arrow refers to the algebraic branch with finite Pleba\'nski--Demia\'nski parameter values.
The dashed frame encloses the $l=0$ twisting family and its further subclasses.
The $a\to0$ arrow connects the subclass with vanishing effective NUT-like parameter $l_{\mathrm{eff}}$ to the non-twisting family.
}
\label{fig:solution-relations}
\end{figure}
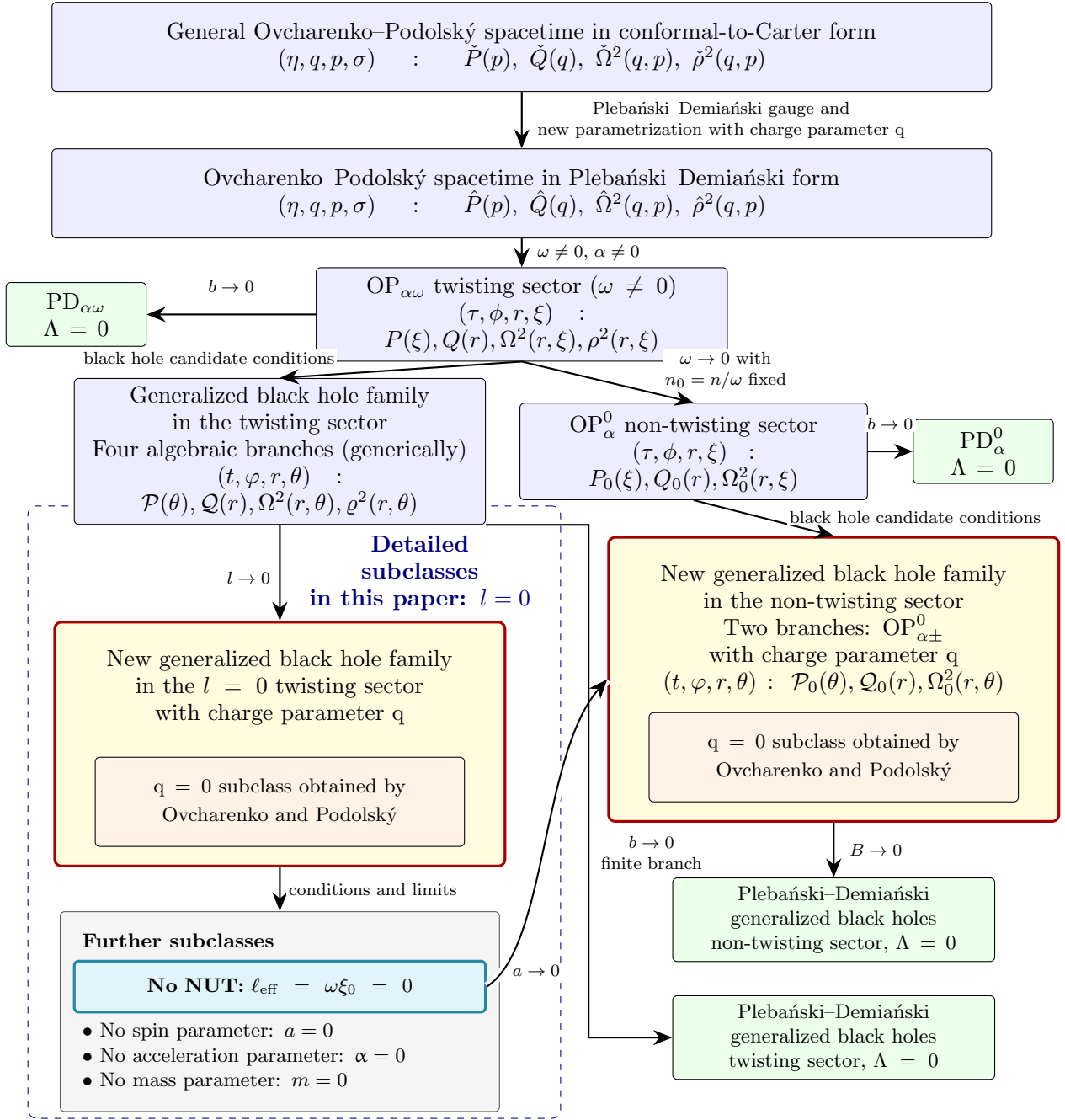

Figure~\ref{fig:solution-relations} summarizes the relations among the metric forms, sectors, and subclasses considered in this paper.
It traces the construction of the generalized black hole families from the general Ovcharenko--Podolsk\'y solution and displays their connections to the previously studied subclasses and the aligned Pleba\'nski--Demia\'nski limits.

The remainder of the paper follows the construction outlined in Fig.~\ref{fig:solution-relations}.
In Sec.~\ref{Sec. review}, we review the general Ovcharenko--Podolsk\'y solution in the conformal-to-Carter form and describe its transformation to the Pleba\'nski--Demia\'nski gauge.
In Sec.~\ref{Sec. OP-PD}, we introduce our parametrization, establish the aligned limit, and construct the twisting and non-twisting sectors by introducing the acceleration and twist parameters and taking the appropriate limit.
In Sec.~\ref{Sec. OP-GP}, we give the Griffiths--Podolsk\'y representation, derive the black hole candidate conditions, and examine the finite Pleba\'nski--Demia\'nski branch of the resulting algebraic equations.
In Sec.~\ref{Sec. family}, we give explicit expressions for the non-twisting and $l=0$ twisting families in dimensionful variables and analyze their properties and further limiting spacetimes.
Section~\ref{Sec. summary} summarizes our results and discusses open questions.
Appendix~\ref{App. NP} specifies our Newman--Penrose conventions.
Appendix~\ref{App. genuine GP} presents the genuine Griffiths--Podolsk\'y forms of the two explicit families and their apparent mass and charge parameters.
It also establishes the correspondence of the non-accelerating effective-NUT-free twisting subclass with the Kerr--Newman--Bertotti--Robinson family and gives its electromagnetic flux charges in our parameters.

\newpage

\section{Review of the Ovcharenko--Podolsk\'y spacetime}
\label{Sec. review}

We review the Ovcharenko--Podolsk\'y spacetime~\cite{Ovcharenko:2025cpm}, a Petrov type~D Einstein--Maxwell solution with a non-null, fully non-aligned electromagnetic field.

\subsection{Ansatz for the Einstein--Maxwell equations}

Under the five geometric conditions stated in the Introduction, the metric can locally be written in the following conformal-to-Carter form without imposing any field equations~\cite{Ovcharenko:2026pow}:
\begin{align}
ds^2
&=\frac{1}{\check\Omega^2(q,p)}\biggl[
-\frac{\check Q(q)}{\check\rho^2(q,p)}
\left(d\eta-\check v(p)d\sigma\right)^2
+\frac{\check\rho^2(q,p)}{\check Q(q)}dq^2
+\frac{\check\rho^2(q,p)}{\check P(p)}dp^2
+\frac{\check P(p)}{\check\rho^2(q,p)}
\left(d\eta+\check u(q)d\sigma\right)^2
\biggr]\,,
\label{CtoC metric}
\end{align}
where
\begin{align}
\check\rho^2(q,p)=\check u(q)+\check v(p)\,,\qquad
\check u(q)=q^2\,,\qquad
\check v(p)=p^2\,.
\label{def rho}
\end{align}
The real metric functions $\check P(p)$, $\check Q(q)$, and $\check\Omega(q,p)$ are to be determined by the Einstein--Maxwell equations.

On a coordinate patch with $\check P>0$, $\check Q>0$, $\check\Omega^2>0$, and $\check\rho^2>0$, we choose the orthonormal tetrad
\begin{align}
e_{(0)}^a
&=\sqrt{\frac{\check\Omega^2}{\check Q\check\rho^2}}
\left(\check u(\del_\eta)^a-(\del_\sigma)^a\right)\,,
&
e_{(1)}^a
&=\sqrt{\frac{\check\Omega^2\check Q}{\check\rho^2}}
(\del_q)^a\,,
\nn
e_{(2)}^a
&=\sqrt{\frac{\check\Omega^2\check P}{\check\rho^2}}
(\del_p)^a\,,
&
e_{(3)}^a
&=\sqrt{\frac{\check\Omega^2}{\check P\check\rho^2}}
\left(\check v(\del_\eta)^a+(\del_\sigma)^a\right)\,.
\end{align}
The associated null tetrad is
\begin{align}
k^a
\coloneqq\frac{1}{\sqrt2}\left(e_{(0)}^a+e_{(1)}^a\right)
&=\sqrt{\frac{\check\Omega^2}{2\check Q\check\rho^2}}
\left(\check u(\del_\eta)^a-(\del_\sigma)^a
+\check Q(\del_q)^a\right)\,,
\label{null k}\\
\ell^a
\coloneqq\frac{1}{\sqrt2}\left(e_{(0)}^a-e_{(1)}^a\right)
&=\sqrt{\frac{\check\Omega^2}{2\check Q\check\rho^2}}
\left(\check u(\del_\eta)^a-(\del_\sigma)^a
-\check Q(\del_q)^a\right)\,,
\label{null ell}\\
m^a
\coloneqq\frac{1}{\sqrt2}\left(e_{(3)}^a+i e_{(2)}^a\right)
&=\sqrt{\frac{\check\Omega^2}{2\check P\check\rho^2}}
\left(\check v(\del_\eta)^a+(\del_\sigma)^a
+i\check P(\del_p)^a\right)\,,
\label{null m}\\
\bar m^a
\coloneqq\frac{1}{\sqrt2}\left(e_{(3)}^a-i e_{(2)}^a\right)
&=\sqrt{\frac{\check\Omega^2}{2\check P\check\rho^2}}
\left(\check v(\del_\eta)^a+(\del_\sigma)^a
-i\check P(\del_p)^a\right)\,.
\label{null m bar}
\end{align}
These vectors satisfy $k^a\ell_a=-1$ and $m^a\bar m_a=1$, with all other inner products vanishing.
The Weyl scalars in this tetrad are
\begin{align}
\Psi_2
&=\frac{\check\Omega^2}{12}
\frac{(q+ip)^2}{q-ip}
\left[
\del_q^2\left(\frac{\check Q}{(q+ip)^3}\right)
+\del_p^2\left(\frac{\check P}{(q+ip)^3}\right)
\right]\,,
\qquad
\Psi_0=\Psi_1=\Psi_3=\Psi_4=0\,.
\label{type D}
\end{align}
Thus, the metric is generically of Petrov type~D, with repeated principal null directions $k^a$ and $\ell^a$.

The spacetime is stationary and axisymmetric, with Killing vector fields $(\del_\eta)^a$ and $(\del_\sigma)^a$.
Following Ref.~\cite{Ovcharenko:2025cpm}, we work in a local gauge in which the electromagnetic potential takes the form
\begin{align}
A_a(q,p)
=A_\eta(q,p)(d\eta)_a+A_\sigma(q,p)(d\sigma)_a\,.
\label{gauge potential}
\end{align}
With $F_{ab}=2\nabla_{[a}A_{b]}$, the Maxwell scalars in this tetrad are
\begin{align}
\phi_0&=\phi_2
=\frac{\check\Omega^2}{2\check\rho^2\sqrt{\check P\check Q}}
\left[
\check Q\left(\check v\del_q A_\eta+\del_q A_\sigma\right)
-i\check P\left(\check u\del_p A_\eta-\del_p A_\sigma\right)
\right]\,,
\label{non-aligned}\\
\phi_1
&=-\frac{\check\Omega^2}{2\check\rho^2}
\left[
-\check u\del_q A_\eta+\del_q A_\sigma
+i\left(\check v\del_p A_\eta+\del_p A_\sigma\right)
\right]\,.
\label{aligned}
\end{align}

To describe a fully non-aligned electromagnetic field, we adopt the ansatz
\begin{align}
\phi_0=\phi_2
=c^\prime\frac{\sqrt{\check Q\check P}}{\check\Omega(q+ip)}\,,
\label{elemag ansatz}
\end{align}
where $c^\prime$ is a nonzero complex constant.
For the conformal-to-Carter metric, Eq.~\eqref{CtoC metric}, this ansatz has been shown not to restrict generality in the non-null, fully non-aligned Einstein--Maxwell sector~\cite{Ovcharenko:2026pow}.
The aligned limit $\abs{c^\prime}\to0$ will be examined in Sec.~\ref{Sec. OP-PD} after introducing the Pleba\'nski--Demia\'nski-adapted parametrization.

\subsection{Solution of the Einstein--Maxwell equations in original conformal-to-Carter form}

The Maxwell equations in the Newman--Penrose formalism are
\begin{equation}
\begin{aligned}
\mr D\phi_1-\bar\delta\phi_0
&=(\pi_s-2\alpha_s)\phi_0+2\rho_s\phi_1-\kappa_s\phi_2\,,\\
\mr D\phi_2-\bar\delta\phi_1
&=-\lambda_s\phi_0+2\pi_s\phi_1+(\rho_s-2\epsilon_s)\phi_2\,,\\
\rD\phi_0-\delta\phi_1
&=(2\gamma_s-\mu_s)\phi_0-2\tau_s\phi_1+\sigma_s\phi_2\,,\\
\rD\phi_1-\delta\phi_2
&=\nu_s\phi_0-2\mu_s\phi_1+(2\beta_s-\tau_s)\phi_2\,,
\end{aligned}
\label{Maxwell eq}
\end{equation}
where $(\mr D,\rD,\delta,\bar\delta)\coloneqq
(k^a\nabla_a,\ell^a\nabla_a,m^a\nabla_a,\bar m^a\nabla_a)$.
For the tetrad given above, the spin coefficients are
\begin{align}
&\kappa_s=\nu_s=\sigma_s=\lambda_s=0\,,\\
&\rho_s=\mu_s
=\frac12\sqrt{\frac{\check Q\check\Omega^2}{2\check\rho^2}}
\left[
\del_q\left(\log\frac{\check\Omega^2}{\check\rho^2}\right)
+i\del_p\left(\log\check\rho^2\right)
\right]\,,\\
&\tau_s=\pi_s
=\frac12\sqrt{\frac{\check P\check\Omega^2}{2\check\rho^2}}
\left[
\del_q\left(\log\check\rho^2\right)
+i\del_p\left(\log\frac{\check\Omega^2}{\check\rho^2}\right)
\right]\,,\\
&\alpha_s=\beta_s
=\frac14\sqrt{\frac{\check P\check\Omega^2}{2\check\rho^2}}
\left[
\del_q\left(\log\check\rho^2\right)
+i\del_p\left(\log\frac{\check P}{\check\Omega^2\check\rho^2}\right)
\right]\,,\\
&\epsilon_s=\gamma_s
=\frac14\sqrt{\frac{\check Q\check\Omega^2}{2\check\rho^2}}
\left[
\del_q\left(\log\frac{\check Q}{\check\Omega^2\check\rho^2}\right)
+i\del_p\left(\log\check\rho^2\right)
\right]\,.
\label{spin coefficients}
\end{align}
For $\Lambda=0$, the Einstein equations take the form
\begin{align}
R=0\,,\qquad
\Phi_{ij}=2\phi_i\bar\phi_j\,,\qquad i,j=0,1,2\,.
\label{Einstein eq}
\end{align}

We first consider the equations for $\Phi_{00}$ and $\Phi_{02}$.
The corresponding Ricci scalars are
\begin{align*}
\Phi_{00}
&=\frac{\check Q\check\Omega}{2\check\rho^2}
\del_q^2\check\Omega\,,
\qquad
\Phi_{02}
=-\frac{\check P\check\Omega}{2\check\rho^2}
\del_p^2\check\Omega\,.
\end{align*}
Substituting the electromagnetic ansatz~\eqref{elemag ansatz} into the Einstein equations for these components gives
\begin{align}
\check Q\del_q^2\check\Omega
=\frac{4\abs{c^\prime}^2}{\check\Omega^3}\check P\check Q
=-\check P\del_p^2\check\Omega\,.
\label{E00 and E02}
\end{align}
Since $\check P$ depends only on $p$ and $\check Q$ only on $q$, these equations imply
\[
\del_q^3(\check\Omega^2)=0\,,\qquad
\del_p^3(\check\Omega^2)=0\,.
\]
Thus, $\check\Omega^2$ is at most quadratic in each of $p$ and $q$.
Substitution into Eq.~\eqref{E00 and E02} further shows that $\check P$ and $\check Q$ are at most quartic in their respective arguments~\cite{Ovcharenko:2025cpm}.
We therefore write
\begin{align}
\check P(p)
&=\check a_0+\check a_1p+\check a_2p^2
+\check a_3p^3+\check a_4p^4\,,
\label{p_exp_1}\\
\check Q(q)
&=\check b_0+\check b_1q+\check b_2q^2
+\check b_3q^3+\check b_4q^4\,,
\label{q_exp_1}\\
\check\Omega^2(q,p)
&=\left(\check c_{00}+\check c_{01}q+\check c_{02}q^2\right)
+\left(\check c_{10}+\check c_{11}q+\check c_{12}q^2\right)p
+\left(\check c_{20}+\check c_{21}q
+\check c_{22}q^2\right)p^2\,,
\label{omega_exp_1}
\end{align}
where all coefficients $\check a_i$, $\check b_i$, and $\check c_{ij}$ are real constants.
The relations among these coefficients will be given below.

\paragraph{Scaling symmetry of the ansatz}

Before determining these coefficient relations, we record the scaling freedom of the metric and electromagnetic ans\"atze, Eqs.~\eqref{CtoC metric} and \eqref{elemag ansatz}.
For a constant $C>0$, introduce new coordinates by
\begin{align}
\eta=C^{-1}\widetilde\eta\,,\qquad
q=C\widetilde q\,,\qquad
p=C\widetilde p\,,\qquad
\sigma=C^{-3}\widetilde\sigma\,.
\label{scaling sym1}
\end{align}
The metric retains its form with the redefined functions
\begin{align}
\widetilde P(\widetilde p)
&=C^{-4}\check P(C\widetilde p)\,,
&
\widetilde Q(\widetilde q)
&=C^{-4}\check Q(C\widetilde q)\,,
\nn
\widetilde\Omega^2(\widetilde q,\widetilde p)
&=\check\Omega^2(C\widetilde q,C\widetilde p)\,,
&
{\widetilde\rho}^2(\widetilde q,\widetilde p)
&={\widetilde q}^2+{\widetilde p}^2\,.
\end{align}
Their polynomial coefficients are therefore
\begin{align}
\widetilde a_i=C^{i-4}\check a_i\,,\qquad
\widetilde b_j=C^{j-4}\check b_j\,,\qquad
\widetilde c_{ij}=C^{i+j}\check c_{ij}\,.
\end{align}
The electromagnetic ansatz retains its form with
\begin{align}
{\widetilde c}^\prime=C^3c^\prime\,.
\end{align}
For $\check c_{11}\neq0$, we may replace $p$ by $-p$, if necessary, so that $\check c_{11}<0$.
This preserves the metric form and, with the corresponding null tetrad and $c^\prime\mapsto\bar c^\prime$, the electromagnetic ansatz.
Choosing $C^2=-2/\check c_{11}$ then sets $\widetilde c_{11}=-2$.
We henceforth restore the original notation for the rescaled quantities.

\paragraph{Scaling symmetry of the Einstein--Maxwell equations}

In addition to the coordinate scaling described above, the source-free Einstein--Maxwell equations with $\Lambda=0$ admit a constant scaling symmetry.
Under
\begin{align}
g_{ab}\longmapsto S^{-2}g_{ab}\,,\qquad S>0\,,
\end{align}
the normalized null-tetrad vectors are multiplied by $S$, and hence
\begin{align}
R_{ab}\longmapsto R_{ab}\,,\qquad
R\longmapsto S^2R\,,\qquad
\Phi_{ij}\longmapsto S^2\Phi_{ij}\,.
\end{align}
The Einstein equations $\Phi_{ij}=2\phi_i\bar\phi_j$ are therefore preserved if
\begin{align}
F_{ab}\longmapsto S^{-1}F_{ab}\,,\qquad
\phi_i\longmapsto S\phi_i\,,\qquad i=0,1,2\,.
\end{align}
The source-free Maxwell equations are also preserved, since the Hodge dual acting on two-forms is unchanged by a conformal rescaling in four dimensions and $S$ is constant.
Thus,
\begin{align}
(g,F)\longmapsto(S^{-2}g,S^{-1}F)
\end{align}
maps an Einstein--Maxwell solution to another solution.

For the metric and electromagnetic ansatz used here, this transformation is implemented by
\begin{align}
\check\Omega\longmapsto S\check\Omega\,,\qquad
c^\prime\longmapsto S^2c^\prime\,,
\end{align}
with the coordinates and $\check P,\check Q$ held fixed \cite{Ovcharenko:2025cpm}.
Consequently, all coefficients of $\check\Omega^2$ transform as $\check c_{ij}\longmapsto S^2\check c_{ij}$.
In what follows, we restrict attention to the sector with $\check c_{00}>0$ and $\check c_{11}\neq 0$.
Within this sector, we combine the constant scaling with the coordinate freedom described above and henceforth impose the normalization
\begin{align}
\check c_{00}=1\,,\qquad \check c_{11}=-2\,.
\end{align}

At this stage, the coordinates, parameters, and line element are dimensionless.
We regard solutions related by the constant scaling above as equivalent up to an overall choice of length scale.
In Sec.~\ref{Sec. family}, we introduce an arbitrary reference length $L$ to restore dimensions and use the resulting dimensionful parametrization to examine further limiting spacetimes.

\subsubsection{Solution with $\check c_{00}=1$ and $\check c_{11}=-2$}

With this normalization, the remaining Einstein--Maxwell equations require~\cite{Ovcharenko:2025cpm}
\begin{align}
    \check c_{20}=-\check c_{02}\,.
    \label{ccs constraint}
\end{align}
The coefficients of $\check P(p)$ and $\check Q(q)$ are then determined by the coefficients of $\check\Omega^2$ as follows:
\begin{align}
    \begin{dcases}
        \check a_0
        =\frac{4\check c_{02}-\check c_{01}^2}
        {16\abs{c^\prime}^2}\\
        \check a_1
        =\frac{\check c_{01}+\check c_{10}\check c_{02}
        +\check c_{12}}{4\abs{c^\prime}^2}\\
        \check a_2
        =-\frac{1+\check c_{02}^2-\check c_{10}\check c_{12}
        +\frac12\check c_{01}\check c_{21}-\check c_{22}}
        {4\abs{c^\prime}^2}\\
        \check a_3
        =\frac{\check c_{10}\check c_{22}
        -\check c_{02}\check c_{12}+\check c_{21}}
        {4\abs{c^\prime}^2}\\
        \check a_4
        =-\frac{\check c_{21}^2+4\check c_{02}\check c_{22}}
        {16\abs{c^\prime}^2}
    \end{dcases}
    \,,\qquad
    \begin{dcases}
        \check b_0
        =\frac{4\check c_{02}+\check c_{10}^2}
        {16\abs{c^\prime}^2}\\
        \check b_1
        =-\frac{\check c_{10}-\check c_{01}\check c_{02}
        +\check c_{21}}{4\abs{c^\prime}^2}\\
        \check b_2
        =\frac{1+\check c_{02}^2+\frac12\check c_{10}\check c_{12}
        -\check c_{01}\check c_{21}-\check c_{22}}
        {4\abs{c^\prime}^2}\\
        \check b_3
        =-\frac{\check c_{01}\check c_{22}
        +\check c_{02}\check c_{21}+\check c_{12}}
        {4\abs{c^\prime}^2}\\
        \check b_4
        =\frac{\check c_{12}^2-4\check c_{02}\check c_{22}}
        {16\abs{c^\prime}^2}
    \end{dcases}
    \,.
    \label{ac-bcs}
\end{align}
The metric functions therefore depend on the six real coefficients 
$(\check c_{01},\check c_{10},\check c_{02},\check c_{12},\check c_{21},\check c_{22})$ and on $\abs{c^\prime}$.
The phase of $c^\prime$ enters in the electromagnetic field.

The remaining Maxwell scalar is
\begin{align}
    \phi_1
    =\frac{1}{4\bar c^\prime}
    \frac{\check\Omega^2}{(q+ip)^2}
    \left[
        \check v\,\del_p
        \left(\frac{\del_q\check\Omega}{p}\right)
        -i\check u\,\del_q
        \left(\frac{\del_p\check\Omega}{q}\right)
    \right]\,.
    \label{phi_1 sol}
\end{align}
A corresponding complex electromagnetic potential, $\bm{\mathcal{A}}=\mathcal A_\eta d\eta +\mathcal A_\sigma d\sigma$, is given by
\begin{align}
    \mathcal A_\eta
    &=\frac{1}{4\bar c^\prime}
    \frac{\del_q\check\Omega-i\del_p\check\Omega}{q+ip}\,,
    \label{Aeta}\\
    \mathcal A_\sigma
    &=\frac{1}{4\bar c^\prime}
    \left[
        \frac{\check u\,\del_q\check\Omega
        +i\check v\,\del_p\check\Omega}{q+ip}
        -\check\Omega
    \right]\,.
    \label{Asigma}
\end{align}
The real potential introduced in Eq.~\eqref{gauge potential} is obtained, up to gauge freedom, as
\begin{align}
    \bm{A}=2\,\operatorname{Re}\bm{\mathcal{A}}\,,
\end{align}
and generates the real Faraday tensor $\bm{F}=d \bm{A}$.

This gives the original Ovcharenko--Podolsk\'y solution in the conformal-to-Carter form.
In the next subsection, we express the same solution in the Pleba\'nski--Demia\'nski gauge without imposing any additional restriction on the solution family.

\subsection{Solution of the Einstein--Maxwell equations in the Pleba\'nski--Demia\'nski gauge}

To relate the Ovcharenko--Podolsk\'y solution to the Pleba\'nski--Demia\'nski family, we shift the two non-Killing coordinates~\cite{Ovcharenko:2025cpm}.
Writing the old coordinates in terms of the new ones as
\begin{align}
    p_{\mathrm{old}}=p+p_0\,,\qquad
    q_{\mathrm{old}}=q+q_0\,,
    \label{pq shift gauge}
\end{align}
with $\eta$ and $\sigma$ unchanged, we define the transformed metric functions by
\begin{align}
    \hat P(p)\coloneqq\check P(p+p_0)\,,\qquad
    \hat Q(q)\coloneqq\check Q(q+q_0)\,,\qquad 
    \hat\Omega^2(q,p)\coloneqq\check\Omega^2(q+q_0,p+p_0)\,.
\end{align}
The shifts leave their polynomial degrees unchanged, so that
\begin{align}
    \hat P(p)=\sum_{i=0}^{4}\hat a_i p^i\,,\qquad
    \hat Q(q)=\sum_{j=0}^{4}\hat b_j q^j\,,\qquad
    \hat\Omega^2(q,p)=\sum_{i=0}^{2}\sum_{j=0}^{2}
    \hat c_{ij}p^i q^j\,.
    \label{PQ Omega ansatz in PD gauge}
\end{align}
The hatted coefficients are determined by expanding the shifted functions in powers of $p$ and $q$.
Choosing real constants $p_0$ and $q_0$ so that
\begin{align}
    \hat c_{10}=0=\hat c_{01}\,,
\end{align}
eliminates the linear terms in $\hat\Omega^2$.
We refer to this choice as the Pleba\'nski--Demia\'nski gauge.
The other metric functions become
\begin{align}
    \hat u(q)&\coloneqq(q+q_0)^2\,,\qquad
    \hat v(p)\coloneqq(p+p_0)^2\,,\nn
    \hat\rho^2(q,p)&=\hat u(q)+\hat v(p)
    =(q+q_0)^2+(p+p_0)^2\,.
    \label{c01=0=c01 gauge}
\end{align}
The metric retains the form of Eq.~\eqref{CtoC metric}, with all checked functions replaced by their hatted counterparts.
The two linear coefficients eliminated from $\hat\Omega^2$ are replaced by the shifts $p_0$ and $q_0$ in $\hat u$, $\hat v$, and $\hat\rho^2$.
This coordinate transformation changes the representation of the solution without imposing an additional restriction on the solution family.

In this gauge, the electromagnetic ansatz becomes
\begin{align}
    \phi_0=\phi_2
    =c^\prime\frac{\sqrt{\hat P\hat Q}}
    {\hat\Omega\left[q+q_0+i(p+p_0)\right]}\,.
    \label{elemag ansatz in PD gauge}
\end{align}

\subsubsection{Solution with $\hat c_{00}=1$ and $\hat c_{11}=-2$}

The coordinate shifts do not in general preserve the normalization of $\check c_{00}$ and $\check c_{11}$.
In the sector with $\hat c_{00}>0$ and $\hat c_{11}\neq0$, we again use the scaling freedoms described above to impose
\begin{align}
    \hat c_{00}=1\,,\qquad \hat c_{11}=-2\,.
\end{align}
The same symbols are used for the rescaled coefficients, shift parameters, and electromagnetic constant $c^\prime$.
With this normalization, the remaining Einstein--Maxwell equations give the constraint
\begin{align}
    (p_0^2+q_0^2)\hat c_{22}
    =-(\hat c_{02}+\hat c_{20})
    +\hat c_{12}p_0+\hat c_{21}q_0
    \label{p0q0 constraint}
\end{align}
and determine the coefficients of $\hat P$ and $\hat Q$ as
\begin{align}
    \begin{dcases}
        \hat a_0=\frac{\hat c_{02}}{4\abs{c^\prime}^2}\\
        \hat a_1=\frac{\hat c_{12}}{4\abs{c^\prime}^2}\\
        \hat a_2=\frac{\hat c_{22}+\hat c_{20}\hat c_{02}-1}
        {4\abs{c^\prime}^2}\\
        \hat a_3=\frac{\hat c_{12}\hat c_{20}+\hat c_{21}}
        {4\abs{c^\prime}^2}\\
        \hat a_4=\frac{4\hat c_{20}\hat c_{22}-\hat c_{21}^2}
        {16\abs{c^\prime}^2}
    \end{dcases}
    \,,\qquad
    \begin{dcases}
        \hat b_0=-\frac{\hat c_{20}}{4\abs{c^\prime}^2}\\
        \hat b_1=-\frac{\hat c_{21}}{4\abs{c^\prime}^2}\\
        \hat b_2=-\hat a_2
        =-\frac{\hat c_{22}+\hat c_{20}\hat c_{02}-1}
        {4\abs{c^\prime}^2}\\
        \hat b_3=-\frac{\hat c_{21}\hat c_{02}+\hat c_{12}}
        {4\abs{c^\prime}^2}\\
        \hat b_4=\frac{\hat c_{12}^2-4\hat c_{02}\hat c_{22}}
        {16\abs{c^\prime}^2}
    \end{dcases}
    \,.
    \label{solution in PD gauge}
\end{align}
The general coefficient relations above are not displayed explicitly in Ref.~\cite{Ovcharenko:2025cpm}, where the subsequent parametrization specializes to $\hat c_{20}=-\hat c_{02}$.
The essential point is that the relation inherited from the original form is Eq.~\eqref{p0q0 constraint}, rather than $\hat c_{20}=-\hat c_{02}$.
Thus, $\hat c_{02}$ and $\hat c_{20}$ need not be opposite.
They are constrained jointly with $p_0$, $q_0$, and $\hat c_{22}$.
For $p_0=q_0=0$, Eq.~\eqref{p0q0 constraint} reduces to $\hat c_{20}=-\hat c_{02}$, as expected.

At fixed $\abs{c^\prime}$, this representation involves the five real coefficients
$(\hat c_{02},\hat c_{20},\hat c_{12},\hat c_{21},\hat c_{22})$
and the two shifts $(p_0,q_0)$, subject to Eq.~\eqref{p0q0 constraint}.
It therefore contains six real parameters generically.
The functions $\hat P$, $\hat Q$, and $\hat\Omega^2$ depend only on these five coefficients and $\abs{c^\prime}$, while the shifts also enter $\hat u$, $\hat v$, and $\hat\rho^2$.
When $\hat c_{22}\neq0$, the constraint can be written as
\begin{align}
    &\left(p_0-\hat{\msf A}\right)^2
    +\left(q_0-\hat{\msf B}\right)^2
    =\hat{\msf R}^{\,2}\,,\nn
    &\hat{\msf A}=\frac{\hat c_{12}}{2\hat c_{22}}\,,\qquad
    \hat{\msf B}=\frac{\hat c_{21}}{2\hat c_{22}}\,,\qquad
    \hat{\msf R}^{\,2}
    =\frac{\hat c_{12}^2+\hat c_{21}^2
    -4\hat c_{22}(\hat c_{02}+\hat c_{20})}{4\hat c_{22}^2}\,.
    \label{p0q0 constraint another form}
\end{align}
For fixed values of the five coefficients and $\hat{\msf R}^{\,2}>0$, the remaining freedom is the position of $(p_0,q_0)$ on this circle.
Real shifts require
\begin{align}
    \hat c_{12}^2+\hat c_{21}^2
    -4\hat c_{22}(\hat c_{02}+\hat c_{20})\geq0\,.
    \label{reality-condition-general-circle}
\end{align}
When equality holds, the circle degenerates to a single point.
In what follows, we assume this inequality and $\hat{c}_{22}\neq 0$.
A complete determination of the admissible parameter and coordinate domains must also account for Lorentzian signature and the positivity of the relevant metric functions.
This analysis is left for future work.

The remaining Maxwell scalar is
\begin{align}
    \phi_1
    &=\frac{1}{4\bar c^\prime}
    \frac{\hat\Omega^2}{\left[q+q_0+i(p+p_0)\right]^2}
    \left[
        \hat v\,\del_p
        \left(\frac{\del_q\hat\Omega}{p+p_0}\right)
        -i\hat u\,\del_q
        \left(\frac{\del_p\hat\Omega}{q+q_0}\right)
    \right]\,.
    \label{phi_1 sol in PD gauge}
\end{align}
A corresponding complex electromagnetic potential,
$\bm{\mathcal A}=\mathcal A_\eta d\eta
+\mathcal A_\sigma d\sigma$, is given by
\begin{align}
    \mathcal A_\eta
    &=\frac{1}{4\bar c^\prime}
    \frac{\del_q\hat\Omega-i\del_p\hat\Omega}
    {q+q_0+i(p+p_0)}\,,
    \label{Aeta in PD gauge}\\
    \mathcal A_\sigma
    &=\frac{1}{4\bar c^\prime}
    \left[
        \frac{\hat u\,\del_q\hat\Omega
        +i\hat v\,\del_p\hat\Omega}
        {q+q_0+i(p+p_0)}
        -\hat\Omega
    \right]\,.
    \label{Asigma in PD gauge}
\end{align}
As before, the real potential is $\bm{A}=2\operatorname{Re}\bm{\mathcal{A}}$, up to gauge freedom.

In Sec.~\ref{Sec. OP-PD}, we first review the earlier Pleba\'nski--Demia\'nski-adapted parametrization, which imposes $\hat c_{20}=-\hat c_{02}$, and then construct a parametrization that retains the independent coefficient freedom identified here.


\section{Ovcharenko--Podolsk\'y spacetime in the Pleba\'nski--Demia\'nski form}
\label{Sec. OP-PD}

Our aim is to establish a correspondence between the coefficients $\set{\hat c_{ij}}$ of the Ovcharenko--Podolsk\'y solution and the parameters
$\set{\hat k,\hat n,\hat\ep,\hat m,\hat s,\hat\lambda}$ of the Pleba\'nski--Demia\'nski family.
This correspondence allows us to draw on the known properties of the Pleba\'nski--Demia\'nski family and helps clarify the physical interpretation of the Ovcharenko--Podolsk\'y parameters after transformation to the Griffiths--Podolsk\'y form.
The electromagnetic field is doubly aligned in the Pleba\'nski--Demia\'nski solution and fully non-aligned in the Ovcharenko--Podolsk\'y solution.
We therefore require the parametrization to admit a smooth $\abs{c^\prime}\to 0$ limit in which both the metric and electromagnetic field reduce to those of the Pleba\'nski--Demia\'nski solution with $\hat\lambda=0$.
In this limit, the Maxwell scalars $\phi_0$ and $\phi_2$ vanish, while $\phi_1$ need not vanish.

\subsection{Relation to Pleba\'nski--Demia\'nski parameters}

The Pleba\'nski--Demia\'nski spacetime is a special case of the conformal-to-Carter class, with $\hat u(q)=q^2$, $\hat v(p)=p^2$, and
\begin{align}
   \hat P_{\mr{PD}}(p)
   &=\hat k+2\hat n p-\hat\ep p^2+2\hat m p^3
   -(\hat k+\hat s+\hat\lambda)p^4\,,
   \label{P of PD}\\
   \hat Q_{\mr{PD}}(q)
   &=\hat k+\hat s-2\hat m q+\hat\ep q^2-2\hat n q^3
   -(\hat k+\hat\lambda)q^4\,,
   \label{Q of PD}\\
   \hat\Omega_{\mr{PD}}(q,p)
   &=1-pq\,.
   \label{Omega of PD}
\end{align}
Here, we define $\hat s\coloneqq\hat e^2+\hat g^2$, where $\hat e$ and $\hat g$ are the electric and magnetic charge parameters, respectively.
The metric depends on these charge parameters only through $\hat s$.
We also define $\hat\lambda\coloneqq\hat\Lambda/3$, with $\hat\Lambda$ denoting the cosmological constant in the dimensionless variables used here.

Since $\hat\Omega_{\mr{PD}}^2=1-2pq+p^2q^2$, this form has $\hat c_{00}=1$ and $\hat c_{11}=-2$.
For the Ovcharenko--Podolsk\'y solution, we continue to use the Pleba\'nski--Demia\'nski gauge $\hat c_{10}=\hat c_{01}=0$ with this normalization, as in Sec.~\ref{Sec. review}.

\subsubsection{Ovcharenko--Podolsk\'y parametrization}

Before considering the full parameter space, we first review the correspondence established by Ovcharenko and Podolsk\'y~\cite{Ovcharenko:2025cpm} for the restricted case $\hat c_{20}=-\hat c_{02}$.
Under this restriction, one has $\hat a_0=\hat b_0$, which naturally corresponds to setting $\hat s=0$ in the Pleba\'nski--Demia\'nski parametrization.
For the non-null, fully non-aligned solutions considered here, which satisfy the generalized Goldberg--Sachs conditions, Van den Bergh's theorem requires $\hat\lambda=0$~\cite{VandenBergh:2016ooq}.
At fixed $c^\prime$, the four independent coefficients $(\hat c_{02},\hat c_{12},\hat c_{21},\hat c_{22})$ are reparametrized in terms of
$(\hat k_\mr{OP},\hat\ep_\mr{OP},\hat m_\mr{OP},\hat n_\mr{OP})$.
The parametrization is chosen so that, as $\abs{c^\prime}\to 0$ with these four parameters held fixed, the solution reduces to the vacuum Pleba\'nski--Demia\'nski spacetime with $\hat s=\hat\lambda=0$.
This limit provides a guide to the physical interpretation of the Ovcharenko--Podolsk\'y parameters.
Introducing the auxiliary quantities $(\delta\hat k_1,\delta\hat k_2,\delta\hat m_\mr{OP}, \delta\hat n_\mr{OP})$, we write their parametrization in the following form:
\begin{align}
    \begin{dcases}
        \hat{a}_0 \equiv \hat{k}_\mr{OP}\\
        \hat{a}_1 \equiv 2 \hat{n}_\mr{OP} -\abs{c^\prime}^2 \delta \hat{n}_\mr{OP}\\
        \hat{a}_2 \equiv -\hat{\ep}_\mr{OP}\\
        \hat{a}_3 \equiv 2\hat{m}_\mr{OP} -\abs{c^\prime}^2 \delta \hat{m}_\mr{OP}\\
        \hat{a}_4 \equiv -\hat{k}_\mr{OP} +\abs{c^\prime}^2 \delta \hat{k}_1
    \end{dcases}\,, \qquad
    \begin{dcases}
        \hat{b}_0 \equiv \hat{k}_\mr{OP}\\
        \hat{b}_1 \equiv -2 \hat{m}_\mr{OP} -\abs{c^\prime}^2 \delta \hat{m}_\mr{OP}\\
        \hat{b}_2 =-\hat{a}_2 = \hat{\ep}_\mr{OP}\\
        \hat{b}_3 \equiv -2\hat{n}_\mr{OP} -\abs{c^\prime}^2 \delta \hat{n}_\mr{OP}\\
        \hat{b}_4 \equiv -\hat{k}_\mr{OP} +\abs{c^\prime}^2 \delta \hat{k}_2
    \end{dcases}\,.
    \label{OP parametrization}
\end{align}
Define $\hat C\coloneqq2\abs{c^\prime}^2\hat k_{\mr{OP}}$.
For $\hat c_{22}\neq0$, the constraint on $p_0$ and $q_0$ takes the form
\begin{align}
    \left[p_0-\hat{\msf{A}}_\mr{OP}\right]^2+\left[q_0-\hat{\msf{B}}_\mr{OP}\right]^2={\hat{\msf{R}}_\mr{OP}}^2\,,
\end{align}
where
\begin{align}
   & \hat{\msf{A}}_\mr{OP} = \frac{2\hat{C}(\hat{n}_\mr{OP}-\hat{C} \hat{m}_\mr{OP})}{(1+{\hat{C}}^2)((1+4{\hat{C}}^2)\hat{k}_\mr{OP} -2\hat{C} \hat{\ep}_\mr{OP})}\,,\\
   & \hat{\msf{B}}_\mr{OP} = \frac{2\hat{C}(\hat{m}_\mr{OP}+\hat{C} \hat{n}_\mr{OP})}{(1+{\hat{C}}^2)((1+4{\hat{C}}^2)\hat{k}_\mr{OP} -2\hat{C} \hat{\ep}_\mr{OP})}\,,\\
   & \hat{\msf{R}}_\mr{OP} = \frac{2\hat{C}\sqrt{{\hat{m}_\mr{OP}}^2+{\hat{n}_\mr{OP}}^2}}{\sqrt{(1+{\hat{C}}^2)} [(1+4{\hat{C}}^2)\hat{k}_\mr{OP} -2\hat{C} \hat{\ep}_\mr{OP}]}\,.
   \label{OP const R}
\end{align}
For $\hat{\msf R}_\mr{OP}\neq0$, the constraint leaves one angular degree of freedom.
We parametrize the shifts by $\beta^\mr{OP}$ as
\begin{align}
    &p_0 = \hat{\msf{A}}_\mr{OP} +\hat{\msf{R}}_\mr{OP} \sin \beta^\mr{OP}\\
    &q_0 = \hat{\msf{B}}_\mr{OP} +\hat{\msf{R}}_\mr{OP} \cos \beta^\mr{OP}
\end{align}
Furthermore, since ${\hat{\msf{A}}_\mr{OP}}^2+{\hat{\msf{B}}_\mr{OP}}^2={\hat{\msf{R}}_\mr{OP}}^2$, there exists a particular value $\beta^\mr{OP}_0$ for which $(q_0,p_0)=(0,0)$:
\begin{align}
   \begin{dcases}
    \sin \beta^\mr{OP}_0=-\frac{\hat{\msf{A}}_\mr{OP}}{\hat{\msf{R}}_\mr{OP}}=-\frac{\hat{n}_\mr{OP}-{\hat{C}}\hat{m}_\mr{OP}}{\sqrt{1+{\hat{C}}^2}\sqrt{{\hat{m}_\mr{OP}}^2+{\hat{n}_\mr{OP}}^2}}\\
    \cos \beta^\mr{OP}_0=-\frac{\hat{\msf{B}}_\mr{OP}}{\hat{\msf{R}}_\mr{OP}}=-\frac{\hat{m}_\mr{OP}+{\hat{C}}\hat{n}_\mr{OP}}{\sqrt{1+{\hat{C}}^2}\sqrt{{\hat{m}_\mr{OP}}^2+{\hat{n}_\mr{OP}}^2}}
   \end{dcases} \,,
   \text{i.e.,} \,\quad 
   \tan \beta^\mr{OP}_0 = \frac{\hat{n}_\mr{OP}-\hat{C}\hat{m}_\mr{OP}}{\hat{m}_\mr{OP}+\hat{C} \hat{n}_\mr{OP}}\,.
   \label{OP special angle}
\end{align}
The parameter $\beta^\mr{OP}$ does not appear in $\hat\Omega^2$, $\hat P$, or $\hat Q$.
It enters the metric through the shifts $p_0$ and $q_0$ in $\hat u$, $\hat v$, and thus $\hat\rho^2$.
In particular,
\begin{align}
    \hat\rho^2
    =(q+q_0)^2+(p+p_0)^2
    =
    \left(q+\hat{\msf B}_\mr{OP}
    +\hat{\msf R}_\mr{OP}\cos\beta^\mr{OP}\right)^2
    +
    \left(p+\hat{\msf A}_\mr{OP}
    +\hat{\msf R}_\mr{OP}\sin\beta^\mr{OP}\right)^2\,.
\end{align}
This completes our review of the Ovcharenko--Podolsk\'y parametrization.

\subsubsection{Our parametrization}

We now remove the restriction $\hat c_{20}=-\hat c_{02}$ and express $(\hat c_{02},\hat c_{20},\hat c_{12},\hat c_{21},\hat c_{22})$ in terms of the five Pleba\'nski--Demia\'nski-type parameters $(\hat k,\hat\ep,\hat m,\hat n,\hat s)$.
As in the preceding subsection, $\hat\lambda=0$.
To this end, we introduce the auxiliary quantities $(\delta\hat k,\delta\hat m,\delta\hat n,\delta\hat s)$ and choose the following coefficient parametrization, whose $\abs{c^\prime}\to0$ limit has the Pleba\'nski--Demia\'nski form:
\begin{align}
      \begin{dcases}
        \hat{a}_0 \equiv \hat{k}\\
        \hat{a}_1 \equiv 2 \hat{n} \\
        \hat{a}_2 \equiv -\hat{\ep}\\
        \hat{a}_3 \equiv 2\hat{m} + \abs{c^\prime}^2 \delta \hat{m}\\
        \hat{a}_4 \equiv -(\hat{k}+{\hat{s}}) +\abs{c^\prime}^2 \delta \hat{s}
    \end{dcases}\,,\qquad
     \begin{dcases}
        \hat{b}_0 \equiv \hat{k}+\hat{s}\\
        \hat{b}_1 \equiv -2 \hat{m} \\
        \hat{b}_2 = -\hat{a}_2 = \hat{\ep}\\
        \hat{b}_3 \equiv -2\hat{n} + \abs{c^\prime}^2 \delta \hat{n}\\
        \hat{b}_4 \equiv -\hat{k} +\abs{c^\prime}^2 \delta \hat{k}
    \end{dcases}\,.
    \label{F parametrization}
\end{align}
Comparing this with Eq.~\eqref{solution in PD gauge}, we obtain
\begin{align}
      \begin{dcases}
     \hat{c}_{02}= 4\abs{c^\prime}^2 \hat{k}\\
     \hat{c}_{20}= -4\abs{c^\prime}^2 (\hat{k}+\hat{s})\\
     \hat{c}_{12}=8\abs{c^\prime}^2 \hat{n}\\
     \hat{c}_{21}=8\abs{c^\prime}^2 \hat{m}\\
     \hat{c}_{22}=1-4\abs{c^\prime}^2\left[\hat{\ep}-4\abs{c^\prime}^2\hat{k}(\hat{k}+\hat{s}) \right]\\
    \end{dcases}\,,
    \label{general OP-PD cs}
\end{align}
{\small
\begin{align}
    \begin{dcases}
        \hat{a}_0=\hat{k}\\
        \hat{a}_1=2\hat{n}\\
        \hat{a}_2=-\hat{\ep}\\
        \hat{a}_3=2\hat{m}-8\abs{c^\prime}^2(\hat{k}+\hat{s})\hat{n}\\
        \hat{a}_4=-(\hat{k}+\hat{s})+4\abs{c^\prime}^2\left[\hat{\ep}(\hat{k}+\hat{s})-\hat{m}^2 -4\abs{c^\prime}^2\hat{k}(\hat{k}+\hat{s})^2\right]
    \end{dcases}\,,\quad
\begin{dcases}
        \hat{b}_0=\hat{k}+\hat{s}\\
        \hat{b}_1=-2\hat{m}\\
        \hat{b}_2=\hat{\ep}\\
        \hat{b}_3=-2\hat{n}-8\abs{c^\prime}^2\hat{k}\hat{m}\\
        \hat{b}_4=-\hat{k}+4\abs{c^\prime}^2\left[\hat{\ep}\hat{k}+\hat{n}^2 -4\abs{c^\prime}^2\hat{k}^2(\hat{k}+\hat{s})\right]
    \end{dcases}\,.
    \label{general OP-PD as and bs}
\end{align}}

For fixed $c^\prime\neq 0$, these relations give an invertible change of parameters.
In particular,
\begin{align}
    \hat s=-\frac{\hat c_{02}+\hat c_{20}}{4\abs{c^\prime}^2}\,,
\end{align}
so the earlier restriction $\hat c_{20}=-\hat c_{02}$ corresponds precisely to $\hat s=0$.
At this stage, $\hat s$ is a real parameter subject to the reality
and coordinate-domain conditions. Its interpretation as a squared
charge will be established in the aligned limit below.

For $\hat c_{22}\neq0$, the quantities $\hat{\msf A}, \hat{\msf B}$, and $\hat{\msf R}$ appearing in the circle constraint \eqref{p0q0 constraint another form} can be written as
\begin{align}
   & \hat{\msf{A}}= \frac{4\abs{c^\prime}^2\hat{n}}{1-4\abs{c^\prime}^2\left[\hat{\ep}-4\abs{c^\prime}^2\hat{k}(\hat{k}+\hat{s})\right]}\,,\\
   & \hat{\msf{B}}= \frac{4\abs{c^\prime}^2\hat{m}}{1-4\abs{c^\prime}^2\left[\hat{\ep}-4\abs{c^\prime}^2\hat{k}(\hat{k}+\hat{s})\right]}\,,\\
   & \hat{\msf{R}}= \frac{2\abs{c^\prime}\sqrt{\hat{s} +4\abs{c^\prime}^2 \left[\hat{m}^2 +\hat{n}^2 -\hat{\ep}\hat{s}+4\abs{c^\prime}^2 \hat{s}\hat{k}(\hat{k}+\hat{s}) \right]     } }{1-4\abs{c^\prime}^2\left[\hat{\ep}-4\abs{c^\prime}^2\hat{k}(\hat{k}+\hat{s})\right]}
\,. \label{p0q0 Constraint in PD parameter}
\end{align}
Reality condition of $\hat{\msf{R}}$ requires
\begin{align}
 &\hat s+4\abs{c^\prime}^2
 \left[\hat m^2+\hat n^2-\hat\ep\hat s
 +4\abs{c^\prime}^2\hat s\hat k(\hat k+\hat s)\right]\geq0\,,
 \label{reality-condition-PD-parameters}
\end{align}
which is equivalent to
\begin{align}
    \hat s\hat c_{22}+4\abs{c^\prime}^2
    (\hat m^2+\hat n^2)\geq0\,.
\end{align}
In the preceding Ovcharenko--Podolsk\'y parametrization, the square root in Eq.~(\ref{OP const R}) involves $\hat m_\mr{OP}^{\,2}+\hat n_\mr{OP}^{\,2}$, which is nonnegative for real parameters.
By contrast, Eq.~\eqref{reality-condition-PD-parameters} must be imposed for general values of the five real parameters used here.

The shifts $p_0$ and $q_0$ can be parametrized by a single angle $\hat\beta$ as
\begin{align}
    &p_0 = \hat{\msf{A}} +\hat{\msf{R}} \sin \hat{\beta}\,,\qquad 
    q_0 = \hat{\msf{B}} +\hat{\msf{R}} \cos \hat{\beta}\,.
    \label{p0q0 representation}
\end{align}
Unlike in Ovcharenko--Podolsk\'y parametrization, one now has
\begin{align}
    \hat{\msf R}^{\,2}-\hat{\msf A}^{\,2}-\hat{\msf B}^{\,2}
    =\frac{4\abs{c^\prime}^2\hat s}{\hat c_{22}}\,.
    \label{non beta0 in general}
\end{align}
Thus, for $c^\prime\neq0$ and $\hat s\neq0$, the constraint circle does not pass through the origin, and no value of $\hat\beta$ gives $p_0=q_0=0$.\footnote{
For $\hat s=0$ and $\hat m^2+\hat n^2\neq0$, the choice
\begin{align}
    \sin\hat\beta_0
    =-\frac{\hat n}{\sqrt{\hat m^2+\hat n^2}}\,,\qquad 
    \cos\hat\beta_0
    =-\frac{\hat m}{\sqrt{\hat m^2+\hat n^2}}
\end{align}
gives $p_0=q_0=0$.
These expressions are simpler than those in Eq.~\eqref{OP special angle}.
}
The parameter $\hat\beta$ does not appear in $\hat\Omega^2$, $\hat P$, or $\hat Q$.
It enters the metric through $p_0$ and $q_0$ in $\hat u$, $\hat v$, and $\hat\rho^2$. 
In particular,
\begin{align}
    \hat{\rho}^2=(q_0+q)^2+(p_0+p)^2
    = (q+\hat{\msf{B}}+\hat{\msf{R}} \cos \hat{\beta})^2+(p+\hat{\msf{A}} + \hat{\msf{R}} \sin \hat{\beta})^2\,.
\end{align}
The remaining metric functions are given by
\begin{align}
  \hat P(p)&=\hat{k}+2\hat{n}p -\hat{\ep}p^2
+\left[2\hat{m}-8\abs{c^\prime}^2(\hat{k}+\hat{s})\hat{n}\right]p^3\nn
  &\quad-\left[\hat{k}+\hat{s}-4 \abs{c'}^2 \left(\hat{\ep}(\hat{k}+\hat{s})-\hat{m}^2 -4 \abs{c'}^2 \hat{k}(\hat{k}+\hat{s})^2\right)\right]p^4\,,
  \label{P-OP-PD0}\\
  \hat Q(q)&=\hat{k}+\hat{s} -2\hat{m}q +\hat{\ep}q^2
  -\left[2\hat{n}+8\abs{c^\prime}^2\hat{k}\hat{m}\right]q^3\nn
  &\quad-\left[\hat{k}-4 \abs{c'}^2 \left(\hat{\ep}\hat{k}+\hat{n}^2 -4 \abs{c'}^2 \hat{k}^2(\hat{k}+\hat{s})\right)\right]q^4\,,
  \label{Q-OP-PD0}\\
  \hat\Omega^2(q,p)&=(1-pq)^2 + 4 \abs{c^\prime}^2 \left[\hat{k}q^2 -(\hat{k}+\hat{s})p^2\right.
  \left.+2 (\hat{n}q + \hat{m}p)pq 
   -\left(\hat{\ep}-4\abs{c^\prime}^2 \hat{k}(\hat{k}+\hat{s})\right)q^2p^2\right]\,.
   \label{Omega-OP-PD0}
\end{align}

\paragraph{Metric functions in our parametrization}
The general Ovcharenko--Podolsk\'y metric in Pleba\'nski--Demia\'nski gauge is
\begin{align}
        ds^2= \frac{1}{\hat{\Omega}^2(q,p)}\left[-\frac{\hat{Q}(q)}{\hat{\rho}^2(q,p)}\left(d\eta - \hat v(p) d\sigma\right)^2
    +\frac{\hat{\rho}^2(q,p)}{\hat{Q}(q)}dq^2 + \frac{\hat{\rho}^2(q,p)}{\hat{P}(p)}dp^2 
    +\frac{\hat{P}(p)}{\hat{\rho}^2(q,p)} \left(d\eta+\hat u(q) d\sigma\right)^2
    \right]\,.
    \label{OP-PD in F parametrization}
\end{align}
To make the aligned limit explicit, we equivalently write the metric functions as
\begin{align}
  &\hat{P}= \hat{P}_\mr{PD} +\abs{c'}^2 \delta \hat{P}\,,\nn
  &\delta \hat{P} =  -4  p^3 \left[ 2(\hat{k}+\hat{s})\hat{n} -\left(\hat{\ep}(\hat{k}+\hat{s})-\hat{m}^2 -4 \abs{c'}^2 \hat{k}(\hat{k}+\hat{s})^2\right)p\right]\,,
  \label{P-OP-PD0-2}\\
  &\hat{Q}=\hat{Q}_\mr{PD} +\abs{c'}^2 \delta \hat{Q}\,,\nn
  &\delta \hat{Q} = -4 q^3 \left[ 2\hat{k}\hat{m} -\left(\hat{\ep}\hat{k}+\hat{n}^2 -4 \abs{c'}^2 \hat{k}^2(\hat{k}+\hat{s})\right)q\right]\,,
  \label{Q-OP-PD0-2}\\
  &\hat{\Omega}^2 = \hat{\Omega}^2_\mr{PD} +\abs{c'}^2 \delta \hat{\Omega}^2 \,,\nn
  &\delta \hat{\Omega}^2 =  4 \left[\hat{k}q^2 -(\hat{k}+\hat{s})p^2 +2 (\hat{n}q + \hat{m}p)pq 
   -\left(\hat{\ep}-4\abs{c^\prime}^2 \hat{k}(\hat{k}+\hat{s})\right)q^2p^2\right]\,,
   \label{Omega-OP-PD0-2}
\end{align}
where $\hat P_\mr{PD}$, $\hat Q_\mr{PD}$, and $\hat\Omega_\mr{PD}$ are given in \cref{P of PD,Q of PD,Omega of PD} with $\hat\lambda=0$.
The remaining functions are $\hat u=(q+q_0)^2$, $\hat v=(p+p_0)^2$, and
\begin{align}
  \hat{\rho}^2 = (q+q_0)^2+(p+p_0)^2\,,
  \label{rho-OP-PD0}
\end{align}
\begin{align}
 & p_0= \frac{2\abs{c'}}{1-4 \abs{c'}^2 \left[\hat{\ep}-4 \abs{c'}^2\hat{k}(\hat{k}+\hat{s})\right]}
  \left[2\abs{c'}\hat{n}+\sqrt{\hat{s} +4\abs{c^\prime}^2 \left[\hat{m}^2 +\hat{n}^2 -\hat{\ep}\hat{s}+4\abs{c^\prime}^2 \hat{s}\hat{k}(\hat{k}+\hat{s}) \right] }\sin \hat{\beta} \right]
  \,,\label{p0 explicit}\\
   & q_0= \frac{2\abs{c'}}{1-4 \abs{c'}^2 \left[\hat{\ep}-4 \abs{c'}^2\hat{k}(\hat{k}+\hat{s})\right]}
  \left[2\abs{c'}\hat{m}+\sqrt{\hat{s} +4\abs{c^\prime}^2 \left[\hat{m}^2 +\hat{n}^2 -\hat{\ep}\hat{s}+4\abs{c^\prime}^2 \hat{s}\hat{k}(\hat{k}+\hat{s}) \right] } \cos \hat{\beta}\right]
  \,.\label{q0 explicit}
\end{align}

\paragraph{Electromagnetic fields in our parametrization}
Writing $c^\prime=\abs{c^\prime}\e{i\gamma^\prime}$, the Maxwell scalars are
\begin{align}
  &\phi_0= \phi_2 = \abs{c'}\e{i\gamma^\prime}\frac{\sqrt{\hat{P}\hat{Q}}}{\hat{\Omega}\left(q+q_0+i(p+p_0)\right)}
  \label{phi0-PD0}\,,\\
   & \phi_1 =\frac{\e{i\gamma^\prime}}{4\abs{c'}}  \frac{\hat{\Omega}^2}{\left[q+q_0+i(p+p_0)\right]^2}
  \left[(p+p_0)^2 \del_p \left(\frac{\del_q \hat{\Omega}}{p+p_0}\right)
    -i(q+q_0)^2 \del_q \left(\frac{\del_p \hat{\Omega}}{q+q_0}\right)\right]
    \label{phi1-PD0}\,.
\end{align}

\paragraph{The $\abs{c^\prime}\to0$ limit}

We take $\abs{c^\prime}\to0$ at fixed $(\hat k,\hat\ep,\hat m,\hat n,\hat s,\hat\beta,\gamma^\prime)$.
A real aligned limit requires $\hat s\geq 0$, since the radicand in Eq.~\eqref{p0q0 Constraint in PD parameter} tends to $\hat s$.
We therefore assume $\hat s\geq0$ in this paragraph.
The quantities $\hat{\msf A}$, $\hat{\msf B}$, and
$\hat{\msf R}$ tend to zero, so $p_0,q_0\to0$ and the metric
reduces to the Pleba\'nski--Demia\'nski metric with $\hat\lambda=0$.

The non-aligned Maxwell scalars satisfy $\phi_0=\phi_2\to0$.
The expression in square brackets in the aligned Maxwell scalar Eq.~\eqref{phi1-PD0} can be rewritten as
\begin{align}
 &\left[p+p_0-i(q+q_0)\right]\del_p\del_q\hat\Omega
 -\del_q\hat\Omega+i\del_p\hat\Omega
 =-p_0+iq_0+O(\abs{c^\prime}^2)\,.
\end{align}
Here, we have used $\hat\Omega=1-pq+O(\abs{c^\prime}^2)$.
Consequently,
\begin{align}
    \phi_1
    &=\frac{\e{i\gamma^\prime}}{4\abs{c^\prime}}
    \left(\frac{\hat\Omega_\mr{PD}}{q+ip}\right)^2
    (-p_0+iq_0)+O(\abs{c^\prime})\,.
\end{align}
From the expressions of the shift Eqs.~(\ref{p0 explicit}) and (\ref{q0 explicit}), $\abs{c'}\to 0$ limit yields
\begin{align}
    \frac{p_0}{\abs{c^\prime}}
    \to  2\sqrt{\hat s}\sin\hat\beta\,,\qquad 
    \frac{q_0}{\abs{c^\prime}}
    \to 2\sqrt{\hat s}\cos\hat\beta\,,
\end{align}
and hence
\begin{align}
    \phi_1
    \to 
    \frac{i\sqrt{\hat s}\,\e{i(\hat\beta+\gamma^\prime)}}{2}
    \left(\frac{\hat\Omega_\mr{PD}}{q+ip}\right)^2
    =\frac{\hat e+i\hat g}{2}
    \left(\frac{\hat\Omega_\mr{PD}}{q+ip}\right)^2\,,
\end{align}
where
\begin{align}
\hat\gamma\coloneqq\hat\beta+\gamma^\prime+\frac\pi2\,,\qquad 
    \hat e\coloneqq\sqrt{\hat s}\cos\hat\gamma\,,\qquad 
    \hat g\coloneqq\sqrt{\hat s}\sin\hat\gamma\,.
\end{align}
The limiting Maxwell scalars are therefore
\begin{align}
 \phi_0 = \phi_2 \to  \phi_0^\mr{PD}=\phi_2^\mr{PD}=0\,,\qquad
    \phi_1 \to \phi_1^\mr{PD}
    =\frac{\hat e+i\hat g}{2}
    \left(\frac{\hat\Omega_\mr{PD}}{q+ip}\right)^2\,.
    \label{PD elemag field}
\end{align}
Thus, both the metric and electromagnetic field reduce smoothly to those of the Pleba\'nski--Demia\'nski solution with $\hat\lambda=0$ and $\hat e^2+\hat g^2=\hat s$.
For $\hat s=0$, the electromagnetic field vanishes and the limit is vacuum.

\paragraph{Relation to the Ovcharenko--Podolsk\'y parametrization}

For $\hat s=0$, the two parametrizations are related by
\begin{align}
    \hat k=\hat k_\mr{OP}\,,\qquad
    \hat\ep=\hat\ep_\mr{OP}\,,\qquad
    \hat n=\frac{\hat n_\mr{OP}-\hat C\hat m_\mr{OP}}
    {1+\hat C^2}\,,\qquad
    \hat m=\frac{\hat m_\mr{OP}+\hat C\hat n_\mr{OP}}
    {1+\hat C^2}\,,\qquad
    \hat C=2\abs{c^\prime}^2\hat k_\mr{OP}\,.
    \label{parametrization-map-s0}
\end{align}
The inverse relations are
\begin{align}
    \hat m_\mr{OP}=\hat m-\hat C\hat n\,,\qquad
    \hat n_\mr{OP}=\hat n+\hat C\hat m\,.
    \label{inv map for para OP}
\end{align}
This one-to-one correspondence shows that the earlier Ovcharenko--Podolsk\'y parametrization is recovered in our $\hat s=0$ subclass.

\subsection{Ovcharenko--Podolsk\'y spacetime with twist and acceleration parameters}

The representation obtained above describes the twisting sector, for which the imaginary part of the spin coefficient $\mu_s$ is generically nonzero.
To include non-twisting limits, we introduce the acceleration and twist parameters $\alpha$ and $\omega$, following the construction of the improved Pleba\'nski--Demia\'nski form~\cite{Griffiths:2005qp}, denoted by $\mr{PD}_{\alpha\omega}$.
The coordinate transformation is invertible for nonzero $\alpha$ and $\omega$ but may become degenerate in the limits of interest.
We now apply this construction to the Ovcharenko--Podolsk\'y solution in the parametrization obtained above.

\subsubsection{Twisting sector}

The metric is given by \cref{OP-PD in F parametrization,P-OP-PD0-2,Q-OP-PD0-2,Omega-OP-PD0-2,rho-OP-PD0,p0 explicit,q0 explicit}.
For the explicit coordinate transformation below, we take $\alpha>0$ and $\omega>0$.
We introduce the coordinates $(\xi,r,\tau,\phi)$ by
\begin{align}
     p=\sqrt{\alpha\omega}\,\xi\,,\qquad
     q=\sqrt{\frac{\alpha}{\omega}}\,r\,,\qquad
     \eta=\sqrt{\frac{\omega}{\alpha}}\,\tau\,,\qquad
     \sigma=\sqrt{\frac{\omega}{\alpha^3}}\,\phi\,,
     \label{PD0 to PDalphaomega}
\end{align}
and define
\begin{align}
    &\hat P(p(\xi))\eqqcolon\alpha^2P(\xi)\,,\qquad
    \hat Q(q(r))\eqqcolon\left(\frac{\alpha}{\omega}\right)^2Q(r)\,,\qquad
    \hat\Omega^2(q(r),p(\xi))\equiv\Omega^2(r,\xi)\,,\nn
    &\rho^2(r,\xi)\coloneqq(r+r_0)^2+\omega^2(\xi+\xi_0)^2\,,\qquad
    p_0\eqqcolon\sqrt{\alpha\omega}\,\xi_0\,,\qquad
    q_0\eqqcolon\sqrt{\frac{\alpha}{\omega}}\,r_0\,.
    \label{metric functions in OP-PDalphaomega}
\end{align}
The metric then takes the form 
\begin{align}
    ds^2
    &=\frac{1}{\Omega^2(r,\xi)}\biggl[
    -\frac{Q(r)}{\rho^2(r,\xi)}
    \left[d\tau-\omega(\xi+\xi_0)^2d\phi\right]^2
    +\frac{P(\xi)}{\rho^2(r,\xi)}
    \left[\omega d\tau+(r+r_0)^2d\phi\right]^2\nn
    &\hspace{20mm}
    +\rho^2(r,\xi)\left(\frac{dr^2}{Q(r)}
    +\frac{d\xi^2}{P(\xi)}\right)\biggr]\,.
    \label{OP-PDalphaomega}
\end{align}
This is the metric of $\mr{OP}_{\alpha\omega}$.

We also make the parameter redefinitions
\begin{align}
    \hat k\eqqcolon\alpha^2k\,,\qquad
    \hat\ep\eqqcolon\frac{\alpha}{\omega}\ep\,,\qquad
    \hat m+i\hat n\eqqcolon\sqrt{\frac{\alpha^3}{\omega^3}}(m+in)\,,\qquad
    \hat s\eqqcolon\left(\frac{\alpha}{\omega}\right)^2\mr q^2\,,\qquad
    c^\prime\eqqcolon\sqrt{\frac{\omega}{\alpha^3}}b\,.
    \label{parameters in OP-PDalphaomega}
\end{align}
The notation $\mr{q}^2$ denotes a real parameter of either sign, related to $\hat{s}$ by the redefinition above.
We do not require $\mr{q}$ itself to be real, but it may be purely imaginary.
At finite $\abs{b}$, the metric and Maxwell field depend on this parameter through $\mr{q}^2$, and negative values are not excluded provided that the shifts and fields remain real on a Lorentzian coordinate domain.
By contrast, the standard aligned limit discussed below, in which the shifts vanish and $\mr{q}^2=e^2+g^2$ in the limiting Pleba\'nski--Demia\'nski solution, requires $\mr{q}^2\geq0$.
We impose this non-negativity condition only when discussing that limit.
A complete determination of the admissible parameter and coordinate domains, including regularity and the physical interpretation of the solutions, is beyond the scope of this paper.

By using these parameters, the metric functions can be expressed as follows.
\begin{align}
    &P(\xi)=P_\mr{PD}(\xi)+\abs{b}^2\delta P(\xi)\,,
    \label{def P in OP-PDalphaomega}\\
    &P_\mr{PD}(\xi)=k+2\omega^{-1}n\xi-\ep\xi^2
    +2\alpha m\xi^3-\alpha^2(\omega^2k+\mr q^2)\xi^4\,,
    \label{P of PDalphaomega}\\
    &\delta P(\xi)=-8\omega^{-1}n(\omega^2k+\mr q^2)\xi^3
    +4\left[\ep(\omega^2k+\mr q^2)-m^2
    -4\abs{b}^2k(\omega^2k+\mr q^2)^2\right]\xi^4\,,
    \label{delta P in OP-PDalphaomega}\\
    &Q(r)=Q_\mr{PD}(r)+\abs{b}^2\delta Q(r)\,,
    \label{def Q in OP-PDalphaomega}\\
    &Q_\mr{PD}(r)=\omega^2k+\mr q^2-2mr+\ep r^2
    -2\alpha\omega^{-1}nr^3-\alpha^2kr^4\,,
    \label{Q of PDalphaiomega}\\
    &\delta Q(r)=-8kmr^3
    +4\left[\ep k+\omega^{-2}n^2
    -4\abs{b}^2k^2(\omega^2k+\mr q^2)\right]r^4\,,
    \label{delta Q in PDalphaomega}\\
    &\Omega^2(r,\xi)=\Omega_\mr{PD}^2(r,\xi)
    +\abs{b}^2\delta\Omega^2(r,\xi)\,,
    \label{def Omega in OP-PDalphaomega}\\
    &\Omega_\mr{PD}^2(r,\xi)=(1-\alpha r\xi)^2
    =1-2\alpha r\xi+\alpha^2r^2\xi^2\,,
    \label{Omega of PDalphaomega}\\
    &\delta\Omega^2(r,\xi)=4kr^2-4(\omega^2k+\mr q^2)\xi^2
    +8\left(\omega^{-1}nr+m\xi\right)r\xi
    -4\left[\ep-4\abs{b}^2k(\omega^2k+\mr q^2)\right]r^2\xi^2\,.
    \label{delta Omega in PDalphaomega}
\end{align}
Under these transformations, the constraint on $p_0$ and $q_0$ becomes
\begin{align}
    (\omega\xi_0-\msf A)^2+(r_0-\msf B)^2=\msf R^2\,,
\end{align}
where
\begin{align}
    &\msf A=\frac{4\abs{b}^2n}
    {\alpha^2-4\abs{b}^2\left[\ep-4\abs{b}^2k(\omega^2k+\mr q^2)\right]}\,,\qquad
    \msf B=\frac{4\abs{b}^2m}
    {\alpha^2-4\abs{b}^2\left[\ep-4\abs{b}^2k(\omega^2k+\mr q^2)\right]}\,,
    \label{AB tilde}\\
    &\msf R=\frac{2\abs {b}\sqrt{\alpha^2\mr q^2
    +4\abs{b}^2\left[m^2+n^2-\ep\mr q^2
    +4\abs{b}^2\mr q^2k(\omega^2k+\mr q^2)\right]}}
    {\alpha^2-4\abs{b}^2\left[\ep-4\abs{b}^2k(\omega^2k+\mr q^2)\right]}\,.
    \label{p0q0 Constraint in PD alpha omega parameter}
\end{align}
As in the preceding subsection, we assume that the common denominator is nonzero and that the radicand is nonnegative, retaining the sign of the denominator in $\msf{R}$.
The shifts can therefore be parametrized by the angle $\beta\coloneqq\hat\beta$ as
\begin{align}
    \omega\xi_0=\msf A+\msf R\sin\beta\,,\qquad
    r_0=\msf B+\msf R\cos\beta\,.
    \label{xi0 r0 constraint}
\end{align}

The transformed principal null tetrad is
\begin{align}
    k^a
    &=\sqrt{\frac{\Omega^2}{2Q\rho^2}}
    \left[
    (r+r_0)^2(\del_{\tau})^a
    -\omega(\del_{\phi})^a+Q(\del_r)^a
    \right]\,,
    \nn
    \ell^a
    &=\sqrt{\frac{\Omega^2}{2Q\rho^2}}
    \left[
    (r+r_0)^2(\del_{\tau})^a
    -\omega(\del_{\phi})^a-Q(\del_r)^a
    \right]\,,
    \nn
    m^a
    &=\sqrt{\frac{\Omega^2}{2P\rho^2}}
    \left[
    \omega(\xi+\xi_0)^2(\del_{\tau})^a
    +(\del_{\phi})^a+iP(\del_{\xi})^a
    \right]\,,
    \nn
    \bar{m}^a
    &=\sqrt{\frac{\Omega^2}{2P\rho^2}}
    \left[
    \omega(\xi+\xi_0)^2(\del_{\tau})^a
    +(\del_{\phi})^a-iP(\del_{\xi})^a
    \right]\,.
    \label{tetrad:OP-alpha-omega}
\end{align}
Then, the Maxwell scalars take the form
\begin{align}
    \phi_0=\phi_2
    &=\frac{b\sqrt{PQ}}
    {\Omega\left[r+r_0+i\omega(\xi+\xi_0)\right]}\,,
    \label{phi0 PDalphaomega}\\
    \phi_1
    &=\frac{1}{4\bar b}\frac{\Omega^2}
    {\left[r+r_0+i\omega(\xi+\xi_0)\right]^2}
    \left[
    \omega(\xi+\xi_0)^2\del_\xi
    \left(\frac{\del_r\Omega}{\xi+\xi_0}\right)
    -i(r+r_0)^2\del_r
    \left(\frac{\del_\xi\Omega}{r+r_0}\right)
    \right]\,.
    \label{phi1 PDalphaomega}
\end{align}

\paragraph{The $\abs{b}  \to0$ limit}
For this standard real aligned limit, we assume $\mr{q}^2\geq0$ and take $\mr{q}$ to be its nonnegative square root.
We take $\abs{b}  \to0$ at fixed $(\alpha,\omega,k,\ep,m,n,\mr {q}^2,\beta,\gamma)$, where $b=\abs{b}  \e{i\gamma}$ and $\gamma=\gamma^\prime$.
In this limit, $\msf A$, $\msf B$, and $\msf R$ all vanish, and the metric functions reduce to those of the $\mr{PD}_{\alpha\omega}$ spacetime with vanishing cosmological constant.
Choosing the branch for which $\Omega\to\Omega_\mr{PD}=1-\alpha r\xi$, the electromagnetic field satisfies $\phi_0=\phi_2\to0$, while
\begin{align}
    \phi_1
    &\to\frac{\mr q\e{i\gamma}(-\sin\beta+i\cos\beta)}{2}
    \left(\frac{1-\alpha r\xi}{r+i\omega\xi}\right)^2
    =\frac{e+ig}{2}
    \left(\frac{\Omega_\mr{PD}}{r+i\omega\xi}\right)^2\,,
\end{align}
where
\begin{align}
    e=\mr q\cos(\beta+\gamma+\pi/2)\,,\qquad
    g=\mr q\sin(\beta+\gamma+\pi/2)\,.
\end{align}
The electromagnetic field of the $\mr{PD}_{\alpha\omega}$ spacetime is given by
\begin{align}
    \phi_0^{\mr{PD}_{\alpha\omega}}
    &=\phi_2^{\mr{PD}_{\alpha\omega}}=0\,,\qquad
    \phi_1^{\mr{PD}_{\alpha\omega}}
    =\frac{e+ig}{2}
    \left(\frac{\Omega_\mr{PD}}{r+i\omega\xi}\right)^2\,.
\end{align}
Thus, both the metric and electromagnetic field reduce smoothly to those of the $\mr{PD}_{\alpha\omega}$ spacetime with $\Lambda=0$ and $e^2+g^2=\mr q^2$.

\subsubsection{Non-twisting sector}

Following the constructions of Griffiths and Podolsk\'y for $\mr{PD}_{\alpha\omega}$~\cite{Griffiths:2005qp} and of Ovcharenko and Podolsk\'y for the previously studied subclass of $\mr{OP}_{\alpha\omega}$~\cite{Ovcharenko:2025cpm,Ovcharenko:2026byw}, we introduce
\begin{align}
    n_0\coloneqq\frac{n}{\omega}\,.
\end{align}
Here $n$ is the twisting-sector parameter defined in Eq.~\eqref{parameters in OP-PDalphaomega}.
We take $\omega\to0$ at fixed $(\alpha,k,\ep,m,n_0,\mr{q}^2,b)$, with $\alpha>0$, and choose the shifts so that $\xi_0$ remains finite.
Then $\omega\xi_0\to0$ and $r_0\to\tilde{r}_0$.
For example, choosing $\beta=0$ or $\beta=\pi$ in Eq.~\eqref{xi0 r0 constraint} gives the two branches described below.

For $\mr{PD}_{\alpha\omega}$, this $\omega \to 0$ limit yields the non-twisting $\mr{PD}_{\alpha}^{0}$ spacetime, which belongs to the Robinson--Trautman class and has a doubly aligned electromagnetic field.
Applying the same limiting procedure to $\mr{OP}_{\alpha\omega}$ gives a Petrov type~D Robinson--Trautman spacetime with a fully non-aligned electromagnetic field, which we denote by $\mr{OP}_{\alpha}^{0}$.

Its metric is
\begin{align}
    ds^2
    &=\frac{1}{\Omega_0^2(r,\xi)}\left[
    -\frac{Q_0(r)}{(r+\tilde{r}_0)^2}d\tau^2
    +(r+\tilde{r}_0)^2\left(
    \frac{dr^2}{Q_0(r)}+\frac{d\xi^2}{P_0(\xi)}
    +P_0(\xi)d\phi^2\right)\right]\,,
    \label{OP-PDalpha}
\end{align}
where
\begin{align}
    &P_0(\xi)=P_0^{\mr{PD}}(\xi)+\abs{b}^2\delta P_0(\xi)\,,\nn
    &P_0^{\mr{PD}}(\xi)
    =k+2n_0\xi-\ep\xi^2+2\alpha m\xi^3
    -\alpha^2\mr{q}^2\xi^4\,,\nn
    &\delta P_0(\xi)
    =-8n_0\mr{q}^2\xi^3
    +4\left[\ep\mr{q}^2-m^2
    -4\abs{b}^2k\mr{q}^4\right]\xi^4\,,
    \label{P in PDalpha0}\\
    &Q_0(r)=Q_0^{\mr{PD}}(r)+\abs{b}^2\delta Q_0(r)\,,\nn
    &Q_0^{\mr{PD}}(r)
    =\mr{q}^2-2mr+\ep r^2-2\alpha n_0r^3-\alpha^2kr^4\,,\nn
    &\delta Q_0(r)
    =-8kmr^3
    +4\left[\ep k+n_0^2-4\abs{b}^2k^2\mr{q}^2\right]r^4\,,
    \label{Q in PDalpha0}\\
    &\Omega_0^2(r,\xi)
    =\Omega_{\mr{PD}}^2(r,\xi)+\abs{b}^2\delta\Omega_0^2(r,\xi)\,,\nn
    &\Omega_{\mr{PD}}(r,\xi)=1-\alpha r\xi\,,\nn
    &\delta\Omega_0^2(r,\xi)
    =4kr^2-4\mr{q}^2\xi^2+8(n_0r+m\xi)r\xi
    -4\left[\ep-4\abs{b}^2k\mr{q}^2\right]r^2\xi^2\,.
    \label{Omega in PDalpha0}
\end{align}
The shift $\tilde{r}_0$ satisfies
\begin{align}
    (\tilde{r}_0-\msf{B}_0)^2=\msf{R}_0^2\,,
\end{align}
with
\begin{align}
    \msf{B}_0
    &=\frac{4\abs{b}^2m}
    {\alpha^2-4\abs{b}^2\left[\ep-4\abs{b}^2k\mr{q}^2\right]}\,,
    \label{AB tilde omega0}\\
    \msf{R}_0
    &=\frac{2\abs{b}\sqrt{\alpha^2\mr{q}^2
    +4\abs{b}^2\left[m^2-\ep\mr{q}^2
    +4\abs{b}^2k\mr{q}^4\right]}}
    {\alpha^2-4\abs{b}^2\left[\ep-4\abs{b}^2k\mr{q}^2\right]}\,.
    \label{p0q0 Constraint in PD alpha parameter}
\end{align}
We assume that the common denominator is nonzero and that the radicand is nonnegative, retaining the sign of the denominator in $\msf{R}_0$.
The two branches are
\begin{align}
    \tilde{r}_0=\tilde{r}_{0\pm}\,,\qquad
    \tilde{r}_{0\pm}\coloneqq\msf{B}_0\pm\abs{\msf{R}_0}\,.
    \label{xi0 r0 constraint in omega0}
\end{align}
We refer to the branch with $\tilde{r}_0=\tilde{r}_{0+}$ as $\mr{OP}_{\alpha+}^{0}$ and to that with $\tilde{r}_0=\tilde{r}_{0-}$ as $\mr{OP}_{\alpha-}^{0}$.
They coincide when $\msf{R}_0=0$.
We refer to this degenerate subclass as $\mr{OP}_{\alpha0}^0$.

For either branch, the limiting principal null tetrad is\footnote{
In this tetrad, the spin coefficients satisfy
\begin{align}
    \kappa_s&=\nu_s=\sigma_s=\lambda_s=0\,,\quad 
    \rho_s=\mu_s
    =\sqrt{\frac{\Omega_0^2Q_0}
    {2(r+\tilde{r}_0)^2}}
    \left(
    \del_r\log\Omega_0-\frac{1}{r+\tilde{r}_0}
    \right)\,,\quad 
    \tau_s=\pi_s
    =i\sqrt{\frac{\Omega_0^2P_0}
    {2(r+\tilde{r}_0)^2}}\,
    \del_{\xi}\log\Omega_0\,.
    \label{tetrad:non-twisting-spin-coefficients}
\end{align}
Thus, on generic expanding patches, $\mr{OP}_{\alpha}^0$ belongs locally to the class integrated by Van den Bergh and Carminati~\cite{VandenBergh:2020lvf}, whose defining conditions were stated in Sec.~\ref{Sec. Intro}.
}
\begin{align}
    k^a
    &=\sqrt{\frac{\Omega_0^2}
    {2Q_0(r+\tilde{r}_0)^2}}
    \left[
    (r+\tilde{r}_0)^2(\del_{\tau})^a
    +Q_0(\del_r)^a
    \right]\,,\quad 
    \ell^a
    =\sqrt{\frac{\Omega_0^2}
    {2Q_0(r+\tilde{r}_0)^2}}
    \left[
    (r+\tilde{r}_0)^2(\del_{\tau})^a
    -Q_0(\del_r)^a
    \right]\,,
    \nn
    m^a
    &=\sqrt{\frac{\Omega_0^2}
    {2P_0(r+\tilde{r}_0)^2}}
    \left[
    (\del_{\phi})^a+iP_0(\del_{\xi})^a
    \right]\,,\quad 
    \bar{m}^a
    =\sqrt{\frac{\Omega_0^2}
    {2P_0(r+\tilde{r}_0)^2}}
    \left[
    (\del_{\phi})^a-iP_0(\del_{\xi})^a
    \right]\,.
    \label{tetrad:OP-alpha-zero}
\end{align}
Then, the Maxwell scalars are
\begin{align}
    \phi_0 &=\phi_2
    =\frac{b\sqrt{P_0Q_0}}{\Omega_0(r+\tilde{r}_0)}\,,
    \label{phi0 PDalpha}\\
    \phi_1
    &=-\frac{i\Omega_0^2}{4\bar{b}}\,
    \del_r\left(\frac{\del_{\xi}\Omega_0}{r+\tilde{r}_0}\right)\,.
    \label{phi1 PDalpha}
\end{align}

\paragraph{The $\abs{b}\to0$ limit}
For this standard real aligned limit, we assume $\mr{q}^2\geq0$ and take $\mr{q}$ to be its nonnegative square root.
We take $\abs{b}\to0$ at fixed $(\alpha,k,\ep,m,n_0,\mr{q}^2,\gamma)$, and choose the branch for which $\Omega_0\to\Omega_{\mr{PD}}=1-\alpha r\xi$.
Both shifts $\tilde{r}_{0\pm}$ tend to zero, and the metric reduces to that of $\mr{PD}_{\alpha}^{0}$ with $\Lambda=0$.
Moreover,
\begin{align}
    \frac{\tilde{r}_{0\pm}}{\abs{b}}
    \longrightarrow\pm\frac{2\mr{q}}{\alpha}\,,
\end{align}
so the electromagnetic field satisfies
\begin{align}
    \phi_0=\phi_2&\to 0 \,,\qquad
    \phi_1\to
    \pm\frac{i\mr{q}\e{i\gamma}}{2}
    \left(\frac{\Omega_{\mr{PD}}}{r}\right)^2
    =\frac{e+ig}{2}
    \left(\frac{\Omega_{\mr{PD}}}{r}\right)^2
    = \phi_1^{\mr{PD}_{\alpha}^0}\,.
\end{align}
Here, we have defined
\begin{align}
    e=\mp\mr{q}\sin\gamma\,,\qquad
    g=\pm\mr{q}\cos\gamma\,.
\end{align}
The upper and lower signs correspond to the $\mr{OP}_{\alpha+}^{0}$ and $\mr{OP}_{\alpha-}^{0}$ branches, respectively.
Thus, both the metric and electromagnetic field reduce smoothly to those of $\mr{PD}_{\alpha}^{0}$ with $\Lambda=0$ and $e^2+g^2=\mr{q}^2$.

\paragraph{Relation to the Ovcharenko--Podolsk\'y parametrization}

For $\mr{q}=0$, the correspondence with the previously used Ovcharenko--Podolsk\'y parametrization is
\begin{align}
    k=k_{\mr{OP}}\,,\quad
    \ep=\ep_{\mr{OP}}\,,\quad
    m=m_{\mr{OP}}\,,\quad
    n_0=n_0^{\mr{OP}}-\alpha C m_{\mr{OP}}\,,\quad
    b=\alpha c_{\mr{OP}}\,,\quad
    C=2\abs{c_{\mr{OP}}}^2k_{\mr{OP}}\,,
    \label{relation to OP and F parameter in omega0}
\end{align}
where
\begin{align}
    n_0^{\mr{OP}}\coloneqq\frac{n_{\mr{OP}}}{\omega}\,.
\end{align}
Here $n_{\mr{OP}}$ is related to the hatted parameter in Eq.~\eqref{inv map for para OP} by $\hat n_{\mr{OP}}=(\alpha/\omega)^{3/2}n_{\mr{OP}}$.
The combination $n_0^{\mr{OP}}$ is held fixed in the non-twisting limit.


\section{Ovcharenko--Podolsk\'y spacetime in the Griffiths--Podolsk\'y form}
\label{Sec. OP-GP}

Starting from the $\mr{OP}_{\alpha\omega}$ solution, we introduce the parameters $(a,l)$ through a further coordinate transformation to obtain the Griffiths--Podolsk\'y form, following the construction used for the Pleba\'nski--Demia\'nski family~\cite{Griffiths:2005qp,Ovcharenko:2025cpm}.
To select generalized black holes, we subsequently take $\phiv$ to be an angular coordinate and choose the angular metric function so that $x=\pm1$ correspond to the two axes.
Writing this function as $\tilde{\mc{P}}(x)=\sum_{i=0}^{4}a_i x^i$, we impose
\begin{align}
    a_1+a_3=0\,,\qquad
    a_0+a_2+a_4=0\,,\qquad
    a_0=1\,.
\end{align}
The first two conditions place roots of $\tilde{\mc{P}}$ at $x=\pm1$, while the third fixes its normalization.
We refer to these as the \emph{black hole candidate conditions}.
We first give the general metric in the Griffiths--Podolsk\'y form, before imposing these conditions.

\subsection{General Ovcharenko--Podolsk\'y spacetime in the twisting sector}
\subsubsection{Metric in the Griffiths--Podolsk\'y form}
For $a\neq0$, we introduce the coordinates $(t,r,x,\phiv)$ by
\begin{align}
    \xi&=\frac{a}{\omega}x+\frac{l}{\omega}\,,\qquad
    \tau=t-\frac{(a+l+\omega\xi_0)^2}{a}\phiv\,,\qquad
    \phi=-\frac{\omega}{a}\phiv\,,
    \label{PDalphaomega to GP}
\end{align}
with
\begin{align}
    &P(\xi(x))\eqqcolon\left(\frac{a}{\omega}\right)^2
    \tilde{\mc{P}}(x)\,,\qquad
    Q(r)\equiv\mc{Q}(r)\,,\qquad
    \Omega^2(r,\xi(x))\eqqcolon\Omega^2(r,x)\,,\nn
    &\rhov^2(r,x)\coloneqq(r+r_0)^2+(ax+l+\omega\xi_0)^2\,.
\end{align}
The metric then takes the form
\begin{align}
    ds^2
    &=\frac{1}{\Omega^2(r,x)}\biggl[
    -\frac{\mc{Q}(r)}{\rhov^2(r,x)}
    \left(dt-\left[a(1-x^2)+2(l+\omega\xi_0)(1-x)\right]d\phiv\right)^2\nn
    &\qquad \qquad \quad \,\, +\rhov^2(r,x)
    \left(\frac{dr^2}{\mc{Q}(r)}
    +\frac{dx^2}{\tilde{\mc{P}}(x)}\right)
    +\frac{\tilde{\mc{P}}(x)}{\rhov^2(r,x)}
    \left(a\,dt-\left[(r+r_0)^2+(a+l+\omega\xi_0)^2\right]
    d\phiv\right)^2\biggr]\,.
    \label{OP-GP form}
\end{align}
The radial function $\mc{Q}$ is unchanged by the transformation and is given by
\begin{align}
    &\mc{Q}(r)=\mc{Q}_{\mr{GP}}(r)+\abs{b}^2\delta\mc{Q}(r)\,,\qquad
    \mc{Q}_{\mr{GP}}(r)\equiv\sum_{j=0}^{4}b_j^{\mr{GP}}r^j\,,\qquad
    \delta\mc{Q}(r)\equiv\sum_{j=0}^{4}\delta b_j r^j\,,\nn
    &\mc{Q}_{\mr{GP}}(r)
    =\omega^2k+\mr{q}^2-2mr+\ep r^2
    -2\alpha\omega^{-1}nr^3-\alpha^2kr^4\,,\nn
    &\delta\mc{Q}(r)
    =-8kmr^3+4\left[\ep k+\omega^{-2}n^2
    -4\abs{b}^2k^2(\omega^2k+\mr{q}^2)\right]r^4\,.
    \label{bOP-GP}
\end{align}
The conformal factor is
\begin{align}
    &\Omega^2(r,x)
    =\Omega^2_{\mr{GP}}(r,x)+\abs{b}^2\delta\Omega^2(r,x)\,,\qquad
    \Omega^2_{\mr{GP}}(r,x)\equiv\sum_{i,j=0}^{2}c_{ij}^\mr{GP}x^i r^j\,,\qquad
    \delta\Omega^2(r,x)\equiv\sum_{i,j=0}^{2}\delta c_{ij}x^i r^j\,,\nn
    &\Omega^2_{\mr{GP}}(r,x)
    =\left[1-\frac{\alpha}{\omega}(ax+l)r\right]^2\nn
    &\hspace{19mm}
    =1-\frac{2\alpha l}{\omega}r-\frac{2\alpha a}{\omega}xr
    +\frac{\alpha^2l^2}{\omega^2}r^2
    +\frac{2\alpha^2al}{\omega^2}r^2x
    +\frac{\alpha^2a^2}{\omega^2}r^2x^2\,,
\end{align}
where
\begin{align}
 \delta c_{00}&=-4l^2(k+\omega^{-2}\mr q^2)\,,&
 \delta c_{10}&=-8al(k+\omega^{-2}\mr q^2)\,,\nn
 \delta c_{01}&=8\omega^{-2}l^2m\,,&
 \delta c_{11}&=16\omega^{-2}lam\,,\nn
 \delta c_{20}&=-4a^2(k+\omega^{-2}\mr q^2)\,,&
 \delta c_{02}&=4\left[k+2\omega^{-2}ln-\omega^{-2}l^2\ep\right]
 +16\abs{b}^2l^2k(k+\omega^{-2}\mr q^2)\,,\nn
 \delta c_{21}&=8\omega^{-2}a^2m\,,&
 \delta c_{12}&=8\omega^{-2}\left[an-al\ep\right]
 +32\abs{b}^2alk(k+\omega^{-2}\mr q^2)\,,\nn
 \delta c_{22}&=-4\omega^{-2}a^2\ep
 +16\abs{b}^2 a^2k(k+\omega^{-2}\mr q^2)\,.
 \label{delta Omega in OP-GP}
\end{align}
For the function $\tilde{\mc{P}}(x)$, we write
\begin{align}
    &\tilde{\mc{P}}(x)
    =\tilde{\mc{P}}_{\mr{GP}}(x)
    +\abs{b}^2\delta\tilde{\mc{P}}(x)\,,\nn
    &\tilde{\mc{P}}_{\mr{GP}}(x)
    \equiv\sum_{i=0}^{4}a_i^{\mr{GP}}x^i\,,\qquad
    \delta\tilde{\mc{P}}(x)
    \equiv\sum_{i=0}^{4}\delta a_i x^i\,,\qquad
    a_i=a_i^{\mr{GP}}+\abs{b}^2\delta a_i\,.
    \label{def of aOP-GP}
\end{align}
The coefficients of the Pleba\'nski--Demia\'nski part are
\begin{align}
    a_4^{\mr{GP}}
    &=-\alpha^2a^2(k+\omega^{-2}\mr{q}^2)\,,\nn
    a_3^{\mr{GP}}
    &=2\alpha\omega^{-1}am
    -4\alpha^2al(k+\omega^{-2}\mr{q}^2)\,,\nn
    a_2^{\mr{GP}}
    &=-\ep+6\alpha\omega^{-1}lm
    -6\alpha^2l^2(k+\omega^{-2}\mr{q}^2)\,,\nn
    a_1^{\mr{GP}}
    &=2a^{-1}\left[n-l\ep+3\alpha\omega^{-1}l^2m
    -2\alpha^2l^3(k+\omega^{-2}\mr{q}^2)\right]\,,\nn
    a_0^{\mr{GP}}
    &=a^{-2}\left[\omega^2k+2ln-l^2\ep+2\alpha\omega^{-1}l^3m
    -\alpha^2l^4(k+\omega^{-2}\mr{q}^2)\right]\,,
    \label{aGPs}
\end{align}
and the remaining coefficients are
\begin{align}
    \delta a_4
    &=4a^2\left[(k+\omega^{-2}\mr{q}^2)
    \left(\ep-4k\abs{b}^2(k\omega^2+\mr{q}^2)\right)
    -\omega^{-2}m^2\right]\,,\nn
    \delta a_3
    &=8a\left[2l\left[(k+\omega^{-2}\mr{q}^2)
    \left(\ep-4k\abs{b}^2(k\omega^2+\mr{q}^2)\right)
    -\omega^{-2}m^2\right]
    -n(k+\omega^{-2}\mr{q}^2)\right]\,,\nn
    \delta a_2
    &=24l\left[l\left[(k+\omega^{-2}\mr{q}^2)
    \left(\ep-4k\abs{b}^2(k\omega^2+\mr{q}^2)\right)
    -\omega^{-2}m^2\right]
    -n(k+\omega^{-2}\mr{q}^2)\right]\,,\nn
    \delta a_1
    &=8a^{-1}l^2\left[2l\left[(k+\omega^{-2}\mr{q}^2)
    \left(\ep-4k\abs{b}^2(k\omega^2+\mr{q}^2)\right)
    -\omega^{-2}m^2\right]
    -3n(k+\omega^{-2}\mr{q}^2)\right]\,,\nn
    \delta a_0
    &=4a^{-2}l^3\left[l\left[(k+\omega^{-2}\mr{q}^2)
    \left(\ep-4k\abs{b}^2(k\omega^2+\mr{q}^2)\right)
    -\omega^{-2}m^2\right]
    -2n(k+\omega^{-2}\mr{q}^2)\right]\,.
    \label{delta as}
\end{align}
The coefficients $\{a_i^{\mr{GP}}\}$, $\{b_j^{\mr{GP}}\}$, and $\{c_{ij}^{\mr{GP}}\}$ coincide with those of the corresponding metric functions in the Griffiths--Podolsk\'y form of the Pleba\'nski--Demia\'nski spacetime with $\Lambda=0$~\cite{Griffiths:2005qp}.
Moreover, since $(r_0,\xi_0)\to(0,0)$ as $\abs{b}\to0$, the metric reduces smoothly to the charged Pleba\'nski--Demia\'nski metric in this limit.

\subsubsection{Electromagnetic field in the Griffiths--Podolsk\'y form}
The principal null tetrad in the Griffiths--Podolsk\'y coordinates is
\begin{align}
    k^a
    &=\sqrt{\frac{\Omega^2}{2\mc{Q}\rhov^2}}
    \left[
    \bigl((r+r_0)^2+(a+l+\omega\xi_0)^2\bigr)(\del_t)^a
    +\mc{Q}(\del_r)^a+a(\del_{\phiv})^a
    \right]\,,
    \nn
    \ell^a
    &=\sqrt{\frac{\Omega^2}{2\mc{Q}\rhov^2}}
    \left[
    \bigl((r+r_0)^2+(a+l+\omega\xi_0)^2\bigr)(\del_t)^a
    -\mc{Q}(\del_r)^a+a(\del_{\phiv})^a
    \right]\,,
    \nn
    m^a
    &=-\sqrt{\frac{\Omega^2}{2\tilde{\mc{P}}\rhov^2}}
    \left[
    \bigl(a(1-x^2)+2(l+\omega\xi_0)(1-x)\bigr)(\del_t)^a
    +(\del_{\phiv})^a-i\tilde{\mc{P}}(\del_x)^a
    \right]\,,
    \nn
    \bar{m}^a
    &=-\sqrt{\frac{\Omega^2}{2\tilde{\mc{P}}\rhov^2}}
    \left[
    \bigl(a(1-x^2)+2(l+\omega\xi_0)(1-x)\bigr)(\del_t)^a
    +(\del_{\phiv})^a+i\tilde{\mc{P}}(\del_x)^a
    \right]\,.
    \label{tetrad:GP-x}
\end{align}
Then, the Maxwell scalars are\footnote{
For $a<0$ with $\omega>0$, we can reverse the signs of $m^a$ and $\bar{m}^a$ relative to the directly transformed tetrad, so that the prefactor in Eq.~\eqref{phi0 in GP} is proportional to $a/\omega$ rather than $\abs{a}/\omega$.
}
\begin{align}
    \phi_0&=\phi_2
    =\frac{ab}{\Omega\omega}\,
    \frac{\sqrt{\tilde{\mc{P}}\mc{Q}}}
    {r+r_0+i(ax+l+\omega\xi_0)}\,,
    \label{phi0 in GP}\\
    \phi_1
    &=\frac{\omega}{4a\bar{b}}\,
    \frac{\Omega^2}
    {\left[r+r_0+i(ax+l+\omega\xi_0)\right]^2}\nn
    &\qquad\times
    \left[
    (ax+l+\omega\xi_0)^2\del_{x}
    \left(\frac{\del_{r}\Omega}{ax+l+\omega\xi_0}\right)
    -i(r+r_0)^2\del_{r}
    \left(\frac{\del_{x}\Omega}{r+r_0}\right)
    \right]\,.
    \label{phi1 0n GP}
\end{align}
Taking $\abs{b}\to0$ with the remaining parameters held fixed with $\mr{q}^2 \geq 0$, and choosing the branch for which $\Omega\to\Omega_{\mr{GP}}=1-\frac{\alpha}{\omega}(ax+l)r$, we obtain
\begin{align}
    \phi_0&=\phi_2\to0\,,\\
    \phi_1
    &\to
    \frac{\mr{q}\e{i\gamma}(-\sin\beta+i\cos\beta)}{2}
    \left(\frac{\Omega_{\mr{GP}}}{r+i(ax+l)}\right)^2
    =\frac{e+ig}{2}
    \left(\frac{\Omega_{\mr{GP}}}{r+i(ax+l)}\right)^2
    =\phi_1^\mr{GP} \,.
\end{align}
Here, as in the preceding section, we have used the expressions 
\begin{align}
    e&=\mr{q}\cos\left(\beta+\gamma+\frac{\pi}{2}\right)\,,
    \qquad
    g=\mr{q}\sin\left(\beta+\gamma+\frac{\pi}{2}\right)\,.
\end{align}
Thus, the electromagnetic field also reduces smoothly to that of the $\mr{PD}_{\alpha\omega}$ spacetime with $\Lambda=0$ in the Griffiths--Podolsk\'y form.

\subsection{Ovcharenko--Podolsk\'y generalized black hole spacetime family}
\subsubsection{Twisting sector}

We now impose the black hole candidate conditions introduced above,
\begin{align}
    a_1+a_3&=0\,,\label{a GP condition odd}\\
    a_0+a_2+a_4&=0\,,\label{a GP condition even}\\
    a_0&=1\,.\label{a0 GP condition}
\end{align}
The angular metric function then factorizes as
\begin{align}
    \tilde{\mc{P}}(x)=(1-x^2)(1-a_3x-a_4x^2)\,.
\end{align}
We first consider the generic sector in which the denominators below are nonzero.
In this sector, Eqs.~\eqref{a GP condition odd} and \eqref{a GP condition even} form a nondegenerate linear system for $(\ep,n)$ and determine them uniquely in terms of $k$ and the remaining parameters.
Degenerate cases must instead be treated directly from the black hole candidate conditions.
We obtain
\begin{align}
    \ep&=\frac{\msf{N}_\ep}{\msf{D}_\ep}\,,\qquad
    \msf{N}_\ep=\msf{N}_{\ep}^{0}
    +\msf{N}_{\ep}^{1}\abs{b}^2
    +\msf{N}_{\ep}^{2}\abs{b}^4
    +\msf{N}_{\ep}^{3}\abs{b}^6\,,
    \label{def sol ep}
\end{align}
where
\begin{align}
    \msf{D}_\ep
    &=(a^2-l^2)
    \bigl[\omega^2-4\abs{b}^2(a-l)^2(\mr{q}^2+k\omega^2)\bigr]
    \bigl[\omega^2-4\abs{b}^2(a+l)^2(\mr{q}^2+k\omega^2)\bigr]\,,\nn
    \msf{N}_{\ep}^{0}
    &=\omega^2\bigl[k\omega^4+4l\alpha(a^2-l^2)m\omega
    -\alpha^2(a^2-l^2)(a^2+3l^2)(\mr{q}^2+k\omega^2)\bigr]\,,\nn
    \msf{N}_{\ep}^{1}
    &=-4\omega^2(a^2+3l^2)
    \bigl[(a^2-l^2)m^2+k\omega^2(\mr{q}^2+k\omega^2)\bigr]
    +4\alpha^2(a^2-l^2)^3(\mr{q}^2+k\omega^2)^2\,,\nn
    \msf{N}_{\ep}^{2}
    &=16(a^2-l^2)(\mr{q}^2+k\omega^2)
    \bigl[(a^2-l^2)^2m^2
    -k\omega^2(a^2+3l^2)(\mr{q}^2+k\omega^2)\bigr]\,,\nn
    \msf{N}_{\ep}^{3}
    &=64k(a^2-l^2)^3(\mr{q}^2+k\omega^2)^3\,.
    \label{sol ep}
\end{align}
Similarly,
\begin{align}
    n&=\frac{\msf{N}_n}{\msf{D}_n}\,,\qquad
    \msf{N}_n=\msf{N}_{n}^{0}
    +\msf{N}_{n}^{1}\abs{b}^2
    +\msf{N}_{n}^{2}\abs{b}^4\,,
    \label{def sol n}
\end{align}
with
\begin{align}
    \msf{D}_n
    &=\msf{D}_{\ep}\,,\nn
    \msf{N}_{n}^{0}
    &=\omega^2\bigl[kl\omega^4-\alpha(a^2-l^2)^2m\omega
    +l\alpha^2(a^2-l^2)^2(\mr{q}^2+k\omega^2)\bigr]\,,\nn
    \msf{N}_{n}^{1}
    &=4l\omega^2\bigl[(a^2-l^2)^2m^2
    -2k\omega^2(a^2+l^2)(\mr{q}^2+k\omega^2)\bigr]
    +4\alpha m\omega(a^2-l^2)^3(\mr{q}^2+k\omega^2)\,,\nn
    \msf{N}_{n}^{2}
    &=16kl(a^2-l^2)^2(\mr{q}^2+k\omega^2)^2\omega^2\,.
    \label{sol n}
\end{align}
Substitution into $a_0$ gives
\begin{align}
    a_0&=\frac{\msf{N}_{a_0}}{\msf{D}_{a_0}}\,,\qquad
    \msf{N}_{a_0}=\msf{N}_{a_0}^{0}
    +\msf{N}_{a_0}^{1}\abs{b}^2
    +\msf{N}_{a_0}^{2}\abs{b}^4
    +\msf{N}_{a_0}^{3}\abs{b}^6\,,
    \label{def sol a0}
\end{align}
where
\begin{align}
    \msf{D}_{a_0}&=\msf{D}_{\ep}=\msf{D}_{n}\,,\nn
    \msf{N}_{a_0}^{0}
    &=\omega^2\bigl[k\omega^4-2l\alpha(a^2-l^2)m\omega
    +3l^2\alpha^2(a^2-l^2)(\mr{q}^2+k\omega^2)\bigr]\,,\nn
    \msf{N}_{a_0}^{1}
    &=4\omega^2\bigl[3l^2(a^2-l^2)m^2
    -k\omega^2(2a^2+3l^2)(\mr{q}^2+k\omega^2)\bigr]\nn
    &\quad+8l\alpha m\omega(a^4-4a^2l^2+3l^4)(\mr{q}^2+k\omega^2)
    -4l^2\alpha^2(a^2-l^2)^2(\mr{q}^2+k\omega^2)^2\,,\nn
    \msf{N}_{a_0}^{2}
    &=16(\mr{q}^2+k\omega^2)\bigl[-l^2(a^2-l^2)^2m^2
    +k\omega^2(a^4+3l^4)(\mr{q}^2+k\omega^2)\bigr]\,,\nn
    \msf{N}_{a_0}^{3}
    &=-64kl^2(a^2-l^2)^2(\mr{q}^2+k\omega^2)^3\,.
    \label{sol a0}
\end{align}
The remaining black hole candidate condition, Eq.~\eqref{a0 GP condition}, therefore reduces in this generic sector to
\begin{align}
    \msf{P}(k)&\coloneqq\msf{D}_{a_0}-\msf{N}_{a_0}=0\,,\qquad
    \msf{P}(k)=\msf{P}_0+\msf{P}_1k+\msf{P}_2k^2
    +\msf{P}_3k^3+\msf{P}_4k^4\,,
    \label{quartic k condition}
\end{align}
with
\begin{align}
    \msf{P}_0
    &=\omega^2(a^2-l^2)
    \bigl(\omega^2+2l\alpha m\omega-3l^2\alpha^2\mr{q}^2\bigr)\nn
    &\quad-4\abs{b}^2(a^2-l^2)\bigl[
    \omega^2\{2(a^2+l^2)\mr{q}^2+3l^2m^2\}
    +2l\alpha m\omega(a^2-3l^2)\mr{q}^2
    -l^2\alpha^2(a^2-l^2)\mr{q}^4\bigr]\nn
    &\quad+16\abs{b}^4(a^2-l^2)^2
    \bigl[l^2m^2+(a^2-l^2)\mr{q}^2\bigr]\mr{q}^2\,,\nn
    \msf{P}_1
    &=-\omega^4\bigl[\omega^2+3l^2\alpha^2(a^2-l^2)\bigr]\nn
    &\quad+4\abs{b}^2\omega^2\bigl[
    \omega^2\{(2a^2+3l^2)\mr{q}^2-2(a^4-l^4)\}
    -2l\alpha m\omega(a^4-4a^2l^2+3l^4)
    +2l^2\alpha^2(a^2-l^2)^2\mr{q}^2\bigr]\nn
    &\quad+16\abs{b}^4\omega^2\bigl[
    l^2(a^2-l^2)^2m^2+2(a^2-l^2)^3\mr{q}^2
    -(a^4+3l^4)\mr{q}^4\bigr]\nn
    &\quad+64\abs{b}^6l^2(a^2-l^2)^2\mr{q}^6\,,\nn
    \msf{P}_2
    &=4\abs{b}^2\omega^4\bigl[
    (2a^2+3l^2)\omega^2+l^2(a^2-l^2)^2\alpha^2\bigr]\nn
    &\quad-16\abs{b}^4\omega^4\bigl[
    2(a^4+3l^4)\mr{q}^2-(a^2-l^2)^3\bigr]
    +192\abs{b}^6\omega^2l^2(a^2-l^2)^2\mr{q}^4\,,\nn
    \msf{P}_3
    &=16\abs{b}^4\omega^4\bigl[
    12l^2(a^2-l^2)^2\mr{q}^2\abs{b}^2
    -(a^4+3l^4)\omega^2\bigr]\,,\nn
    \msf{P}_4
    &=64\abs{b}^6l^2(a^2-l^2)^2\omega^6\,.
    \label{quartic k coefficients}
\end{align}
For generic parameter values, Eq.~\eqref{quartic k condition} has four roots over the complex numbers, counted with multiplicity, and gives four algebraic branches of $(k,\ep,n)$ whenever the denominators are nonzero.
The reality and metric-signature conditions must be imposed separately.
This quartic structure contrasts with the fifth-order condition obtained in the earlier Ovcharenko--Podolsk\'y parametrization~\cite{Ovcharenko:2025cpm}.

We now examine the Pleba\'nski--Demia\'nski limit of the black hole candidate conditions.
To identify a finite limiting solution, it is sufficient to take $\abs{b}\to0$ in the determining equations before solving for $(k,\ep,n)$.
With $(\alpha,\omega,a,l,m,\mr{q}^2)$ held fixed, the finite limiting values obey
\begin{align}
    \ep&=\frac{\omega^2k}{a^2-l^2}
    +4\alpha\frac{l}{\omega}m
    -(a^2+3l^2)\frac{\alpha^2}{\omega^2}
    (\omega^2k+\mr{q}^2)\,,\\
    n&=\frac{\omega^2kl}{a^2-l^2}
    -\alpha\frac{a^2-l^2}{\omega}m
    +(a^2-l^2)l\frac{\alpha^2}{\omega^2}
    (\omega^2k+\mr{q}^2)\,,
    \label{PD limit ep n}
\end{align}
obtained from Eqs.~\eqref{def sol ep} and \eqref{def sol n}, while Eq.~\eqref{a0 GP condition} becomes
\begin{align}
    \left(\frac{\omega^2}{a^2-l^2}+3\alpha^2l^2\right)k
    =1+2\alpha\frac{l}{\omega}m
    -3\alpha^2\frac{l^2}{\omega^2}\mr{q}^2\,.
    \label{PD limit k}
\end{align}
These are precisely the conditions for the Pleba\'nski--Demia\'nski generalized black holes with $\Lambda=0$ and $e^2+g^2=\mr{q}^2$, given in Eqs.~(14)--(16) of Ref.~\cite{Griffiths:2005qp}.

The relation between this limiting solution and the roots at $\abs{b}\neq0$ can be established directly from the polynomial.
Set $u\coloneqq\abs{b}^2$ and write $\msf{P}(k;u)$ for Eq.~\eqref{quartic k condition}, displaying its dependence on $u$.
For $a^2\neq l^2$ and $\omega\neq0$, let $k_{\mr{PD}}$ be the finite value determined by Eq.~\eqref{PD limit k}, assuming
\begin{align}
    \omega^2+3\alpha^2l^2(a^2-l^2)\neq0\,.
    \label{PD root nondegeneracy}
\end{align}
At $u=0$, the polynomial reduces to
\begin{align}
    \msf{P}(k;0)
    =-\omega^4\left[\omega^2+3\alpha^2l^2(a^2-l^2)\right]
    (k-k_{\mr{PD}})\,.
    \label{quartic finite PD limit}
\end{align}
Thus, $k_{\mr{PD}}$ is a simple root.
The implicit function theorem gives a unique local branch
\begin{align}
    k(u)=k_{\mr{PD}}+O(u)\,.
    \label{finite PD branch}
\end{align}
For real fixed parameters, this branch is real for sufficiently small $u\geq0$.
Since the common denominator of $\ep$ and $n$ tends to $(a^2-l^2)\omega^4\neq0$ along this branch, they also approach their finite Pleba\'nski--Demia\'nski values.
With $\beta$ and the phase $\gamma$ of $b$ held fixed, the results of the preceding subsection then show that both the metric and electromagnetic field reduce smoothly, to those of the charged Pleba\'nski--Demia\'nski spacetime in the Griffiths--Podolsk\'y form with $\Lambda=0$.

In the generic case in which Eq.~\eqref{quartic k condition} is quartic for $u\neq0$, the other three roots become unbounded as $u\to0$.
Taking the limit of the determining equations therefore identifies only the finite Pleba\'nski--Demia\'nski branch.
The reality and metric admissibility of the remaining branches, and whether they admit finite spacetime limits after additional coordinate or parameter rescalings, require separate analysis and are beyond the scope of the paper.
This behavior is analogous to the distinction made in Ref.~\cite{Ovcharenko:2025cpm} between the branch with a finite Pleba\'nski--Demia\'nski limit and the remaining branches of the earlier Ovcharenko--Podolsk\'y parametrization.

\paragraph{Metric and electromagnetic field}

For the full family satisfying the black hole candidate conditions, we now introduce $x=\cos\theta$ and take $\phiv\sim\phiv+2\pi\mc{C}$, where $\mc{C}>0$ is the conicity parameter.
We define the angular function $\mc{P}(\theta)$ by
\begin{align}
    \tilde{\mc{P}}(\cos\theta)
    &=\sin^2\theta\,\mc{P}(\theta)\,,\qquad
    \mc{P}(\theta)=1-a_3\cos\theta-a_4\cos^2\theta\,.
    \label{angular function theta}
\end{align}
Here $a_3$ and $a_4$ include the non-aligned corrections and are evaluated on a chosen solution of the black hole candidate conditions.
The conformal factor is evaluated at $x=\cos\theta$, and we write
\begin{align}
    \rhov^2(r,\theta)=(r+r_0)^2+(a\cos\theta+l+\omega\xi_0)^2\,.
\end{align}
The metric is
\begin{align}
  \!\!  ds^2&=\frac{1}{\Omega^2(r,\theta)}\biggl[
    -\frac{\mc{Q}(r)}{\rhov^2(r,\theta)}
    \left(dt-\left[a\sin^2\theta
    +2(l+\omega\xi_0)(1-\cos\theta)\right]d\phiv\right)^2\nn
    &\qquad\qquad  +\rhov^2(r,\theta)\left(\frac{dr^2}{\mc{Q}(r)}
    +\frac{d\theta^2}{\mc{P}(\theta)}\right)
    +\frac{\mc{P}(\theta)}{\rhov^2(r,\theta)}\sin^2\theta
    \left(a\,dt-\left[(r+r_0)^2+(a+l+\omega\xi_0)^2\right]
    d\phiv\right)^2\biggr]\,.
    \label{OP-generalized black holes}
\end{align}
For $0<\theta<\pi$, the null tetrad becomes
\begin{align}
    k^a
    &=\sqrt{\frac{\Omega^2}{2\mc{Q}\rhov^2}}
    \left[
    \bigl((r+r_0)^2+(a+l+\omega\xi_0)^2\bigr)(\del_t)^a
    +\mc{Q}(\del_r)^a+a(\del_{\phiv})^a
    \right]\,,
    \\
    \ell^a
    &=\sqrt{\frac{\Omega^2}{2\mc{Q}\rhov^2}}
    \left[
    \bigl((r+r_0)^2+(a+l+\omega\xi_0)^2\bigr)(\del_t)^a
    -\mc{Q}(\del_r)^a+a(\del_{\phiv})^a
    \right]\,,
    \\
    m^a
    &=-\sqrt{\frac{\Omega^2}{2\mc{P}\rhov^2}}
    \frac{1}{\sin\theta}
    \Bigl[
    \bigl(a\sin^2\theta
    +2(l+\omega\xi_0)(1-\cos\theta)\bigr)(\del_t)^a
    +(\del_{\phiv})^a
    +i\mc{P}\sin\theta(\del_{\theta})^a
    \Bigr]\,,
    \\
    \bar{m}^a
    &=-\sqrt{\frac{\Omega^2}{2\mc{P}\rhov^2}}
    \frac{1}{\sin\theta}
    \Bigl[
    \bigl(a\sin^2\theta
    +2(l+\omega\xi_0)(1-\cos\theta)\bigr)(\del_t)^a
    +(\del_{\phiv})^a
    -i\mc{P}\sin\theta(\del_{\theta})^a
    \Bigr]\,.
    \label{tetrad:GP-theta}
\end{align}
Thus, the non-aligned Maxwell scalars are
\begin{align}
    \phi_0=\phi_2
    &=\frac{ab}{\omega\Omega}
    \frac{\sqrt{\mc{P}\mc{Q}}\sin\theta}
    {(r+r_0)+i(a\cos\theta+l+\omega\xi_0)}\,,
    \label{phi_0-GPBH}
\end{align}
while the aligned Maxwell scalar is
\begin{align}
    \phi_1&=-\frac{\omega}{4a\bar{b}\sin\theta}
    \frac{\Omega^2}
    {[(r+r_0)+i(a\cos\theta+l+\omega\xi_0)]^2}\nn
    &\quad\times\Bigg[
    (a\cos\theta+l+\omega\xi_0)^2
    \del_{\theta}\left(\frac{\del_{r}\Omega}
    {a\cos\theta+l+\omega\xi_0}\right)
    -i(r+r_0)^2\del_{r}
    \left(\frac{\del_{\theta}\Omega}{r+r_0}\right)\Bigg]\,.
    \label{phi_1-GPBH}
\end{align}
A corresponding complex electromagnetic potential is
$\bm{\mathcal{A}}=\mathcal{A}_t\,dt
+\mathcal{A}_{\phiv}\,d\phiv$, with
\begin{align}
    \mathcal{A}_t
    &=\frac{\omega}{4a\bar{b}}\,
    \frac{\left(a\,\del_{r}\Omega
    +\dfrac{i}{\sin\theta}\del_{\theta}\Omega\right)}
    {(r+r_0)+i(a\cos\theta+l+\omega\xi_0)}\,,
    \nn
    \mathcal{A}_{\phiv}
    &=\frac{\omega}{4a\bar{b}}\Bigg[
    \Omega
    -\frac{(r+r_0)^2+(a+l+\omega\xi_0)^2}
    {(r+r_0)+i(a\cos\theta+l+\omega\xi_0)}
    \,\del_{r}\Omega
    -\frac{i\left[a\sin^2\theta
    +2(l+\omega\xi_0)(1-\cos\theta)\right]}
    {\left[(r+r_0)+i(a\cos\theta+l+\omega\xi_0)\right]\sin\theta}
    \,\del_{\theta}\Omega
    \Bigg]\,.
    \label{A-GPBH}
\end{align}
As before, the real electromagnetic potential is $\bm{A}=2\operatorname{Re}\bm{\mathcal{A}}$, up to gauge freedom.

\paragraph{Conicity}

The conicity parameter $\mc{C}$ can be chosen to remove the conical singularity at either axis. We assume that $\mc{P}(\theta)>0$ for $0\leq\theta\leq\pi$ and that $\Omega^2$ and $\rhov^2$ are finite and nonzero near the axis segments under consideration.
Define
\begin{align}
    \mc{P}_{+}&\coloneqq\mc{P}(0)=1-a_3-a_4\,,\\
    \mc{P}_{-}&\coloneqq\mc{P}(\pi)=1+a_3-a_4\,,
    \label{axis-Ppm}
\end{align}
where $+$ and $-$ denote the north and south axes, respectively.

The coordinates in Eq.~\eqref{OP-generalized black holes} are adapted to the north axis.
At fixed $t$ and $r$, the limiting circumference-to-radius ratio is
\begin{align}
    \lim_{\theta\to0}
    \frac{\text{circumference}}{\text{radius}}
    =\lim_{\theta\to0}
    \frac{2\pi\mc{C}\sqrt{g_{\phiv\phiv}}}
    {\theta\sqrt{g_{\theta\theta}}}
    =2\pi\mc{C}\mc{P}_{+}\,.
    \label{north-axis-circumference}
\end{align}
Thus, the north axis is free of a conical singularity when
\begin{align}
    \mc{C}=\mc{C}_{+}\coloneqq\frac{1}{\mc{P}_{+}}\,.
    \label{north-axis-normalization}
\end{align}

To examine the south axis, define the effective NUT-like parameter
\begin{align}
    \ell_{\mathrm{eff}}\coloneqq l+\omega\xi_0\,.
    \label{effective NUT parameter}
\end{align}
For $\ell_{\mathrm{eff}}\neq0$, the original time coordinate is not adapted to this axis.
We instead use the local time coordinates
\begin{align}
    t_{+}=t\,,\qquad
    t_{-}=t-4\ell_{\mathrm{eff}}\phiv\,.
    \label{axis-time-patches}
\end{align}
In the south-axis coordinates, the one-forms in the metric become
\begin{align}
    dt-\left[a\sin^2\theta
    +2\ell_{\mathrm{eff}}(1-\cos\theta)\right]d\phiv
    &=dt_{-}-\left[a\sin^2\theta
    -2\ell_{\mathrm{eff}}(1+\cos\theta)\right]d\phiv\,,\nn
    a\,dt-\left[(r+r_0)^2+(a+\ell_{\mathrm{eff}})^2\right]d\phiv
    &=a\,dt_{-}-\left[(r+r_0)^2+(a-\ell_{\mathrm{eff}})^2\right]d\phiv\,.
\end{align}
The local circumference-to-radius ratio, evaluated at fixed $t_{-}$ and $r$, is therefore
\begin{align}
    \lim_{\theta\to\pi}
    \frac{\text{circumference}}{\text{radius}}
    =2\pi\mc{C}\mc{P}_{-}\,.
    \label{south-axis-circumference}
\end{align}
The conical singularity at the south axis is removed by choosing
\begin{align}
    \mc{C}=\mc{C}_{-}\coloneqq\frac{1}{\mc{P}_{-}}\,.
    \label{south-axis-normalization}
\end{align}
This is the same local axis analysis as for the Pleba\'nski--Demia\'nski metric, with $l$ replaced by $\ell_{\mathrm{eff}}$ and $r$ by $r+r_0$
\cite{Griffiths:2005se,Podolsky:2021zwr}.

With either local axis-adapted time coordinate, the conical deficits are
\begin{align}
    \delta_{\pm}=2\pi\left(1-\mc{C}\mc{P}_{\pm}\right)\,.
    \label{axis-conicity}
\end{align}
In particular, regularizing the north axis leaves
\begin{align}
    \left.\delta_{-}\right|_{\mc{C}=\mc{C}_{+}}
    &=2\pi\left(1-\frac{\mc{P}_{-}}{\mc{P}_{+}}\right)
    =-\frac{4\pi a_3}{1-a_3-a_4}\,,
    \label{south-axis-deficit}
\end{align}
whereas regularizing the south axis leaves
\begin{align}
    \left.\delta_{+}\right|_{\mc{C}=\mc{C}_{-}}
    &=2\pi\left(1-\frac{\mc{P}_{+}}{\mc{P}_{-}}\right)
    =\frac{4\pi a_3}{1+a_3-a_4}\,.
    \label{north-axis-deficit}
\end{align}
A negative deficit denotes a conical excess.
A single azimuthal period can remove the conical singularities at both axes if and only if $a_3=0$, in which case $\mc{C}_{+}=\mc{C}_{-}=(1-a_4)^{-1}$.
This condition does not remove the NUT-like effect associated with $\ell_{\mathrm{eff}}\neq0$.

Two standard interpretations of this NUT-like effect are the Bonnor~\cite{Bonnor:1969ala} and Misner~\cite{Misner:1963fr} interpretations.
In the former, time is nonperiodic and the Misner string is retained and interpreted as a line source of angular momentum.\footnote{
Closed timelike curves can also occur without periodic time
identification.
For Taub--NUT spacetime, however, it is argued that their presence need not rule out its physical relevance~\cite{Clement:2015cxa}.
}
In the latter, the string is removed by patching the two local time coordinates with a compatible periodic identification of time, which produces closed timelike curves in regions where the identified time direction is timelike.

\subsubsection{Non-twisting sector}

In the non-twisting sector, the parameters $a$ and $l$ need not be introduced.
Starting from the metric in Eq.~\eqref{OP-PDalpha}, we impose the same black hole candidate conditions on the angular function $P_0$:
\begin{align}
    P_0(1)=P_0(-1)=0\,,\qquad P_0(0)=1\,.
    \label{non-twisting black hole conditions}
\end{align}
These are equivalent to the coefficient conditions \eqref{a GP condition odd}--\eqref{a0 GP condition}, now applied to $P_0$.
Since $P_0(0)=k$, the normalization fixes $k=1$.
For generic parameter values, the remaining two conditions determine $\ep$ and $n_0$ uniquely.
Their explicit expressions will be given in Sec.~\ref{Sec. family}.

\paragraph{Metric and electromagnetic field}

We introduce
\begin{align}
    \tau=t\,,\qquad \phi=\phiv\,,\qquad \xi=\cos\theta\,,\qquad
    \phiv\sim\phiv+2\pi\mc{C}\,,
\end{align}
where $\mc{C}>0$ is the conicity parameter.
The black hole candidate conditions give the factorization
\begin{align}
    P_0(\xi)=(1-\xi^2)\left[1+2n_0\xi+(1-\ep)\xi^2\right]\,.
\end{align}
We therefore define
\begin{align}
    P_0(\cos\theta)
    &=\sin^2\theta\,\mc{P}_0(\theta)\,,\qquad
    \mc{P}_0(\theta)=1+2n_0\cos\theta+(1-\ep)\cos^2\theta\,,\nn
    \mc{Q}_0(r)&\coloneqq Q_0(r)\,,\qquad
    \Omega_0(r,\theta)\coloneqq\Omega_0(r,\cos\theta)\,.
    \label{non-twisting GP metric functions}
\end{align}
The metric takes the form
\begin{align}
    ds^2
    &=\frac{1}{\Omega_0^2(r,\theta)}\biggl[
    -\frac{\mc{Q}_0(r)}{(r+\tilde{r}_0)^2}dt^2
    +(r+\tilde{r}_0)^2\left(
    \frac{dr^2}{\mc{Q}_0(r)}
    +\frac{d\theta^2}{\mc{P}_0(\theta)}
    +\mc{P}_0(\theta)\sin^2\theta\,d\phiv^2
    \right)\biggr]\,.
    \label{non-twisting GP metric}
\end{align}
Here $\tilde{r}_0$ is either of the two shifts $\tilde{r}_{0\pm}$ defined in Eq.~\eqref{xi0 r0 constraint in omega0}, corresponding to $\mr{OP}_{\alpha\pm}^0$.
The metric functions and shifts are understood to satisfy Eq.~\eqref{non-twisting black hole conditions}.

The principal null tetrad is
\begin{align}
    k^a
    &=\sqrt{\frac{\Omega_0^2}
    {2\mc{Q}_0(r+\tilde{r}_0)^2}}
    \left[
    (r+\tilde{r}_0)^2(\del_t)^a
    +\mc{Q}_0(\del_r)^a
    \right]\,,\quad 
    \ell^a
    =\sqrt{\frac{\Omega_0^2}
    {2\mc{Q}_0(r+\tilde{r}_0)^2}}
    \left[
    (r+\tilde{r}_0)^2(\del_t)^a
    -\mc{Q}_0(\del_r)^a
    \right]\,,
    \nn
    m^a
    &=\sqrt{\frac{\Omega_0^2}
    {2\mc{P}_0(r+\tilde{r}_0)^2}}
    \left[
    \frac{1}{\sin\theta}(\del_{\phiv})^a
    -i\mc{P}_0(\del_{\theta})^a
    \right]\,,\quad 
    \bar{m}^a
    =\sqrt{\frac{\Omega_0^2}
    {2\mc{P}_0(r+\tilde{r}_0)^2}}
    \left[
    \frac{1}{\sin\theta}(\del_{\phiv})^a
    +i\mc{P}_0(\del_{\theta})^a
    \right]\,.
    \label{tetrad:non-twisting-GP}
\end{align}
Then, the Maxwell scalars become
\begin{align}
    \phi_0&=\phi_2
    =\frac{b\sqrt{\mc{P}_0\mc{Q}_0}\sin\theta}
    {\Omega_0(r+\tilde{r}_0)}\,,
    \label{non-twisting GP phi0}\\
    \phi_1
    &=\frac{i\Omega_0^2}{4\bar{b}\sin\theta}
    \del_r\left(
    \frac{\del_{\theta}\Omega_0}{r+\tilde{r}_0}
    \right)\,.
    \label{non-twisting GP phi1}
\end{align}
A corresponding complex electromagnetic potential is
\begin{align}
    \mathcal{A}_{\mu}dx^{\mu}
    =\frac{1}{4\bar{b}}\left[
    \frac{i\del_{\theta}\Omega_0}
    {(r+\tilde{r}_0)\sin\theta}\,dt
    +\left((r+\tilde{r}_0)\del_r\Omega_0-\Omega_0\right)d\phiv
    \right]\,.
    \label{non-twisting GP potential}
\end{align}
As before, the real potential is $\bm{A}=2\operatorname{Re}\bm{\mathcal{A}}$, up to gauge freedom.

\paragraph{Conicity}

The metric has no cross term of $dt\,d\phiv$, and the same time coordinate is adapted to both axes.
There is therefore no NUT-like mismatch between the two axis patches and no associated Misner string.
Conical singularities may nevertheless remain, and the local conicity analysis of the preceding subsection applies without a change of time coordinate.
Under the same local regularity assumptions, define
\begin{align}
    \mc{P}_{0+}&\coloneqq\mc{P}_0(0)=2-\ep+2n_0\,,\qquad
    \mc{P}_{0-}\coloneqq\mc{P}_0(\pi)=2-\ep-2n_0\,.
    \label{non-twisting general axis values}
\end{align}
The conical deficits are
\begin{align}
    \delta_{0\pm}=2\pi\left(1-\mc{C}\mc{P}_{0\pm}\right)\,.
    \label{non-twisting general conicity}
\end{align}
Thus, choosing $\mc{C}=\mc{P}_{0+}^{-1}$ or $\mc{C}=\mc{P}_{0-}^{-1}$ regularizes the north or south axis, respectively.
Both axes can be regularized simultaneously if and only if $\mc{P}_{0+}=\mc{P}_{0-}$, equivalently $n_0=0$, with $\mc{C}=(2-\ep)^{-1}$.

In Sec.~\ref{Sec. family}, we give the explicit solution of the black hole candidate conditions and express the resulting metric functions and shifts in a dimensionful parametrization, before analyzing their properties and further limiting spacetimes.


\section{Specific subclasses of the generalized black hole spacetime family}
\label{Sec. family}

In this section, we give explicit expressions for the non-twisting and $l=0$ twisting generalized black hole families obtained in Sec.~\ref{Sec. OP-GP}.
Using dimensionful parametrizations, we analyze their black hole and acceleration horizons, extremality conditions, and conicity, and examine special subclasses and further limiting spacetimes.
The parameter names follow the terminology of the Pleba\'nski--Demia\'nski family.
The corresponding physical interpretations hold in appropriate limits, but not necessarily in general.

\paragraph{Dimensions}

The coordinates, parameters, and line element used so far are dimensionless.
As explained in Sec.~\ref{Sec. review}, we introduce an arbitrary reference length $L$ to restore dimensions, without counting it as an independent solution parameter.
For the Griffiths--Podolsk\'y form, the replacements
\begin{align}
    t&\to Lt\,,\qquad
    r\to Lr\,,\qquad
    r_0\to Lr_0\,,\qquad
    a\to La\,,\qquad
    l\to Ll\,,\qquad
    \omega\xi_0\to L\omega\xi_0\,,\nn
    \rhov^2&\to L^2\rhov^2\,,\qquad
    \mc{Q}\to L^2\mc{Q}\,,
    \label{GP dimensional scaling}
\end{align}
give $ds^2\to L^2ds^2$, while the angular coordinates, the conicity parameter, $\mc{P}$, and $\Omega^2$ remain dimensionless.
In the non-twisting sector, $\tilde{r}_0$ and $\mc{Q}_0$ scale in the same way as $r_0$ and $\mc{Q}$, respectively.
For each family, we first give the explicit metric functions and electromagnetic field in dimensionless variables.
We then determine the corresponding parameter rescalings and introduce a dimensionful parametrization before analyzing its properties and further limiting spacetimes.
After restoring dimensions, we retain the same symbols unless a new parameter is explicitly introduced.

\subsection{Non-twisting sector with charge parameter $\mr{q}$}

We first give the explicit form of the non-twisting families $\mr{OP}_{\alpha\pm}^{0}$, retaining the charge parameter $\mr{q}$.
We introduce
\begin{align}
    B\coloneqq2\abs{b}>0\,,\qquad
    b=\frac{B}{2}\e{i\gamma}\,.
    \label{non-twisting B definition}
\end{align}
The black hole candidate conditions \eqref{non-twisting black hole conditions} give
\begin{align}
    k&=1\,,\nn
    \ep&=\frac{1-B^2m^2-\mr{q}^2(\alpha^2+B^4\mr{q}^2)}
    {1-B^2\mr{q}^2}
    =1-\frac{\alpha^2\mr{q}^2+B^2(m^2-\mr{q}^2)+B^4\mr{q}^4}
    {1-B^2\mr{q}^2}\,,\nn
    n_0&=-\frac{m\alpha}{1-B^2\mr{q}^2}\,.
    \label{GBH condition for nonomega}
\end{align}
We assume $1-B^2\mr{q}^2\neq0$ and introduce the quantities
\begin{align}
    I_0\coloneqq1-\frac{B^2m^2}{1-B^2\mr{q}^2}\,,\qquad
    J_0\coloneqq\frac{\alpha^2}{B^2(1-B^2\mr{q}^2)}-1\,.
    \label{non-twisting I0 J0}
\end{align}

\paragraph{Metric and electromagnetic field}

The metrics of $\mr{OP}_{\alpha\pm}^{0}$ are
\begin{align}
    ds_{\pm}^2
    &=\frac{1}{\Omega_0^2(r,\theta)}\biggl[
    -\frac{\mc{Q}_0(r)}{(r+\tilde{r}_{0\pm})^2}dt^2
    +(r+\tilde{r}_{0\pm})^2\left(
    \frac{dr^2}{\mc{Q}_0(r)}
    +\frac{d\theta^2}{\mc{P}_0(\theta)}
    +\mc{P}_0(\theta)\sin^2\theta\,d\phiv^2
    \right)\biggr]\,,
    \label{non-twisting explicit metric}
\end{align}
where
\begin{align}
    \mc{P}_0(\theta)
    &=1-\frac{2m\alpha}{1-B^2\mr{q}^2}\cos\theta
    +\left(1-I_0+B^2\mr{q}^2J_0\right)\cos^2\theta\,,\nn
    \mc{Q}_0(r)
    &=(\mr{q}^2-2mr+I_0r^2)(1-B^2J_0r^2)\,,\nn
    \Omega_0^2(r,\theta)
    &=(1-\alpha r\cos\theta)^2
    +B^2(r^2-\mr{q}^2\cos^2\theta)
    +2B^2mr\cos\theta
    \left(\cos\theta-\frac{\alpha r}{1-B^2\mr{q}^2}\right)\nn
    &\quad-\left[I_0-B^2\mr{q}^2(1+J_0)\right]
    B^2r^2\cos^2\theta\,.
    \label{non-twisting explicit metric functions}
\end{align}
With the branch convention $\tilde{r}_{0\pm}=\msf{B}_0\pm\abs{\msf{R}_0}$ introduced in Eq.~\eqref{xi0 r0 constraint in omega0}, the shifts are
\begin{align}
    \tilde{r}_{0\pm}
    =\frac{m}{1-I_0+J_0}
    \pm\frac{\sqrt{1-I_0+B^2\mr{q}^2J_0}}
    {B\abs{1-I_0+J_0}}\,.
    \label{eq:r0_pm}
\end{align}
Here we assume $1-I_0+J_0\neq0$ and $1-I_0+B^2\mr{q}^2J_0\geq0$.
When the radicand vanishes, the two branches coincide.
We denote this degenerate subclass by $\mr{OP}_{\alpha0}^{0}$ and examine it below.

The corresponding Maxwell scalars are
\begin{align}
    \phi_0^{\pm}&=\phi_2^{\pm}
    =\frac{B\e{i\gamma}\sqrt{\mc{P}_0\mc{Q}_0}\sin\theta}
    {2\Omega_0(r+\tilde{r}_{0\pm})}\,,
    \label{phi0 OP0}\\
    \phi_1^{\pm}
    &=\frac{i\e{i\gamma}\Omega_0^2}{2B\sin\theta}
    \del_{r}\left(
    \frac{\del_{\theta}\Omega_0}{r+\tilde{r}_{0\pm}}
    \right)\,.
    \label{phi1 OP0}
\end{align}
A corresponding complex electromagnetic potential is
\begin{align}
    \mathcal{A}_{\mu}^{\pm}dx^{\mu}
    =\frac{\e{i\gamma}}{2B}\left[
    \left((r+\tilde{r}_{0\pm})\del_{r}\Omega_0-\Omega_0\right)d\phiv
    +\frac{i\del_{\theta}\Omega_0}
    {(r+\tilde{r}_{0\pm})\sin\theta}\,dt
    \right]\,.
    \label{non-twisting explicit potential}
\end{align}
As before, the real potential is $\bm{A}^{\pm}=2\operatorname{Re}\bm{\mathcal{A}}^{\pm}$, up to gauge freedom.

\paragraph{Dimensions}

The explicit metric functions show that the parameter rescalings consistent with Eq.~\eqref{GP dimensional scaling} are
\begin{align}
    B\to L^{-1}B\,,\qquad
    m\to Lm\,,\qquad
    \mr{q}^2\to L^2\mr{q}^2\,,\qquad
    \alpha\to L^{-1}\alpha\,.
    \label{non-twisting parameter dimensions}
\end{align}
Thus, $m$ and $\mr{q}$ have dimensions of length, while $B$ and $\alpha$ have dimensions of inverse length.
The quantity $\gamma$ remains dimensionless.
These rescalings preserve the functional form of the expressions above, with $\mc{Q}_0$ having dimensions of length squared and $\tilde{r}_{0\pm}$ having dimensions of length.
In the remainder of this subsection, we use the resulting dimensionful parametrization.

\paragraph{Horizons and extremality}

The first factor of $\mc{Q}_0$ gives the black hole horizon radii
\begin{align}
    r_b^{\pm}=\frac{m\pm\sqrt{m^2-I_0\mr{q}^2}}{I_0}\,,
    \label{non-twisting black hole horizons}
\end{align}
provided that $I_0\neq0$ and $m^2\geq I_0\mr{q}^2$.
In a black hole domain with $I_0>0$ in which both roots represent regular horizons, $r_b^{+}$ and $r_b^{-}$ are the outer and inner horizons, respectively.
For $J_0>0$, the second factor gives the acceleration horizon radii
\begin{align}
    r_a^{\pm}=\pm\frac{1}{B\sqrt{J_0}}\,.
    \label{non-twisting acceleration horizons}
\end{align}
The coordinate locations of these roots are common to the two branches $\mr{OP}_{\alpha\pm}^{0}$, since $\mc{Q}_0$ does not depend on the choice of $\tilde{r}_{0\pm}$.
The extremality condition associated with coalescence of the two black hole roots is
\begin{align}
    m^2=I_0\mr{q}^2
    \quad\Longleftrightarrow\quad
    m^2=(1-B^2\mr{q}^2)\mr{q}^2\,,
    \label{non-twisting extremality}
\end{align}
in which case
\begin{align}
    I_0=1-B^2\mr{q}^2\,.
\end{align}

\paragraph{Conicity}

Using the notation of Eq.~\eqref{non-twisting general axis values}, the north- and south-axis values are
\begin{align}
    \mc{P}_{0\pm}
    =2-I_0+B^2\mr{q}^2J_0
    \mp\frac{2m\alpha}{1-B^2\mr{q}^2}\,.
    \label{non-twisting-axis-values}
\end{align}
Under the same local regularity assumptions as in Sec.~\ref{Sec. OP-GP}, including $\mc{P}_0(\theta)>0$ for $0\leq\theta\leq\pi$, we choose $\mc{C}=\mc{P}_{0+}^{-1}$ to regularize the north axis.
The remaining conical deficit on the south axis is
\begin{align}
    \delta_{0-}
    =2\pi\left(1-\frac{\mc{P}_{0-}}{\mc{P}_{0+}}\right)\,.
    \label{non-twisting-south-deficit}
\end{align}
Alternatively, $\mc{C}=\mc{P}_{0-}^{-1}$ regularizes the south axis.
The condition for simultaneously regularizing both axes is $\mc{P}_{0+}=\mc{P}_{0-}$, which here is equivalent to $m\alpha=0$.

The genuine Griffiths--Podolsk\'y forms of $\mr{OP}_{\alpha\pm}^{0}$, obtained by the radial translations $r'_{\pm}=r+\tilde{r}_{0\pm}$, are given in Appendix~\ref{App. genuine GP non-twisting}.
There we introduce the apparent mass and charge parameters of the two branches and discuss the distinction between $\mr{q}$ and the electromagnetic flux charges.

\subsubsection{Degenerate subclass: $\mr{OP}_{\alpha0}^{0}$}

The two branches coincide when the radicand in Eq.~\eqref{eq:r0_pm} vanishes:
\begin{align}
    1-I_0+B^2\mr{q}^2J_0=0
    \quad\Longleftrightarrow\quad
    \alpha^2\mr{q}^2
    +B^2\left[m^2-(1-B^2\mr{q}^2)\mr{q}^2\right]=0\,.
    \label{OP00-condition}
\end{align}
This gives the degenerate subclass $\mr{OP}_{\alpha0}^{0}$ introduced above.
The common shift is
\begin{align}
    \tilde{r}_{0+}=\tilde{r}_{0-}
    =\tilde{r}_{00}
    \coloneqq\frac{m}{1-I_0+J_0}
    =\frac{m}{J_0(1-B^2\mr{q}^2)}\,.
    \label{non-twisting degenerate shift}
\end{align}
The metric functions simplify to
\begin{align}
    \mc{P}_0(\theta)
    &=1-\frac{2m\alpha}{1-B^2\mr{q}^2}\cos\theta\,,\nn
    \mc{Q}_0(r)
    &=(\mr{q}^2-2mr+I_0r^2)(1-B^2J_0r^2)\,,\nn
    \Omega_0^2(r,\theta)
    &=(1-\alpha r\cos\theta)^2
    +B^2\biggl[
    r^2-\mr{q}^2\cos^2\theta
    +2mr\cos\theta
    \left(\cos\theta-\frac{\alpha r}{1-B^2\mr{q}^2}\right)\nn
    &\hspace{42mm}
    -(1-B^2\mr{q}^2)r^2\cos^2\theta
    \biggr]\,.
    \label{non-twisting degenerate metric functions}
\end{align}
The Maxwell scalars and complex electromagnetic potential follow from Eqs.~\eqref{phi0 OP0}, \eqref{phi1 OP0}, and \eqref{non-twisting explicit potential} by setting $\tilde{r}_{0\pm}=\tilde{r}_{00}$.

\paragraph{Non-accelerating limit: $\alpha\to0$}

We take $\alpha\to0$ within this degenerate subclass, keeping $B$ and $\mr{q}$ fixed and adjusting $m$ to satisfy Eq.~\eqref{OP00-condition}.
Then $J_0\to-1$, and the limiting parameters obey
\begin{align}
    m^2=(1-B^2\mr{q}^2)\mr{q}^2\,,\qquad
    I_0=1-B^2\mr{q}^2\,.
    \label{non-twisting degenerate nonaccelerating condition}
\end{align}
This condition coincides with the extremality condition~\eqref{non-twisting extremality}.
The quadratic factor of $\mc{Q}_0$ has a double zero at
\begin{align}
    r_\ast\coloneqq\frac{m}{I_0}
    =-\tilde{r}_{00}
    =\frac{m}{1-B^2\mr{q}^2}\,.
    \label{non-twisting degenerate double root}
\end{align}
The limiting metric functions are
\begin{align}
    \mc{P}_0(\theta)&=1\,,\nn
    \mc{Q}_0(r)&=I_0(r-r_\ast)^2(1+B^2r^2)\,,\nn
    \Omega_0^2(r,\theta)
    &=1+B^2r^2-B^2I_0(r-r_\ast)^2\cos^2\theta\,.
    \label{non-twisting degenerate nonaccelerating functions}
\end{align}
In this limit, the double zero of $\mc{Q}_0$ coincides with the double zero of $(r-r_\ast)^2$ in the metric.
Consequently, the corresponding factors cancel in the norm of the
static Killing vector:
\begin{align}
    g_{tt}
    =-\frac{I_0(1+B^2r^2)}{\Omega_0^2(r,\theta)}
    \to  -I_0\neq0
    \qquad\text{as}\quad r\to r_\ast\,.
    \label{non-twisting degenerate Killing norm}
\end{align}
Thus, this Killing horizon candidate is not an extremal Killing horizon.

\subsubsection{Vanishing charge parameter: $\mr{q}=0$}

Setting $\mr{q}=0$ recovers the previously studied non-twisting Ovcharenko--Podolsk\'y family~\cite{Ovcharenko:2025cpm,Ovcharenko:2026byw}.
In our parametrization, the metric functions are
\begin{align}
    \mc{P}_0(\theta)
    &=1-2m\alpha\cos\theta+B^2m^2\cos^2\theta\,,\nn
    \mc{Q}_0(r)
    &=\left[-2mr+(1-B^2m^2)r^2\right]
    \left[1-(\alpha^2-B^2)r^2\right]\,,\nn
    \Omega_0^2(r,\theta)
    &=(1-\alpha r\cos\theta)^2+B^2\biggl[
    r^2+2mr\cos\theta(\cos\theta-\alpha r)
    -(1-B^2m^2)r^2\cos^2\theta\biggr]\,.
    \label{non-twisting q0 metric functions}
\end{align}
The shift constraint reduces to
\begin{align}
    (\tilde{r}_0-\msf{B}_0)^2=\msf{B}_0^2\,,\qquad
    \msf{B}_0
    =\frac{B^2m}{\alpha^2-B^2(1-B^2m^2)}\,,
    \label{non-twisting q0 shift constraint}
\end{align}
since $\msf{R}_0^2=\msf{B}_0^2$.
With the branch convention adopted above,
\begin{align}
    \tilde{r}_{0\pm}
    &=\msf{B}_0\pm\abs{\msf{B}_0}\,,\nn
    \left\{\tilde{r}_{0+}\,,\,\tilde{r}_{0-}\right\}
    &=\left\{0\,,\,\frac{2B^2m}{\alpha^2-B^2(1-B^2m^2)}\right\}\,.
    \label{r0 condition}
\end{align}
For $\msf{B}_0>0$, the nonzero-shift solution belongs to $\mr{OP}_{\alpha+}^{0}$ and the zero-shift solution to $\mr{OP}_{\alpha-}^{0}$.
These assignments are reversed when $\msf{B}_0<0$.
For $\alpha>0$, the metric functions and the two shift values agree with the Griffiths--Podolsk\'y representation in Sec.~III.B of Ref.~\cite{Ovcharenko:2026byw}, before its further parameter redefinitions, with $B=2\alpha\abs{c_{\mr{OP}}}$ as follows from Eq.~\eqref{relation to OP and F parameter in omega0}.
For $\mr{q}=0$ and $B>0$, Eq.~\eqref{OP00-condition} reduces to $m=0$.
The two shifts then vanish and the branches coincide in $\mr{OP}_{\alpha0}^{0}$.

\subsubsection{Charged $C$-metric limit: $B\to0$}

We take $B\to0$ with $(\alpha,m,\mr{q}^2,\gamma)$ fixed and $\alpha>0,\mr{q}^2 \geq 0$.
Using
\begin{align}
    B^2J_0=\frac{\alpha^2}{1-B^2\mr{q}^2}-B^2\to\alpha^2\,,\qquad
    I_0\to1\,,\qquad \tilde{r}_{0\pm}\to0\,,
\end{align}
the metric functions reduce to
\begin{align}
    \mc{P}_0(\theta)
    &\to1-2\alpha m\cos\theta
    +\alpha^2\mr{q}^2\cos^2\theta\,,\nn
    \mc{Q}_0(r)
    &\to(\mr{q}^2-2mr+r^2)(1-\alpha^2r^2)\,,\nn
    \Omega_0^2(r,\theta)&\to(1-\alpha r\cos\theta)^2\,.
    \label{non-twisting charged C metric limit}
\end{align}
These give the charged $C$-metric, namely the non-twisting Pleba\'nski--Demia\'nski generalized black hole with $\Lambda=0$~\cite{Griffiths:2005qp,Griffiths:2009dfa}.
The Maxwell scalars have the aligned limit
\begin{align}
    \phi_0^{\pm}=\phi_2^{\pm}&\to0\,,\qquad
    \phi_1^{\pm}\to
    \pm\frac{i\mr{q}\e{i\gamma}}{2}
    \left(\frac{1-\alpha r\cos\theta}{r}\right)^2
    =\frac{e+ig}{2}
    \left(\frac{1-\alpha r\cos\theta}{r}\right)^2\,,
\end{align}
where $e=\mp\mr{q}\sin\gamma$ and $g=\pm\mr{q}\cos\gamma$ for the two branches, respectively, so that $e^2+g^2=\mr{q}^2$.
The subsequent $\alpha\to0$ limit gives the Reissner--Nordstr\"om solution.

\subsection{Twisting sector with vanishing NUT-like parameter $l$}

We next give the explicit form of the $l=0$ twisting family, retaining the charge parameter $\mr{q}$.
Setting $l=0$ simplifies the black hole candidate conditions \eqref{a GP condition odd}--\eqref{a0 GP condition}.
In particular, the constant coefficient of the angular function is
\begin{align}
    a_0=\frac{\omega^2k}{a^2}\,,
    \label{twisting l0 a0}
\end{align}
so the normalization $a_0=1$ fixes $k=a^2/\omega^2$.
We introduce
\begin{align}
    B\coloneqq2a\abs{b}\,,\qquad
    b=\frac{B}{2a}\e{i\gamma}\,.
    \label{twisting B definition}
\end{align}
Here $a\neq0$ and $b\neq0$.
With this definition, $B$ has the same sign as $a$.
The three black hole candidate conditions give
\begin{align}
    k&=\frac{a^2}{\omega^2}\,,\nn
    \ep&=I-B^2(a^2+\mr{q}^2)J\,,\nn
    n&=-\frac{m\alpha a^2\omega}
    {\omega^2-B^2(a^2+\mr{q}^2)}\,,
    \label{twisting l0 candidate solution}
\end{align}
where
\begin{align}
    I&\coloneqq1-\frac{B^2m^2}{\omega^2-B^2(a^2+\mr{q}^2)}\,,\qquad
    J\coloneqq\frac{\alpha^2a^2}
    {B^2\left[\omega^2-B^2(a^2+\mr{q}^2)\right]}
    -\frac{1}{\omega^2}\,.
    \label{twisting l0 I J general omega}
\end{align}
These expressions apply when $\omega^2-B^2(a^2+\mr{q}^2)\neq0$.
Using the remaining parametrization freedom, we henceforth choose $\omega=1$~\cite{Griffiths:2009dfa,Ovcharenko:2025cpm}.
Then
\begin{align}
    I&=1-\frac{B^2m^2}{1-B^2(a^2+\mr{q}^2)}\,,\qquad
    J=\frac{\alpha^2a^2}
    {B^2\left[1-B^2(a^2+\mr{q}^2)\right]}-1\,.
    \label{twisting l0 I J}
\end{align}
Although $l=0$, the effective NUT-like parameter $\ell_{\mathrm{eff}}=\omega\xi_0$ need not vanish.

\paragraph{Metric and electromagnetic field}

The metric takes the form
\begin{align}
    ds^2
    &=\frac{1}{\Omega^2(r,\theta)}\biggl[
    -\frac{\mc{Q}(r)}{\rhov^2(r,\theta)}
    \left(dt-\left[a\sin^2\theta
    +2\omega\xi_0(1-\cos\theta)\right]d\phiv\right)^2\nn
    &\quad+\rhov^2(r,\theta)\left(
    \frac{dr^2}{\mc{Q}(r)}+\frac{d\theta^2}{\mc{P}(\theta)}\right)
    +\frac{\mc{P}(\theta)\sin^2\theta}{\rhov^2(r,\theta)}
    \left(a\,dt-\left[(r+r_0)^2+(a+\omega\xi_0)^2\right]
    d\phiv\right)^2\biggr]\,,
    \label{non-NUTty OPBH}
\end{align}
with
\begin{align}
    \mc{P}(\theta)
    &=1-\frac{2m\alpha a}{1-B^2(a^2+\mr{q}^2)}\cos\theta
    +\left[1-I+B^2(a^2+\mr{q}^2)J\right]\cos^2\theta\,,\nn
    \mc{Q}(r)
    &=\left(a^2+\mr{q}^2-2mr+Ir^2\right)
    \left(1-B^2Jr^2\right)\,,\nn
    \Omega^2(r,\theta)
    &=(1-\alpha ar\cos\theta)^2+B^2\biggl[
    r^2-(a^2+\mr{q}^2)\cos^2\theta
    +2mr\cos\theta\left(\cos\theta
    -\frac{\alpha ar}{1-B^2(a^2+\mr{q}^2)}\right)\nn
    &\hspace{42mm}
    -\left[I-B^2(a^2+\mr{q}^2)(1+J)\right]
    r^2\cos^2\theta\biggr]\,,\nn
    \rhov^2(r,\theta)
    &=(r+r_0)^2+(a\cos\theta+\omega\xi_0)^2\,.
    \label{twisting l0 metric functions}
\end{align}
The shifts are
\begin{align}
    r_0
    &=\frac{m+\sqrt{Km^2+J\mr{q}^2}\cos\beta}{1-I+J}\,,\nn
    \omega\xi_0
    &=\frac{1}{1-I+J}\left[
    -\frac{m\alpha a^2}{1-B^2(a^2+\mr{q}^2)}
    +\sqrt{Km^2+J\mr{q}^2}\sin\beta\right]\,,
    \label{twisting l0 shifts}
\end{align}
where
\begin{align}
    K
    &\coloneqq\frac{1+JB^2a^2}{1-B^2(a^2+\mr{q}^2)}\nn
    &=\frac{1-B^2a^2}{1-B^2(a^2+\mr{q}^2)}
    +\frac{\alpha^2a^4}{\left[1-B^2(a^2+\mr{q}^2)\right]^2}\,.
    \label{twisting l0 K}
\end{align}
We assume $1-I+J\neq0$ and $Km^2+J\mr{q}^2\geq0$.

The Maxwell scalars are
\begin{align}
    \phi_0&=\phi_2
    =\frac{B\e{i\gamma}}{2\Omega}
    \frac{\sqrt{\mc{P}\mc{Q}}\sin\theta}
    {(r+r_0)+i(a\cos\theta+\omega\xi_0)}\,,
    \label{phi_0-nonlBH}\\
    \phi_1
    &=-\frac{\e{i\gamma}}{2B\sin\theta}
    \frac{\Omega^2}
    {\left[(r+r_0)+i(a\cos\theta+\omega\xi_0)\right]^2}\nn
    &\quad\times\Bigg[
    (a\cos\theta+\omega\xi_0)^2
    \del_{\theta}\left(
    \frac{\del_{r}\Omega}{a\cos\theta+\omega\xi_0}\right)
    -i(r+r_0)^2\del_{r}\left(
    \frac{\del_{\theta}\Omega}{r+r_0}\right)\Bigg]\,.
    \label{phi_1-nonlBH}
\end{align}
A corresponding complex electromagnetic potential is
$\bm{\mathcal{A}}=\mathcal{A}_t\,dt
+\mathcal{A}_{\phiv}\,d\phiv$, with
\begin{align}
    \mathcal{A}_t
    &=\frac{\e{i\gamma}}{2B}\,
    \frac{\left(a\,\del_{r}\Omega
    +\dfrac{i}{\sin\theta}\del_{\theta}\Omega\right)}
    {(r+r_0)+i(a\cos\theta+\omega\xi_0)}\,,
    \nn
    \mathcal{A}_{\phiv}
    &=\frac{\e{i\gamma}}{2B}\Bigg[
    \Omega
    -\frac{(r+r_0)^2+(a+\omega\xi_0)^2}
    {(r+r_0)+i(a\cos\theta+\omega\xi_0)}
    \,\del_{r}\Omega
    -\frac{i\left[a\sin^2\theta
    +2\omega\xi_0(1-\cos\theta)\right]}
    {\left[(r+r_0)+i(a\cos\theta+\omega\xi_0)\right]\sin\theta}
    \,\del_{\theta}\Omega
    \Bigg]\,.
    \label{A-nonlBH}
\end{align}
As before, the real electromagnetic potential is $\bm{A}=2\operatorname{Re}\bm{\mathcal{A}}$, up to gauge freedom.

\paragraph{Dimensions}

The explicit metric functions show that the parameter rescalings consistent with Eq.~\eqref{GP dimensional scaling} are
\begin{align}
    B\to L^{-1}B\,,\qquad
    m\to Lm\,,\qquad
    \mr{q}^2\to L^2\mr{q}^2\,,\qquad
    \alpha\to L^{-2}\alpha\,,
    \label{twisting parameter dimensions}
\end{align}
in addition to $a\to La$.
Thus, the parameter $\alpha$ used in this representation has dimensions of inverse length squared.
We instead introduce the acceleration parameter
\begin{align}
    \upalpha\coloneqq a\alpha\,,
    \label{twisting acceleration parameter}
\end{align}
which scales as $\upalpha\to L^{-1}\upalpha$ and has dimensions of inverse length.
Now, $m, a$ and $\mr{q}$ have dimensions of length, while $B$ and $\upalpha$ have dimensions of inverse length. 
The quantities $\beta$ and $\gamma$ are dimensionless.
In terms of $\upalpha$, we have
\begin{align}
    J
    &=\frac{\upalpha^2}
    {B^2\left[1-B^2(a^2+\mr{q}^2)\right]}-1\,,\nn
    K
    &=\frac{1+JB^2a^2}{1-B^2(a^2+\mr{q}^2)}\nn
    &=\frac{1-B^2a^2}{1-B^2(a^2+\mr{q}^2)}
    +\frac{a^2\upalpha^2}
    {\left[1-B^2(a^2+\mr{q}^2)\right]^2}\,.
    \label{twisting l0 dimensionful J K}
\end{align}
The affected metric functions and angular shift become
\begin{align}
    \mc{P}(\theta)
    &=1-\frac{2m\upalpha}{1-B^2(a^2+\mr{q}^2)}\cos\theta
    +\left[1-I+B^2(a^2+\mr{q}^2)J\right]\cos^2\theta\,,\nn
    \Omega^2(r,\theta)
    &=(1-\upalpha r\cos\theta)^2+B^2\biggl[
    r^2-(a^2+\mr{q}^2)\cos^2\theta
    +2mr\cos\theta\left(\cos\theta
    -\frac{\upalpha r}{1-B^2(a^2+\mr{q}^2)}\right)\nn
    &\hspace{42mm}
    -\left[I-B^2(a^2+\mr{q}^2)(1+J)\right]
    r^2\cos^2\theta\biggr]\,,\nn
    \omega\xi_0
    &=\frac{1}{1-I+J}\left[
    -\frac{m\upalpha a}{1-B^2(a^2+\mr{q}^2)}
    +\sqrt{Km^2+J\mr{q}^2}\sin\beta\right]\,.
    \label{twisting l0 dimensionful functions}
\end{align}
The expressions for $I$, $r_0$, $\mc{Q}$, $\rhov^2$, and the electromagnetic field retain their functional form, with $J$ and $K$ given by Eq.~\eqref{twisting l0 dimensionful J K}.
In the remainder of this subsection, we use the resulting dimensionful parametrization.

\paragraph{Horizons and extremality}

The first factor of $\mc{Q}$ gives the black hole horizon radii
\begin{align}
    r_b^{\pm}
    =\frac{m\pm\sqrt{m^2-I(a^2+\mr{q}^2)}}{I}\,,
    \label{twisting black hole horizons}
\end{align}
provided that $I\neq0$ and $m^2\geq I(a^2+\mr{q}^2)$.
In a black hole domain with $I>0$ in which both roots represent regular horizons, $r_b^{+}$ and $r_b^{-}$ are the outer and inner horizons, respectively.
For $J>0$, the second factor gives the acceleration horizon radii
\begin{align}
    r_a^{\pm}=\pm\frac{1}{\abs{B}\sqrt{J}}\,.
    \label{twisting acceleration horizons}
\end{align}
The extremality condition is
\begin{align}
    m^2=I(a^2+\mr{q}^2)
    \quad\Longleftrightarrow\quad
    m^2=\left[1-B^2(a^2+\mr{q}^2)\right](a^2+\mr{q}^2)\,,
    \label{twisting extremality}
\end{align}
in which case
\begin{align}
    I=1-B^2(a^2+\mr{q}^2)\,.
\end{align}

\paragraph{Conicity}

The local conicity analysis in Sec.~\ref{Sec. OP-GP} applies with $\ell_{\mathrm{eff}}=\omega\xi_0$ and
\begin{align}
    \mc{P}_{\pm}
    \coloneqq 
    \begin{dcases}
        \mc{P}(0) :& \text{for $+$}\\
        \mc{P}(\pi): & \text{for $-$}
    \end{dcases}
    =2-I+B^2(a^2+\mr{q}^2)J
    \mp\frac{2m\upalpha}{1-B^2(a^2+\mr{q}^2)}\,.
    \label{twisting l0 axis values}
\end{align}
Under the same local regularity assumptions, including $\mc{P}(\theta)>0$ for $0\leq\theta\leq\pi$, choosing $\mc{C}=\mc{P}_{+}^{-1}$ regularizes the north axis and leaves
\begin{align}
    \delta_{-}
    =2\pi\left(1-\frac{\mc{P}_{-}}{\mc{P}_{+}}\right)
    \label{twisting l0 south deficit}
\end{align}
on the south axis.
Alternatively, $\mc{C}=\mc{P}_{-}^{-1}$ regularizes the south axis.
For $\ell_{\mathrm{eff}}\neq0$, the south-axis expressions are evaluated in the local time coordinate $t_{-}=t-4\ell_{\mathrm{eff}}\phiv$, as in Eq.~\eqref{axis-time-patches}.
Both conical singularities can be removed with a single azimuthal period if and only if $m\upalpha=0$, under the assumptions above.
This condition does not remove the NUT-like effect associated with $\ell_{\mathrm{eff}}\neq0$.

The genuine Griffiths--Podolsk\'y form of this $l=0$ family, obtained by the radial translation $r'=r+r_0$, is noted in Appendix~\ref{App. genuine GP}.
The apparent mass and charge parameters defined there will be used below when discussing further subclasses.

\subsubsection{Effective-NUT-free subclasses: $\ell_{\mr{eff}}=0$}
\label{twisting effective NUT free subclasses}

Although $l=0$ for the family considered above, the effective NUT-like parameter $\ell_{\mr{eff}}=\omega\xi_0$ is generically nonzero.
Its vanishing requires
\begin{align}
    -\frac{m\upalpha a}{1-B^2(a^2+\mr{q}^2)}
    +\sqrt{Km^2+J\mr{q}^2}\sin\beta=0\,.
    \label{effective-NUT-zero}
\end{align}
We retain the nonzero-denominator and reality conditions stated above.
For $Km^2+J\mr{q}^2>0$, the required angle $\beta_0$ satisfies
\begin{align}
    \sin\beta_0
    =\frac{m\upalpha a}
    {\left[1-B^2(a^2+\mr{q}^2)\right]
    \sqrt{Km^2+J\mr{q}^2}}\,.
    \label{no-NUT condition}
\end{align}
The definition of $K$ gives the identity
\begin{align}
    Km^2+J\mr{q}^2
    -\frac{m^2\upalpha^2a^2}
    {\left[1-B^2(a^2+\mr{q}^2)\right]^2}
    =m^2+(1-I+J)\mr{q}^2\,.
    \label{effective NUT free radicand identity}
\end{align}
Thus, a real choice of $\beta_0$ exists if and only if
\begin{align}
    m^2+(1-I+J)\mr{q}^2\geq0\,.
    \label{effective NUT free reality}
\end{align}
The corresponding shifts obey
\begin{align}
    (1-I+J)r_0^2-2mr_0-\mr{q}^2&=0\,\, \ar 
   r_0= r_0^{\pm}
    =\frac{m \pm \sqrt{m^2+(1-I+J)\mr{q}^2}}{1-I+J}\,.
    \label{effective NUT free shifts}
\end{align}
For a positive discriminant, the two signs correspond to the two possible signs of $\cos\beta_0$.
The branches coincide when the discriminant vanishes.

If $m\upalpha a=0$, Eq.~\eqref{effective-NUT-zero} is satisfied by $\beta=0,\pi$, or by the degenerate case $Km^2+J\mr{q}^2=0$.
In the latter case, $r_0=m/(1-I+J)$ and $\beta$ is redundant.
If $m\upalpha a\neq0$, the radicand cannot vanish in an effective-NUT-free solution.
The two branches can nevertheless coincide when Eq.~\eqref{effective NUT free reality} is saturated.
Then $\abs{\sin\beta_0}=1$.

\subsubsection{Vanishing spin parameter: $a\to0$}
\label{twisting zero spin limit}

We take $a\to0$ at fixed dimensionful parameters
$(m,\upalpha,B,\mr{q}^2,\beta,\gamma)$.
For comparison with the non-twisting convention $B>0$, we take $a\to0^{+}$ with $B>0$.
Writing $\alpha=\upalpha$ on the non-twisting side, the limiting quantities are
\begin{align}
    I&\to I_0=1-\frac{B^2m^2}{1-B^2\mr{q}^2}\,,\qquad
    J\to J_0=\frac{\upalpha^2}{B^2(1-B^2\mr{q}^2)}-1\,,\qquad
    K\to K_0\coloneqq\frac{1}{1-B^2\mr{q}^2}\,.
    \label{twisting zero spin IJK}
\end{align}
We assume $1-B^2\mr{q}^2\neq0$, $1-I_0+J_0\neq0$, and $K_0m^2+J_0\mr{q}^2\geq0$.
The radial and angular shifts tend to
\begin{align}
    r_0&\to r_{0\beta}
    \coloneqq\frac{m+\sqrt{K_0m^2+J_0\mr{q}^2}\cos\beta}
    {1-I_0+J_0}\,,\nn
    \ell_{\mr{eff}}&\to\ell_{\beta}
    \coloneqq\frac{\sqrt{K_0m^2+J_0\mr{q}^2}\sin\beta}
    {1-I_0+J_0}\,.
    \label{twisting zero spin shifts}
\end{align}
The metric functions have the limits
\begin{align}
    \mc{P}(\theta)&\to\mc{P}_0(\theta)
    =1-\frac{2m\upalpha}{1-B^2\mr{q}^2}\cos\theta
    +\left(1-I_0+B^2\mr{q}^2J_0\right)\cos^2\theta\,,\nn
    \mc{Q}(r)&\to\mc{Q}_0(r)
    =(\mr{q}^2-2mr+I_0r^2)(1-B^2J_0r^2)\,,\nn
    \Omega^2(r,\theta)&\to\Omega_0^2(r,\theta)
    =(1-\upalpha r\cos\theta)^2
    +B^2\biggl[r^2-\mr{q}^2\cos^2\theta\nn
    &\hspace{34mm}+2mr\cos\theta\left(\cos\theta
    -\frac{\upalpha r}{1-B^2\mr{q}^2}\right)
    -\left(I_0-B^2\mr{q}^2(1+J_0)\right)
    r^2\cos^2\theta\biggr]\,.
    \label{twisting zero spin metric functions}
\end{align}
These are the functions of the non-twisting family, but the full metric also depends on $\ell_{\beta}$.

\paragraph{Non-twisting limit: $\ell_{\beta}=0$}

For $\beta=0,\pi$, Eq.~\eqref{twisting zero spin shifts} gives $\ell_{\beta}=0$.
Using
\begin{align}
    K_0m^2+J_0\mr{q}^2
    =\frac{1-I_0+B^2\mr{q}^2J_0}{B^2}\,,
\end{align}
the two limiting shifts form precisely the pair $\{\tilde{r}_{0+},\tilde{r}_{0-}\}$ in Eq.~\eqref{eq:r0_pm}.
For $1-I_0+J_0>0$, $\beta=0$ and $\beta=\pi$ give $\tilde{r}_{0+}$ and $\tilde{r}_{0-}$, respectively.
These assignments are reversed when $1-I_0+J_0<0$.
Taking the limit of Eq.~\eqref{non-NUTty OPBH} gives
\begin{align}
    ds^2\to\frac{1}{\Omega_0^2}\biggl[
    -\frac{\mc{Q}_0}{(r+r_{0\beta})^2}dt^2
    +(r+r_{0\beta})^2\left(
    \frac{dr^2}{\mc{Q}_0}+\frac{d\theta^2}{\mc{P}_0}
    +\mc{P}_0\sin^2\theta\,d\phiv^2\right)\biggr]\,.
    \label{twisting to non-twisting metric}
\end{align}
This is the metric of $\mr{OP}_{\alpha\pm}^{0}$.
The Maxwell field likewise agrees with that family after matching the tetrad and electromagnetic phase conventions.\footnote{
With $t$ and $\phiv$ unchanged, the real potential obtained from Eq.~\eqref{A-nonlBH} in this limit agrees with Eq.~\eqref{non-twisting explicit potential} upon setting $\gamma_{\mr{nt}}=\pi-\gamma$, where $\gamma_{\mr{nt}}$ denotes the phase used in the non-twisting potential. 
This is only a change in the phase convention used to represent the same real field.
}
If $K_0m^2+J_0\mr{q}^2=0$, the two shifts coincide and $\ell_{\beta}=0$ for every $\beta$.
The limit then gives $\mr{OP}_{\alpha0}^{0}$.

\paragraph{Residual NUT-like twist: $\ell_{\beta}\neq0$}

For $K_0m^2+J_0\mr{q}^2>0$ and $\sin\beta\neq0$, the spin parameter vanishes but the effective NUT-like parameter remains nonzero.
Defining
\begin{align}
    \Sigma_\beta(r)\coloneqq(r+r_{0\beta})^2+\ell_{\beta}^2\,,
\end{align}
the limiting metric is
\begin{align}
    ds^2=\frac{1}{\Omega_0^2}\biggl[
    &-\frac{\mc{Q}_0}{\Sigma_\beta}
    \left(dt-2\ell_\beta(1-\cos\theta)d\phiv\right)^2
    +\Sigma_\beta\left(
    \frac{dr^2}{\mc{Q}_0}+\frac{d\theta^2}{\mc{P}_0}
    +\mc{P}_0\sin^2\theta\,d\phiv^2\right)\biggr]\,.
    \label{zero spin residual NUT metric}
\end{align}
The NUT-like twist and the mismatch between the two axis-adapted time coordinates therefore persist.
Therefore, $a=0$ does not by itself imply that the solution is non-twisting.
The limiting Maxwell field is obtained from Eqs.~\eqref{phi_0-nonlBH}--\eqref{A-nonlBH} with $a=0$, $r_0=r_{0\beta}$, and $\omega\xi_0=\ell_\beta$.

\subsubsection{Non-accelerating limit: $\upalpha\to0$}
\label{twisting nonaccelerating limit}

We take $\upalpha\to0$ at fixed $(a,m,B,\mr{q}^2,\beta,\gamma)$.
Then $J\to-1$, while $I$ is unchanged, and
\begin{align}
    K\to K'\coloneqq
    \frac{1-B^2a^2}{1-B^2(a^2+\mr{q}^2)}\,.
    \label{twisting nonaccelerating K}
\end{align}
The limiting metric functions and shifts are
\begin{align}
    \mc{P}(\theta)
    &=1+\left[1-I-B^2(a^2+\mr{q}^2)\right]\cos^2\theta\,,\nn
    \mc{Q}(r)
    &=\left(a^2+\mr{q}^2-2mr+Ir^2\right)(1+B^2r^2)\,,\nn
    \Omega^2(r,\theta)
    &=1+B^2r^2
    -B^2\left(a^2+\mr{q}^2-2mr+Ir^2\right)\cos^2\theta\,,\nn
    r_0&=-\frac{m+\sqrt{K'm^2-\mr{q}^2}\cos\beta}{I}\,,\nn
    \omega\xi_0&=-\frac{\sqrt{K'm^2-\mr{q}^2}\sin\beta}{I}\,.
    \label{eq:omegaxi_0b}
\end{align}
Here $I\neq0$ and $K'm^2-\mr{q}^2\geq0$.
The roots of the first factor of $\mc{Q}$ are unchanged, whereas the acceleration horizon factor becomes $1+B^2r^2$ and has no real zeros.
Whether a remaining root represents a regular horizon is subject to the conditions stated above.

\paragraph{Effective-NUT-free subclasses: $\ell_{\mr{eff}}=0$}

The effective NUT-like parameter vanishes for $\beta=0,\pi$, or when $K'm^2-\mr{q}^2=0$.
For the two nondegenerate choices of angle, the shifts are
\begin{align}
    r_{0\pm}=-\frac{m\pm\sqrt{K'm^2-\mr{q}^2}}{I}\,.
    \label{nonaccelerating effective NUT free shifts}
\end{align}
The upper and lower signs correspond to $\beta=0$ and $\beta=\pi$, respectively.

When the radicand vanishes, both values coincide at $r_0=-m/I$ and $\beta$ is redundant.
This degenerate condition is equivalent to
\begin{align}
    (1-B^2a^2)m^2
    =\left[1-B^2(a^2+\mr{q}^2)\right]\mr{q}^2\,.
    \label{nonaccelerating shift merger}
\end{align}

The genuine Griffiths--Podolsk\'y representation is noted in Appendix~\ref{App. genuine GP twisting}.
For this non-accelerating effective-NUT-free subclass, its radial factor is
\begin{align}
    \Delta(r')=I(r')^2-2Mr'+a^2\,,\qquad
    M=m+Ir_0\,,\qquad \mr{Q}^2=0\,,
    \label{nonaccelerating apparent parameters}
\end{align}
where $M$ and $\mr{Q}$ are the apparent mass and charge parameters defined in that appendix.
The relation to the Kerr--Newman--Bertotti--Robinson family very recently reported in Ref.~\cite{Ovcharenko:2026tos}, including the electromagnetic flux charges, is also given there.
In particular, the vanishing of $\mr{Q}^2$ does not imply that the physical flux charges vanish.

\subsubsection{Vanishing mass parameter: $m\to0$}
\label{twisting zero mass parameter}

We take $m\to0$ at fixed $(a,\upalpha,B,\mr{q}^2,\beta,\gamma)$.
Then $I\to1$, and the metric functions become
\begin{align}
    \mc{P}(\theta)
    &=1+B^2(a^2+\mr{q}^2)J\cos^2\theta\,,\nn
    \mc{Q}(r)
    &=(a^2+\mr{q}^2+r^2)(1-B^2Jr^2)\,,\nn
    \Omega^2(r,\theta)
    &=(1-\upalpha r\cos\theta)^2+B^2\biggl[
    r^2-(a^2+\mr{q}^2)\cos^2\theta
    -\left(1-B^2(a^2+\mr{q}^2)(1+J)\right)
    r^2\cos^2\theta\biggr]\,.
    \label{massless parameter metric functions}
\end{align}
For $J\neq0$, the shifts are
\begin{align}
    r_0=\frac{\sqrt{J\mr{q}^2}}{J}\cos\beta\,,\qquad
    \omega\xi_0=\frac{\sqrt{J\mr{q}^2}}{J}\sin\beta\,.
    \label{massless parameter shifts}
\end{align}
Their reality requires $J\mr{q}^2\geq0$.
Thus, $\mr{q}^2>0$ requires $J>0$, whereas $\mr{q}^2<0$ requires $J<0$.
In the latter case a non-accelerating limit, for which $J\to-1$, is possible.
The shift constraint becomes
\begin{align}
    J\left(r_0^2+\ell_{\mr{eff}}^2\right)=\mr{q}^2\,.
    \label{massless parameter shift constraint}
\end{align}
For $J=0$ and $\mr{q}^2\neq0$, it admits no finite shifts.

\paragraph{Effective-NUT-free subclasses: $\ell_{\mr{eff}}=0$}

For $J\mr{q}^2>0$, choosing $\beta=0$ or $\beta=\pi$ removes the effective NUT-like parameter and gives $Jr_0^2=\mr{q}^2$.
For $\mr{q}^2=0$ with $J\neq0$, both shifts vanish and the angle is redundant.
In the non-accelerating case, $J=-1$, the nonzero-shift solutions therefore satisfy
\begin{align}
    m=0\,,\qquad
    \mr{q}^2=-r_0^2<0\,,\qquad
    \ell_{\mr{eff}}=0\,.
    \label{massless genuine KBR condition}
\end{align}
As shown in Appendix~\ref{App. genuine GP twisting}, these include the genuine uncharged Kerr--Bertotti--Robinson solution identified in Ref.~\cite{Ovcharenko:2026tos}.
The apparent mass and charge parameters are $M=r_0$ and $\mr{Q}^2=0$, and both physical electromagnetic flux charges vanish.
In particular, $r_0>\abs{a}$ gives two distinct roots of $\Delta(r')=(r')^2-2r_0r'+a^2$.
The $B\to0$ limit at fixed $(r_0,a)$ gives Kerr with mass parameter $r_0$.
Thus, setting the original mass parameter $m$ to zero does not in general remove the mass associated with a familiar limiting metric.
Moreover, a purely imaginary charge parameter $\mr{q}$ need not imply an unphysical spacetime.

\subsubsection{Vanishing charge parameter: $\mr{q}=0$}
\label{twisting zero charge parameter}

Setting $\mr{q}=0$ gives the previously studied twisting Ovcharenko--Podolsk\'y family, under the parameter correspondence described in Sec.~\ref{Sec. OP-PD}.
In our parametrization,
\begin{align}
    I'&\coloneqq1-\frac{B^2m^2}{1-B^2a^2}\,,\qquad
    J'\coloneqq\frac{\upalpha^2}{B^2(1-B^2a^2)}-1\,,\nn
    K''&\coloneqq\left.K\right|_{\mr{q}^2=0}
    =1+\frac{a^2\upalpha^2}{(1-B^2a^2)^2}
    =1+\frac{(J'+1)B^2a^2}{1-B^2a^2}\,.
    \label{twisting q0 IJK}
\end{align}
The metric functions and shifts become
\begin{align}
    \mc{P}(\theta)
    &=1-\frac{2m\upalpha}{1-B^2a^2}\cos\theta
    +(1-I'+B^2a^2J')\cos^2\theta\,,\nn
    \mc{Q}(r)
    &=(a^2-2mr+I'r^2)(1-B^2J'r^2)\,,\nn
    \Omega^2(r,\theta)
    &=(1-\upalpha r\cos\theta)^2+B^2\biggl[
    r^2-a^2\cos^2\theta
    +2mr\cos\theta\left(\cos\theta
    -\frac{\upalpha r}{1-B^2a^2}\right)\nn
    &\hspace{42mm}
    -\left(I'-B^2a^2(1+J')\right)r^2\cos^2\theta\biggr]\,,\nn
    r_0&=\frac{m+\sqrt{K''m^2}\cos\beta}{1-I'+J'}\,,\nn
    \omega\xi_0
    &=\frac{-m\upalpha a/(1-B^2a^2)
    +\sqrt{K''m^2}\sin\beta}{1-I'+J'}\,.
    \label{eq:omegaxi_0c}
\end{align}

\paragraph{Effective-NUT-free subclasses: $\ell_{\mr{eff}}=0$}

For $m\neq0$, the required angles satisfy
\begin{align}
    \sin\beta_0
    =\frac{m\upalpha a}
    {(1-B^2a^2)\sqrt{K''m^2}}\,.
    \label{q0 effective NUT free angle}
\end{align}
With Eq.~\eqref{twisting q0 IJK}, the square of the right-hand side is
\begin{align}
    \frac{a^2\upalpha^2}{(1-B^2a^2)^2+a^2\upalpha^2}<1\,.
\end{align}
Thus, for $1-B^2a^2\neq0$, two real angles exist.
The corresponding shifts are
\begin{align}
    \left\{r_{0}^+\,,\,r_{0}^-\right\}
    =\left\{0\,,\,\frac{2m}{1-I'+J'}\right\}\,.
    \label{q0 effective NUT free shifts}
\end{align}
Note that $r_0^\pm$ is the $\mr{q} \to 0$ limit of $r_0^\pm$ in Eq.~(\ref{effective NUT free shifts}).
For $m=0$, both shifts vanish.

\paragraph{Non-accelerating limit: $\upalpha\to0$}

In the further limit $\upalpha\to0$, one has $J'\to-1$ and $K''\to1$.
The metric functions reduce to
\begin{align}
    \mc{P}(\theta)
    &=1+(1-I'-B^2a^2)\cos^2\theta\,,\nn
    \mc{Q}(r)
    &=(a^2-2mr+I'r^2)(1+B^2r^2)\,,\nn
    \Omega^2(r,\theta)
    &=1+B^2r^2-B^2(a^2-2mr+I'r^2)\cos^2\theta\,,\nn
    r_0&=-\frac{m+\abs{m}\cos\beta}{I'}\,,\qquad
    \omega\xi_0=-\frac{\abs{m}\sin\beta}{I'}\,.
    \label{q0 nonaccelerating metric functions}
\end{align}
Here $I'\neq0$.
The effective NUT-like parameter vanishes for $\beta=0,\pi$ or $m=0$.

For $m>0$, the choices $\beta=0$ and $\beta=\pi$ give
\begin{align}
\begin{dcases}
     \beta=0:&
    r_0=-\frac{2m}{I'}\,,\qquad M=-m\,,\qquad \mr{Q}^2=0\,,\\
    \beta=\pi:&
    r_0=0\,,\qquad M=m\,,\qquad \mr{Q}^2=0
\end{dcases}
   \,.
\label{KBR parameter branches}
\end{align}
The zero-shift solution is the previously known Kerr--Bertotti--Robinson solution, recently identified as charged and denoted Kerr--BR$_s$ in Ref.~\cite{Ovcharenko:2026tos}.
Both branches generally carry nonzero electromagnetic flux charges, as shown in Appendix~\ref{App. genuine GP twisting}.

\subsubsection{Pleba\'nski--Demia\'nski limit: $B\to0$}
\label{twisting charged PD limit}

For the standard charged aligned limit, we take $B\to0^{+}$ at fixed $(a,m,\upalpha,\mr{q}^2,\beta,\gamma)$ with $a>0$, $\upalpha>0$, and $\mr{q}^2\geq0$.
We choose $\mr{q}$ to be the nonnegative square root of $\mr{q}^2$ in the electromagnetic formulas below.
Then
\begin{align}
    I\to1\,,\qquad
    B^2J\to\upalpha^2\,,\qquad
    r_0\to0\,,\qquad \ell_{\mr{eff}}\to0\,,
\end{align}
and
\begin{align}
    \mc{P}(\theta)
    &\to1-2m\upalpha\cos\theta
    +\upalpha^2(a^2+\mr{q}^2)\cos^2\theta\,,\nn
    \mc{Q}(r)
    &\to(a^2+\mr{q}^2-2mr+r^2)(1-\upalpha^2r^2)\,,\nn
    \Omega^2(r,\theta)&\to(1-\upalpha r\cos\theta)^2\,,\qquad
    \rhov^2(r,\theta)\to r^2+a^2\cos^2\theta\,.
    \label{twisting charged PD metric limit}
\end{align}
These are the metric functions of the accelerating Kerr--Newman solution in the Griffiths--Podolsk\'y form, with $\Lambda=0$ and $e^2+g^2=\mr{q}^2$~\cite{Griffiths:2005qp,Griffiths:2009dfa}.
The Maxwell scalars reduce to
\begin{align}
    &\phi_0=\phi_2\to0\,,\qquad
    \phi_1\to\frac{e+ig}{2}
    \left(\frac{1-\upalpha r\cos\theta}{r+ia\cos\theta}\right)^2\,,\nn
    &e=\mr{q}\cos\left(\beta+\gamma+\frac{\pi}{2}\right)\,,\qquad
    g=\mr{q}\sin\left(\beta+\gamma+\frac{\pi}{2}\right)\,.
    \label{twisting charged PD field limit}
\end{align}
Thus, both the metric and electromagnetic field reduce smoothly to those of the charged Pleba\'nski--Demia\'nski solution with $\Lambda=0$.
Taking $\upalpha\to0$ afterwards gives the Kerr--Newman solution.
The order of these limits is important.
For example, in the effective-NUT-free sector, taking $\upalpha\to0$ first and then $B\to0$ along a smooth real branch with finite radial shift instead yields Kerr with mass $M_0=\sqrt{m^2-\mr{q}^2}$, as discussed around Eq.~\eqref{app:weak-field-Kerr-mass}.


\section{Summary and discussion}
\label{Sec. summary}

We have constructed a Pleba\'nski--Demia\'nski-adapted parametrization of the general Ovcharenko--Podolsk\'y solution and used it to extend the previously studied generalized black hole families.
In the Pleba\'nski--Demia\'nski gauge $\hat{c}_{10}=\hat{c}_{01}=0$, the Einstein--Maxwell equations do not require $\hat{c}_{20}=-\hat{c}_{02}$, which was imposed in the earlier parametrization~\cite{Ovcharenko:2025cpm}.
Retaining the independent coefficient freedom gives an additional charge parameter $\mr{q}$, with the earlier parametrization recovered at $\mr{q}=0$ through an explicit parameter correspondence.
The quantity $\mr{q}^2$ is a real parameter of either sign, so that $\mr{q}$ may be real or purely imaginary.
For $\mr{q}^2\geq0$, we established a standard aligned limit in which the coordinate shifts vanish, $\phi_0,\phi_2\to0$, and $\phi_1$ has a finite limit.
Both the metric and electromagnetic field then reduce smoothly to those of the charged Pleba\'nski--Demia\'nski solution with $\Lambda=0$ and $e^2+g^2=\mr{q}^2$.

Introducing acceleration and twist parameters through coordinate transformations and their degenerate limits, we obtained the twisting $\mr{OP}_{\alpha\omega}$ and non-twisting $\mr{OP}_{\alpha}^{0}$ sectors.
We expressed the solutions in the Griffiths--Podolsk\'y form and imposed the black hole-candidate conditions that allow Killing horizon cross sections with $\mb{S}^2$ topology.
The resulting eight-parameter class extends the previously studied seven-parameter class by one independent parameter.
In the generic twisting sector, two of these conditions determine $\ep$ and $n$ as rational functions of $k$ and the remaining parameters, while the normalization condition gives a quartic equation for $k$.
This defines four algebraic branches before the reality and coordinate-domain conditions are imposed, in contrast to the fifth-order condition in the earlier parametrization.
In the standard real aligned limit, one branch generically approaches finite Pleba\'nski--Demia\'nski values of $(k,\ep,n)$, whereas the other three roots become unbounded with the remaining parameters held fixed.

We gave explicit metrics, Newman--Penrose Maxwell scalars, and complex electromagnetic potentials for the non-twisting and $l=0$ twisting families.
Using dimensionful parametrizations, we analyzed their black hole and acceleration horizons, extremality conditions, and local conicity.
The non-twisting family has two branches $\mr{OP}_{\alpha\pm}^{0}$ distinguished by their radial shifts, which coincide in the degenerate subclass $\mr{OP}_{\alpha0}^{0}$.
In the non-accelerating limit of this subclass, the double zero of the radial function cancels from the norm of the static Killing vector and does not correspond to an extremal Killing horizon.
This observation shows that coalescence of algebraic roots does not necessarily imply the existence of an extremal Killing horizon.

For the $l=0$ twisting family, the effective NUT-like parameter $\ell_{\mr{eff}}=\omega\xi_0$ is generically nonzero.
We obtained the condition selecting $\ell_{\mr{eff}}=0$, the two corresponding radial-shift branches, and their degenerate cases.
Taking $a\to0$ recovers the non-twisting family when the effective NUT-like parameter vanishes in the limit, whereas generic choices of the angular parameter $\beta$ can retain a NUT-like twist.
We also examined the non-accelerating subclasses and those with vanishing mass or charge parameters.
The $\mr{q}=0$ sectors reproduce the previously studied Ovcharenko--Podolsk\'y families, while the standard $B\to0$ limits at fixed nonzero acceleration and $\mr{q}^2\geq0$ give the charged $C$-metric and the accelerating Kerr--Newman solution in the non-twisting and twisting sectors, respectively.

The genuine Griffiths--Podolsk\'y forms presented in Appendix~\ref{App. genuine GP} further clarify the interpretation of the solution parameters.
After absorbing the radial shifts into the coordinates, we introduced apparent mass and charge parameters through the radial metric functions.
For the non-accelerating effective-NUT-free twisting subclass, we established an explicit coordinate and parameter correspondence, including the electromagnetic field, with the Kerr--Newman--Bertotti--Robinson family recently presented in Ref.~\cite{Ovcharenko:2026tos}.
Using the electromagnetic flux charges evaluated in that work, we expressed the physical electric and magnetic charges in our parameters.
These relations show that neither $\mr{q}=0$ nor a vanishing apparent charge parameter $\mr{Q}$ implies vanishing electromagnetic flux charges in this rotating subclass.

Conversely, allowing $\mr{q}^2<0$ includes the genuine uncharged Kerr--Bertotti--Robinson solution identified in the same reference.
In our parametrization, it is obtained for $\upalpha=0$, $\ell_{\mr{eff}}=0$, $m=0$, and $\mr{q}^2=-r_0^2<0$.
Its electromagnetic flux charges vanish, and its $B\to0$ limit at fixed $(a,r_0)$ gives Kerr with mass parameter $r_0$.
Here the radial shift remains finite, unlike in the standard charged Pleba\'nski--Demia\'nski limit.
This example shows that a purely imaginary charge parameter need not imply an unphysical solution and that the original mass parameter $m$ need not coincide with the mass parameter of a limiting spacetime.

We now discuss the remaining questions concerning the parameter domains, global interpretation, and further applications of these solutions.


\paragraph{The general $l\neq0$ sector:}

The general twisting family deserves a more detailed investigation beyond its $l=0$ subclasses.
The quartic condition~\eqref{quartic k condition}, together with the rational expressions for $\ep$ and $n$, provides a direct starting point for studying the real solution branches, including their horizons and electromagnetic fields.
Of particular interest are the geometries suggested by branches~II--V in Ref.~\cite{Ovcharenko:2025cpm}.
For the parameter choices examined there, branches~II and III retain finite radial roots and were reported to possess poles in the electromagnetic field.
On the other hand, branches~IV and V have no real radial roots in those examples.
The parameter correspondence given in Sec.~\ref{Sec. OP-PD} shows that these solutions are contained in the $\mr{q}=0$ sector of our family.
It would be interesting to study their geometric properties in our parametrization and to investigate their $\mr{q}\neq0$ counterparts.

Within the $l\neq0$ sector, the condition $\ell_{\mr{eff}}=l+\omega\xi_0=0$ is also useful because it removes the mismatch between the local time coordinates at the two axes.
It would be interesting to determine the resulting branches and how they extend the effective-NUT-free families obtained here for $l=0$.

Another useful direction is a parametrization in which the limit $a\to0$ at nonzero $l$ can be taken regularly.
The Astorino and improved Astorino forms resolve the corresponding parametrization issue for accelerating NUT solutions in the Pleba\'nski--Demia\'nski family~\cite{Ovcharenko:2024yyu,Ovcharenko:2025fxg}.
An analogous construction for the Ovcharenko--Podolsk\'y family would allow a systematic comparison with the NUT-like twist that already survives the $a\to0$ limit of our $l=0$ family.

\paragraph{Global structure and horizon topology:}

The reality, horizon, and axis conditions obtained in this paper constrain the admissible domains, but do not determine their maximal extensions.
Recent work has clarified this issue for important subclasses.
Zhou et al.~\cite{Zhou:2026tkm} constructed an extension of the previously known Kerr--Bertotti--Robinson solution across $r=\infty$, and Ovcharenko~\cite{Ovcharenko:2026ooh} investigated the conformal structure and extensions of the Kerr--Newman--Bertotti--Robinson family, including its static limit.
These studies show that regular portions of the coordinate surface $r=\infty$ can be crossed and need not represent a conformal boundary.
The correspondence established in Appendix~\ref{App. genuine GP} allows these results to be translated into our parameters in the corresponding subclasses.
For the broader family, the remaining problem is to determine how acceleration and NUT-like twist affect these extensions, the accessible singularities, and the causal roles of the Killing horizons.

The angular domains should be considered as part of the same analysis.
Our black hole candidate conditions select the sector with two angular roots at $x=\pm1$, followed by $x=\cos\theta$ and a periodic azimuthal coordinate.
Relaxing these conditions permits other configurations and coordinate domains to be examined.
It remains to determine whether these admit regular non-compact horizon sections, including planar or hyperbolic cases, or other global identifications.

The choice of global domain also matters for wave propagation.
The hidden symmetries of the conformal-to-Carter class provide separability structures for null geodesic motion and several test field equations~\cite{Gray:2025lwy}.
Zhou et al.~\cite{Zhou:2026tkm} have already studied scalar scattering on an extended Kerr--Bertotti--Robinson region and found growing modes under the boundary conditions considered there.

Extending these studies to the present subclasses would be an important direction for future work.
In particular, it would be interesting to compute the quasinormal mode spectra of test fields and determine how they depend on the external electromagnetic field, acceleration, and NUT-like twist.

\paragraph{Physical charges and thermodynamics:}

The flux charges obtained in Appendix~\ref{App. genuine GP} provide a starting point for extending the analysis of physical charges to the accelerating and NUT-like sectors.
Thermodynamic descriptions of Kerr--Bertotti--Robinson black holes have employed the Christodoulou--Ruffini mass relation \cite{Christodoulou:1971pcn,Hu:2026slp,Kubiznak:2026uro}.
It would be interesting to determine the mass and angular momentum of the present solutions within the covariant phase space formalism \cite{Iyer:1994ys}, with appropriate boundary conditions and without assuming this mass relation.
Such an analysis could also clarify their first law and the thermodynamic roles of the external electromagnetic field and axis defects.

\paragraph{Exceptional sectors:}

The normalizations used here do not cover the cases $\check{c}_{00}=0$ and/or $\check{c}_{11}=0$.
These sectors should be examined directly in the unnormalized Einstein--Maxwell equations.
It remains to determine whether their solutions are represented by limits or alternative parametrizations of the families discussed here.

\paragraph{Solution-generating methods and extremal configurations:}

The parametrization developed here also provides seeds for further
solution-generating constructions.
Harrison transformations have already been used to combine
Bertotti--Robinson and Bonnor--Melvin external fields around static
black holes~\cite{Astorino:2025lih,DiPinto:2026ynu}.
In particular, Ref.~\cite{Astorino:2025lih} introduced a demagnetizing
construction yielding a nontrivial static vacuum geometry, whose
properties were subsequently analyzed in
Ref.~\cite{Astorino:2026okd}.
Further inversion and demagnetizing constructions were studied in
Refs.~\cite{Barrientos:2026shy,Ma:2026otg}.
Applying these methods to the present families could clarify how the
independent charge parameter, acceleration, and NUT-like twist enter
such transformations, and whether the external field can be removed
while retaining intrinsic black hole charges.

Subclasses with an extremal horizon provide another direction.
Multi-black hole configurations with Bertotti--Robinson asymptotics have been constructed from an extremal Reissner--Nordstr\"om seed in that background~\cite{DiPinto:2026rvp,Furugori:2026pdt,Clement:2026vzr}.
The extremality conditions derived here may identify further solutions whose near-horizon geometry and reduced field equations can be examined for an analogous construction.


\section*{Acknowledgements}
H.F. is grateful to Daisuke Yoshida for helpful discussions on the original Ovcharenko--Podolsk\'y solution.
H.F. also thanks Hideki Maeda for kind comments on this work at the workshop ``Singularity 27''.

\appendix
\section{Definitions of Newman--Penrose scalars}
\label{App. NP}

We follow the Newman--Penrose conventions of Griffiths and Podolsk\'y~\cite{Griffiths:2009dfa}.

\subsection*{Null tetrad}

Let $(k^\mu,\ell^\mu,m^\mu,\bar m^\mu)$ be a null tetrad in the signature $(-+++)$, normalized by
\begin{align}
 k^\mu k_\mu=\ell^\mu\ell_\mu=m^\mu m_\mu=\bar m^\mu\bar m_\mu=0,\qquad
 k^\mu\ell_\mu=-1,\qquad m^\mu\bar m_\mu=1.
 \label{app:null-tetrad-normalization}
\end{align}
All other inner products vanish.
Equivalently,
\begin{align}
 g_{\mu\nu}=-2k_{(\mu}\ell_{\nu)}
 +2m_{(\mu}\bar m_{\nu)}.
 \label{app:null-tetrad-metric}
\end{align}

\subsection*{Spin coefficients}
We omit the subscript $s$ on the spin coefficients in this appendix.
They are defined by
\begin{align}
 \kappa&\coloneqq-m^\mu k^\nu\nabla_\nu k_\mu,&
 \nu&\coloneqq\bar m^\mu\ell^\nu\nabla_\nu\ell_\mu,\nn
 \sigma&\coloneqq-m^\mu m^\nu\nabla_\nu k_\mu,&
 \lambda&\coloneqq\bar m^\mu\bar m^\nu\nabla_\nu\ell_\mu,\nn
 \rho&\coloneqq-m^\mu\bar m^\nu\nabla_\nu k_\mu,&
 \mu&\coloneqq\bar m^\mu m^\nu\nabla_\nu\ell_\mu,\nn
 \tau&\coloneqq-m^\mu\ell^\nu\nabla_\nu k_\mu,&
 \pi&\coloneqq\bar m^\mu k^\nu\nabla_\nu\ell_\mu,\nn
 \alpha&\coloneqq-\frac12
 \left(\ell^\mu\bar m^\nu\nabla_\nu k_\mu
 -\bar m^\mu\bar m^\nu\nabla_\nu m_\mu\right),&
 \beta&\coloneqq-\frac12
 \left(\ell^\mu m^\nu\nabla_\nu k_\mu
 -\bar m^\mu m^\nu\nabla_\nu m_\mu\right),\nn
 \epsilon&\coloneqq-\frac12
 \left(\ell^\mu k^\nu\nabla_\nu k_\mu
 -\bar m^\mu k^\nu\nabla_\nu m_\mu\right),&
 \gamma&\coloneqq-\frac12
 \left(\ell^\mu\ell^\nu\nabla_\nu k_\mu
 -\bar m^\mu\ell^\nu\nabla_\nu m_\mu\right).
 \label{app:spin-coefficients}
\end{align}

\subsection*{Curvature and electromagnetic scalars}
The Weyl scalars are
\begin{align}
 \Psi_0&\coloneqq C_{\mu\nu\rho\sigma}k^\mu m^\nu k^\rho m^\sigma,&
 \Psi_1&\coloneqq C_{\mu\nu\rho\sigma}k^\mu\ell^\nu k^\rho m^\sigma,\nn
 \Psi_2&\coloneqq C_{\mu\nu\rho\sigma}k^\mu m^\nu\bar m^\rho\ell^\sigma,&
 \Psi_3&\coloneqq C_{\mu\nu\rho\sigma}k^\mu\ell^\nu\bar m^\rho\ell^\sigma,\nn
 \Psi_4&\coloneqq C_{\mu\nu\rho\sigma}\ell^\mu\bar m^\nu\ell^\rho\bar m^\sigma.
 \label{app:Weyl-scalars}
\end{align}

The Ricci scalars are
\begin{align}
 \Phi_{00}&\coloneqq\frac12R_{\mu\nu}k^\mu k^\nu,&
 \Phi_{22}&\coloneqq\frac12R_{\mu\nu}\ell^\mu\ell^\nu,\nn
 \Phi_{01}&\coloneqq\frac12R_{\mu\nu}k^\mu m^\nu,&
 \Phi_{02}&\coloneqq\frac12R_{\mu\nu}m^\mu m^\nu,\nn
 \Phi_{12}&\coloneqq\frac12R_{\mu\nu}m^\mu\ell^\nu,&
 \Phi_{11}&\coloneqq\frac14R_{\mu\nu}
 \left(k^\mu\ell^\nu+m^\mu\bar m^\nu\right)
 \label{app:curvature-and-electromagnetic-scalars}
\end{align}

Electromagnetic scalars are
\begin{align}
 \phi_0\coloneqq F_{\mu\nu}k^\mu m^\nu,\qquad 
 \phi_1\coloneqq\frac12F_{\mu\nu}
 \left(k^\mu\ell^\nu+\bar m^\mu m^\nu\right),\qquad 
 \phi_2\coloneqq F_{\mu\nu}\bar m^\mu\ell^\nu.
\end{align}


\section{Generalized black holes in genuine Griffiths--Podolsk\'y form}
\label{App. genuine GP}

In this appendix, we note the genuine Griffiths--Podolsk\'y forms of the general non-twisting family and the $l=0$ twisting family considered in Sec.~\ref{Sec. family}.
In each case, we absorb the constant radial shift into the radial coordinate and rewrite the metric functions accordingly.
We use the dimensionful parametrizations introduced in that section.

\subsection{Genuine Griffiths--Podolsk\'y form of the non-twisting sector}
\label{App. genuine GP non-twisting}

For each branch $\mr{OP}_{\alpha\pm}^{0}$, introduce
\begin{align}
    r'_{\pm}\coloneqq r+\tilde{r}_{0\pm}\,,
    \label{non-twisting genuine radial coordinate}
\end{align}
leaving $t$, $\theta$, and $\phiv$ unchanged.
We define
\begin{align}
    \mc{Q}_{0\pm}^{\prime}(r'_{\pm})
    &\coloneqq\frac{\mc{Q}_0(r'_{\pm}-\tilde{r}_{0\pm})}
    {(r'_{\pm})^2}\,,\qquad
    \Omega_{0\pm}^{\prime}(r'_{\pm},\theta)
    \coloneqq\Omega_0(r'_{\pm}-\tilde{r}_{0\pm},\theta)\,.
    \label{non-twisting genuine function definitions}
\end{align}
In particular, $\mc{Q}_{0\pm}^{\prime}$ is dimensionless, unlike the quartic function $\mc{Q}_0$, which has dimensions of length squared.
The metric takes the genuine Griffiths--Podolsk\'y form
\begin{align}
    ds_{\pm}^2
    &=\frac{1}{\bigl(\Omega_{0\pm}^{\prime}\bigr)^2}
    \biggl[
    -\mc{Q}_{0\pm}^{\prime}(r'_{\pm})\,dt^2
    +\frac{(dr'_{\pm})^2}{\mc{Q}_{0\pm}^{\prime}(r'_{\pm})}
    +(r'_{\pm})^2\left(
    \frac{d\theta^2}{\mc{P}_0(\theta)}
    +\mc{P}_0(\theta)\sin^2\theta\,d\phiv^2
    \right)\biggr]\,,
    \label{non-twisting genuine GP metric}
\end{align}
where
\begin{align}
    \mc{Q}_{0\pm}^{\prime}(r'_{\pm})
    &=\left[I_0-\frac{2M_{\pm}}{r'_{\pm}}
    +\frac{Q_{\pm}^2}{(r'_{\pm})^2}\right]
    \left[1-B^2J_0(r'_{\pm}-\tilde{r}_{0\pm})^2\right]\,,
    \label{non-twisting genuine radial function}\\
    \bigl(\Omega_{0\pm}^{\prime}(r'_{\pm},\theta)\bigr)^2
    &=\left[1-\alpha(r'_{\pm}-\tilde{r}_{0\pm})\cos\theta\right]^2
    +B^2\left[(r'_{\pm}-\tilde{r}_{0\pm})^2
    -\mr{q}^2\cos^2\theta\right]\nn
    &\quad+2B^2m(r'_{\pm}-\tilde{r}_{0\pm})\cos\theta
    \left[\cos\theta
    -\frac{\alpha(r'_{\pm}-\tilde{r}_{0\pm})}
    {1-B^2\mr{q}^2}\right]\nn
    &\quad-\left[I_0-B^2\mr{q}^2(1+J_0)\right]B^2
    (r'_{\pm}-\tilde{r}_{0\pm})^2\cos^2\theta\,.
    \label{non-twisting genuine conformal factor}
\end{align}
The angular function $\mc{P}_0(\theta)$ is unchanged and is given in Eq.~\eqref{non-twisting explicit metric functions}.

\paragraph{Apparent mass and charge parameters}

The combinations appearing in the first factor of Eq.~\eqref{non-twisting genuine radial function} are
\begin{align}
    M_{\pm}
    &\coloneqq m+I_0\tilde{r}_{0\pm}\nn
    &=\frac{1+J_0}{1-I_0+J_0}\,m
    \pm\frac{I_0\sqrt{1-I_0+B^2\mr{q}^2J_0}}
    {B\abs{1-I_0+J_0}}\,,
    \label{non-twisting effective mass}\\
    Q_{\pm}^2
    &\coloneqq\mr{q}^2+2m\tilde{r}_{0\pm}
    +I_0\tilde{r}_{0\pm}^{\,2}\,.
    \label{non-twisting effective squared charge}
\end{align}
We refer to these as the apparent mass and charge parameters of the Reissner--Nordstr\"om-type factor.

The shift constraint can equivalently be written as
\begin{align}
    (1-I_0+J_0)\tilde{r}_{0\pm}^{\,2}
    -2m\tilde{r}_{0\pm}-\mr{q}^2=0\,.
    \label{non-twisting effective shift identity}
\end{align}
Consequently,
\begin{align}
    &Q_{\pm}^2
    =(1+J_0)\tilde{r}_{0\pm}^{\,2}
    =\frac{\alpha^2\tilde{r}_{0\pm}^{\,2}}
    {B^2(1-B^2\mr{q}^2)}\,,
    \label{non-twisting effective charge shift relation}\\
    &M_{\pm}^2-I_0Q_{\pm}^2
    =m^2-I_0\mr{q}^2\,.
    \label{non-twisting effective discriminant}
\end{align}
For $I_0\neq0$, the latter identity also gives
\begin{align}
    Q_{\pm}^2=\mr{q}^2+\frac{M_{\pm}^2-m^2}{I_0}\,.
\end{align}

\paragraph{Electromagnetic field and the $\mr{q}=0$ subclass}

In the shifted coordinates, the null tetrad is
\begin{align}
    k^a
    &=\sqrt{\frac{(\Omega_{0\pm}^{\prime})^2}
    {2\mc{Q}_{0\pm}^{\prime}}}
    \left[
    (\del_t)^a
    +\mc{Q}_{0\pm}^{\prime}(\del_{r'_{\pm}})^a
    \right]\,,\quad 
    \ell^a
    =\sqrt{\frac{(\Omega_{0\pm}^{\prime})^2}
    {2\mc{Q}_{0\pm}^{\prime}}}
    \left[
    (\del_t)^a
    -\mc{Q}_{0\pm}^{\prime}(\del_{r'_{\pm}})^a
    \right]\,,
    \nn
    m^a
    &=\sqrt{\frac{(\Omega_{0\pm}^{\prime})^2}
    {2\mc{P}_0(r'_{\pm})^2}}
    \left[
    \frac{1}{\sin\theta}(\del_{\phiv})^a
    -i\mc{P}_0(\del_{\theta})^a
    \right]\,,\quad 
    \bar{m}^a
    =\sqrt{\frac{(\Omega_{0\pm}^{\prime})^2}
    {2\mc{P}_0(r'_{\pm})^2}}
    \left[
    \frac{1}{\sin\theta}(\del_{\phiv})^a
    +i\mc{P}_0(\del_{\theta})^a
    \right]\,.
    \label{tetrad:genuine-non-twisting}
\end{align}
Then, the Maxwell scalars are
\begin{align}
    \phi_0^{\pm}&=\phi_2^{\pm}
    =\frac{B\e{i\gamma}\sin\theta}
    {2\Omega_{0\pm}^{\prime}r'_{\pm}}
    \sqrt{(r'_{\pm})^2\mc{P}_0\mc{Q}_{0\pm}^{\prime}}\,,
    \label{non-twisting genuine phi0}\\
    \phi_1^{\pm}
    &=\frac{i\e{i\gamma}
    \bigl(\Omega_{0\pm}^{\prime}\bigr)^2}
    {2B\sin\theta}\,
    \del_{r'_{\pm}}\left(
    \frac{\del_{\theta}\Omega_{0\pm}^{\prime}}
    {r'_{\pm}}
    \right)\,.
    \label{non-twisting genuine phi1}
\end{align}
A corresponding complex electromagnetic potential is
\begin{align}
    \mathcal{A}_{\mu}^{\pm}dx^{\mu}
    =\frac{\e{i\gamma}}{2B}\biggl[
    \left(r'_{\pm}\del_{r'_{\pm}}\Omega_{0\pm}^{\prime}
    -\Omega_{0\pm}^{\prime}\right)d\phiv
    +\frac{i\del_{\theta}\Omega_{0\pm}^{\prime}}
    {r'_{\pm}\sin\theta}\,dt\biggr]\,.
    \label{non-twisting genuine complex potential}
\end{align}
As before, the real potential is $\bm{A}^{\pm}=2\operatorname{Re}\bm{\mathcal{A}}^{\pm}$, up to gauge freedom.

For $\mr{q}=0$, the two shift values are
\begin{align}
    \left\{\tilde{r}_{0+}\,,\,\tilde{r}_{0-}\right\}
    =\left\{0\,,\,\frac{2B^2m}
    {\alpha^2-B^2(1-B^2m^2)}\right\}\,.
\end{align}
The zero-shift solution has $M=m$ and $Q^2=0$.
For the nonzero-shift solution, Eq.~\eqref{non-twisting effective charge shift relation} reduces to
\begin{align}
    Q_{\pm}^2
    =\frac{\alpha^2\tilde{r}_{0\pm}^{\,2}}{B^2}
    =\frac{4\alpha^2B^2m^2}
    {\left[\alpha^2-B^2(1-B^2m^2)\right]^2}\,.
    \label{non-twisting q0 effective charge}
\end{align}
Thus, a nonzero squared-charge parameter remains for $\alpha\neq0$ and $\tilde{r}_{0\pm}\neq0$, even though $\mr{q}=0$.

For this $\mr{q}=0$ family, Ref.~\cite{Ovcharenko:2026byw} evaluated the physical electric and magnetic charges by integrating the electromagnetic fluxes over a compact surface surrounding the black hole.
With $\phiv\sim\phiv+2\pi\mc{C}$ and the charge conventions of that reference, its Eq.~(57) gives, in our parameters,
\begin{align}
    q_{\mr{e}}^{\pm}
    &=-\mc{C}\frac{\alpha\tilde{r}_{0\pm}}{B}\sin\gamma\,,
    \qquad
    q_{\mr{m}}^{\pm}
    =\mc{C}\frac{\alpha\tilde{r}_{0\pm}}{B}\cos\gamma\,.
    \label{non-twisting q0 physical charges}
\end{align}
Here we have used $B=2\alpha\abs{c_{\mr{OP}}}$ for $\alpha>0$.
Accordingly,
\begin{align}
    \bigl(q_{\mr{e}}^{\pm}\bigr)^2
    +\bigl(q_{\mr{m}}^{\pm}\bigr)^2
    =\mc{C}^2Q_{\pm}^2
    \qquad (\mr{q}=0)\,.
\end{align}
The nonzero-shift branch can therefore carry physical electromagnetic charge even when the parameter $\mr{q}$ vanishes.

\paragraph{Horizons and conicity}

To distinguish the two horizon roots from the two solution branches, let $\varsigma=\pm1$ label the roots and retain the subscript $\pm$ for the branch.
For $I_0\neq0$ and real roots, the black hole horizon coordinates are
\begin{align}
    r_{b,\pm}^{\prime\,\varsigma}
    &=\frac{M_{\pm}+\varsigma
    \sqrt{M_{\pm}^2-I_0Q_{\pm}^2}}{I_0}
    =r_b^{\varsigma}+\tilde{r}_{0\pm}\,.
    \label{non-twisting genuine black hole horizons}
\end{align}
For $J_0>0$, the acceleration horizon coordinates are
\begin{align}
    r_{a,\pm}^{\prime\,\varsigma}
    =\tilde{r}_{0\pm}+\frac{\varsigma}{B\sqrt{J_0}}
    =r_a^{\varsigma}+\tilde{r}_{0\pm}\,.
    \label{non-twisting genuine acceleration horizons}
\end{align}
These expressions represent horizons under the same local regularity and domain conditions as in Sec.~\ref{Sec. family}.
Equation~\eqref{non-twisting effective discriminant} shows that the extremality condition is unchanged:
\begin{align}
    M_{\pm}^2=I_0Q_{\pm}^2
    \quad\Longleftrightarrow\quad
    m^2=I_0\mr{q}^2\,.
\end{align}
As in the main text, a root at $r'_{\pm}=0$ must be examined in the full metric, since factors may cancel and the root need not represent a Killing horizon.

Since only the radial coordinate has been shifted, the angular function, the azimuthal period, and all conicity data in Eqs.~\eqref{non-twisting-axis-values}--\eqref{non-twisting-south-deficit} are unchanged.

\subsection{Genuine Griffiths--Podolsk\'y form of the $l=0$ twisting sector}
\label{App. genuine GP twisting}

We use the dimensionful parametrization of the $l=0$ family introduced in Sec.~\ref{Sec. family}, with $\omega=1$ and acceleration parameter $\upalpha$.
For a chosen value of $\beta$, introduce
\begin{align}
    r'&\coloneqq r+r_0\,,\qquad
    \ell_{\mathrm{eff}}\coloneqq\omega\xi_0\,,\nn
    \rhov^{\prime\,2}(r',\theta)
    &\coloneqq(r')^2+(a\cos\theta+\ell_{\mathrm{eff}})^2\,,\nn
    \mc{Q}'(r')&\coloneqq\mc{Q}(r'-r_0)\,,\qquad
    \Omega'(r',\theta)\coloneqq\Omega(r'-r_0,\theta)\,.
    \label{app:genuine-GP-coordinate-transformation}
\end{align}
The coordinates $t$, $\theta$, and $\phiv$ are unchanged.
The metric becomes
\begin{align}
    ds^2
    &=\frac{1}{(\Omega')^2}\biggl[
    -\frac{\mc{Q}'}{\rhov^{\prime\,2}}
    \left(dt-\left[a\sin^2\theta
    +2\ell_{\mathrm{eff}}(1-\cos\theta)\right]d\phiv\right)^2\nn
    &\qquad \qquad \,\,+\rhov^{\prime\,2}\left(
    \frac{(dr')^2}{\mc{Q}'}
    +\frac{d\theta^2}{\mc{P}(\theta)}\right)
    +\frac{\mc{P}(\theta)\sin^2\theta}{\rhov^{\prime\,2}}
    \left(a\,dt-\left[(r')^2+(a+\ell_{\mathrm{eff}})^2\right]
    d\phiv\right)^2\biggr]\,.
    \label{app:genuine-GP-metric}
\end{align}
The shift $r_0$ no longer appears in $\rhov^{\prime\,2}$, while $\ell_{\mathrm{eff}}$ retains its NUT-like role.
The radial and angular functions are
\begin{align}
    \mc{Q}'(r')
    &=\left[I(r')^2-2Mr'+a^2+\mr{Q}^2\right]
    \left[1-B^2J(r'-r_0)^2\right]\,,
    \label{app:genuine-GP-radial-function}\\
    \mc{P}(\theta)
    &=1-\frac{2m\upalpha}{1-B^2(a^2+\mr{q}^2)}\cos\theta
    +\left[1-I+B^2(a^2+\mr{q}^2)J\right]\cos^2\theta\,,
    \label{app:genuine-GP-angular-function}
\end{align}
and the conformal factor is
\begin{align}
  \!\!\!  (\Omega')^2(r',\theta)
    &=\left[1-\upalpha(r'-r_0)\cos\theta\right]^2
    +B^2\biggl[
    (r'-r_0)^2-(a^2+\mr{q}^2)\cos^2\theta\nn
    &\hspace{52mm} +2m(r'-r_0)\cos\theta
    \left(\cos\theta
    -\frac{\upalpha(r'-r_0)}{1-B^2(a^2+\mr{q}^2)}\right)\nn
    &\hspace{55mm} -\left(I-B^2(a^2+\mr{q}^2)(1+J)\right)
    (r'-r_0)^2\cos^2\theta\biggr]\,.
    \label{app:genuine-GP-conformal-factor}
\end{align}
Here $I$, $J$, $r_0$, and $\ell_{\mathrm{eff}}$ have the definitions and parameter ranges specified in Sec.~\ref{Sec. family}.
These expressions are
\begin{align}
    I&=1-\frac{B^2m^2}{1-B^2(a^2+\mr q^2)}\,,\qquad 
  J =\frac{\upalpha^2}
    {B^2\left[1-B^2(a^2+\mr{q}^2)\right]}-1\,,\nn
  r_0
    &=\frac{m+\sqrt{Km^2+J\mr{q}^2}\cos\beta}{1-I+J}\,,\nn
   \ell_\mr{eff} &= \omega\xi_0
    =\frac{1}{1-I+J}\left[
    -\frac{m\upalpha a}{1-B^2(a^2+\mr{q}^2)}
    +\sqrt{Km^2+J\mr{q}^2}\sin\beta\right]\,,
\end{align}
with 
\begin{align}
         K=\frac{1+JB^2a^2}{1-B^2(a^2+\mr{q}^2)}\,.
\end{align}

\paragraph{Apparent mass and charge parameters}
As in the non-twisting case, the apparent mass and charge parameters in Eq.~\eqref{app:genuine-GP-radial-function} are defined by
\begin{align}
    M&\coloneqq m+Ir_0\,,\nn
    \mr{Q}^2
    &\coloneqq\mr{q}^2+2mr_0+Ir_0^2\,.
    \label{app:twisting-effective-parameters}
\end{align}
Explicitly, we can rewrite the apparent mass parameter as follows
\begin{align}
    M=\frac{(1+J)m+I\sqrt{Km^2+J\mr{q}^2}\cos\beta}{1-I+J}\,.
    \label{app:twisting-effective-mass-explicit}
\end{align}
The circle constraint for $r_0$ and $\omega \xi_0$ can be written as
\begin{align}
    (1-I+J)(r_0^2+\ell_{\mathrm{eff}}^2)
    -2mr_0
    +\frac{2m\upalpha a}{1-B^2(a^2+\mr{q}^2)}\ell_{\mathrm{eff}}
    -\mr{q}^2=0\,.
    \label{app:twisting-shift-identity}
\end{align}
Consequently,
\begin{align}
    &\mr{Q}^2
    =(1+J)r_0^2+(1-I+J)\ell_{\mathrm{eff}}^2
    +\frac{2m\upalpha a}{1-B^2(a^2+\mr{q}^2)}\ell_{\mathrm{eff}}\,,
    \label{app:twisting-charge-shift-identity}\\
    &M^2-I(a^2+\mr{Q}^2)
    =m^2-I(a^2+\mr{q}^2)\,.
    \label{app:twisting-effective-discriminant}
\end{align}
For $I\neq0$, the definitions also imply
\begin{align}
    \mr{Q}^2
    =\mr{q}^2+\frac{M^2-m^2}{I}\,.
\end{align}
In the subclass with $\ell_{\mathrm{eff}}=0$, Eq.~\eqref{app:twisting-charge-shift-identity} simplifies to
\begin{align}
    \mr{Q}^2=(1+J)r_0^2
    =\frac{\upalpha^2r_0^2}{B^2[1-B^2(a^2+\mr{q}^2)]}\,.
    \label{app:twisting-noNUT-effective-charge}
\end{align}
In particular, the coefficient $\mr{Q}^2$ vanishes for $\upalpha=0$ and $\ell_{\mathrm{eff}}=0$.
However, this does not in general imply that the electromagnetic flux charges vanish.

\paragraph{Electromagnetic field}
The principal null tetrad takes the form
\begin{align}
    k^a
    &=\sqrt{\frac{(\Omega')^2}
    {2\mc{Q}'\rhov^{\prime\,2}}}
    \left[
    \bigl((r')^2+(a+\ell_{\mr{eff}})^2\bigr)(\del_t)^a
    +\mc{Q}'(\del_{r'})^a+a(\del_{\phiv})^a
    \right]\,,
    \nn
    \ell^a
    &=\sqrt{\frac{(\Omega')^2}
    {2\mc{Q}'\rhov^{\prime\,2}}}
    \left[
    \bigl((r')^2+(a+\ell_{\mr{eff}})^2\bigr)(\del_t)^a
    -\mc{Q}'(\del_{r'})^a+a(\del_{\phiv})^a
    \right]\,,
    \nn
    m^a
    &=-\sqrt{\frac{(\Omega')^2}
    {2\mc{P}\rhov^{\prime\,2}}}
    \frac{1}{\sin\theta}
    \Bigl[
    \bigl(a\sin^2\theta
    +2\ell_{\mr{eff}}(1-\cos\theta)\bigr)(\del_t)^a
    +(\del_{\phiv})^a
    +i\mc{P}\sin\theta(\del_{\theta})^a
    \Bigr]\,,
    \nn
    \bar{m}^a
    &=-\sqrt{\frac{(\Omega')^2}
    {2\mc{P}\rhov^{\prime\,2}}}
    \frac{1}{\sin\theta}
    \Bigl[
    \bigl(a\sin^2\theta
    +2\ell_{\mr{eff}}(1-\cos\theta)\bigr)(\del_t)^a
    +(\del_{\phiv})^a
    -i\mc{P}\sin\theta(\del_{\theta})^a
    \Bigr]\,.
    \label{tetrad:genuine-twisting}
\end{align}
Then, the Maxwell scalars are
\begin{align}
    \phi_0&=\phi_2
    =\frac{B\e{i\gamma}}{2\Omega'}
    \frac{\sqrt{\mc{P}\mc{Q}'}\sin\theta}
    {r'+i(a\cos\theta+\ell_{\mathrm{eff}})}\,,
    \label{app:twisting-genuine-phi0}\\
    \phi_1
    &=-\frac{\e{i\gamma}}{2B\sin\theta}
    \frac{(\Omega')^2}
    {[r'+i(a\cos\theta+\ell_{\mathrm{eff}})]^2}\nn
    &\qquad\times\Biggl[
    (a\cos\theta+\ell_{\mathrm{eff}})^2
    \del_{\theta}\left(
    \frac{\del_{r'}\Omega'}{a\cos\theta+\ell_{\mathrm{eff}}}\right)
    -i(r')^2\del_{r'}\left(
    \frac{\del_{\theta}\Omega'}{r'}\right)\Biggr]\,.
    \label{app:twisting-genuine-phi1}
\end{align}
A corresponding complex potential is
\begin{align}
    \mathcal{A}_{\mu}dx^{\mu}
    =\frac{\e{i\gamma}}{2B}\Biggl[&
    \del_{r'}\Omega'
    \frac{a\,dt-[(r')^2+(a+\ell_{\mathrm{eff}})^2]d\phiv}
    {r'+i(a\cos\theta+\ell_{\mathrm{eff}})}
    +\Omega'\,d\phiv\nn
    &+\frac{i\del_{\theta}\Omega'}{\sin\theta}
    \frac{dt-[a\sin^2\theta
    +2\ell_{\mathrm{eff}}(1-\cos\theta)]d\phiv}
    {r'+i(a\cos\theta+\ell_{\mathrm{eff}})}\Biggr]\,.
    \label{app:twisting-genuine-potential}
\end{align}
As before, the real potential is $\bm{A}=2\operatorname{Re}\bm{\mathcal{A}}$, up to gauge freedom.

\paragraph{Horizons and conicity}

The radial translation gives black hole horizons and acceleration horizons 
\begin{align}
    r_b^{\prime\,\pm}
    &=\frac{M\pm\sqrt{M^2-I(a^2+\mr{Q}^2)}}{I}
    =r_b^{\pm}+r_0\,,
    \label{app:twisting-genuine-black-hole-horizons}\\
    r_a^{\prime\,\pm}
    &=r_0\pm\frac{1}{\abs{B}\sqrt{J}}
    =r_a^{\pm}+r_0\,.
    \label{app:twisting-genuine-acceleration-horizons}
\end{align}
The first expression assumes $I\neq0$ and a nonnegative discriminant, and the second assumes $J>0$.
The discriminant identity \eqref{app:twisting-effective-discriminant} shows that the algebraic extremality condition is unchanged.
The angular function, azimuthal period, and effective NUT-like parameter are also unchanged, so the conicity arguments remain those given in Sec.~\ref{Sec. family}.

\subsubsection{Non-accelerating effective-NUT-free subclass}

For $\upalpha=0$ and $\ell_{\mathrm{eff}}=0$, one has $J=-1$ and $\mr{Q}^2=0$.
The shift constraint and the definition of the apparent mass become
\begin{align}
    \mr{q}^2+2mr_0+Ir_0^2&=0\,,\qquad
    M=m+Ir_0\,.
    \label{app:nonaccelerating-shift-constraint}
\end{align}
Here $\mr{q}^2$ is allowed to be a real parameter of either sign, as explained in Sec.~\ref{Sec. OP-PD}.
Defining
\begin{align}
    \Delta(r')\coloneqq I(r')^2-2Mr'+a^2\,,
    \label{app:nonaccelerating-Delta}
\end{align}
the metric functions reduce to
\begin{align}
    \rhov^{\prime\,2}&=(r')^2+a^2\cos^2\theta\,,\nn
    \mc{P}(\theta)&=1+B^2(M^2-Ia^2)\cos^2\theta\,,\nn
    \mc{Q}'(r')&=\Delta(r')\left[1+B^2(r'-r_0)^2\right]\,,\nn
    (\Omega')^2(r',\theta)
    &=1+B^2(r'-r_0)^2-B^2\Delta(r')\cos^2\theta\,.
    \label{app:twisting-nonaccelerating-noNUT}
\end{align}
The absence of an apparent charge term in this factor does not imply vanishing electromagnetic flux charges.

While this manuscript was being finalized, Ovcharenko and Podolsk\'y~\cite{Ovcharenko:2026tos} presented a Kerr--Newman--Bertotti--Robinson family and identified its uncharged, genuine Kerr--Bertotti--Robinson subclass.
The relation to our parametrization can be exhibited explicitly.
We distinguish the parameters and coordinates of that reference by a subscript $\mr{OP}$ and define
\begin{align}
    h\coloneqq1+B^2r_0^2>0\,.
    \label{app:KNBR-h}
\end{align}
The coordinate and parameter correspondence is
\begin{align}
    r'&=\sqrt{h}\,r_{\mr{OP}}\,,\qquad
    t=\sqrt{h}\,t_{\mr{OP}}\,,\qquad
    \theta=\theta_{\mr{OP}}\,,\qquad
    \phiv=\phiv_{\mr{OP}}\,,\nn
    a_{\mr{OP}}&=\frac{a}{\sqrt{h}}\,,\qquad
    B_{\mr{OP}}=\frac{B}{\sqrt{h}}\,,\qquad
    m_{\mr{OP}}=\sqrt{h}\,M\,,\nn
    k_{\mr{OP}}&=-Br_0\,,\qquad
    e_{\mr{OP}}^2=a^2(1-I)
    =\frac{a^2B^2m^2}{1-B^2(a^2+\mr{q}^2)}\,.
    \label{app:KNBR-parameter-map}
\end{align}
Here $k_{\mr{OP}}$ is the auxiliary constant used in Ref.~\cite{Ovcharenko:2026tos}, not the Pleba\'nski--Demia\'nski-type parameter denoted by $k$ elsewhere in this paper.
This comparison is made on real branches of the two parametrizations.
The sign of $e_{\mr{OP}}$ is chosen so that Eq.~(2.3) of that reference holds, with $s_{\mr{OP}}$ taken to be the nonnegative square root of the right-hand side of Eq.~(2.4).
Using $h=1+k_{\mr{OP}}^2$, the inverse parameter transformation is given by
\begin{align}
    a&=\sqrt{h}\,a_{\mr{OP}}\,,\qquad
    B=\sqrt{h}\,B_{\mr{OP}}\,,\qquad
    m=\frac{m_{\mr{OP}}+I k_{\mr{OP}}/B_{\mr{OP}}}{\sqrt{h}}\,,\nn
    r_0&=-\frac{k_{\mr{OP}}}{\sqrt{h}\,B_{\mr{OP}}}\,,\qquad
    I=1-\frac{e_{\mr{OP}}^2}{h\,a_{\mr{OP}}^2}\,, \qquad 
    M=\frac{m_{\mr{OP}}}{\sqrt{h}}\,,\qquad 
    \mr{q}^2 =\frac{I k_{\mr{OP}}^2
    +2m_{\mr{OP}}k_{\mr{OP}}B_{\mr{OP}}}
    {h\,B_{\mr{OP}}^2}\,.
    \label{app:KNBR-inverse-parameter-map}
\end{align}
These expressions assume $a_{\mr{OP}}B_{\mr{OP}}\neq0$.
Together with Eq.~\eqref{app:KNBR-parameter-map} and the sign convention for $e_{\mr{OP}}$ stated above, they establish a one-to-one correspondence on the common real, nondegenerate parameter domain.

To verify the metric correspondence, introduce the functions
\begin{align}
    \mathcal{I}_{\mr{OP}}(r_{\mr{OP}})
    &\coloneqq
    (1+k_{\mr{OP}}B_{\mr{OP}}r_{\mr{OP}})^2
    +B_{\mr{OP}}^2r_{\mr{OP}}^2\,,\nn
    \Delta_{\mr{OP}}(r_{\mr{OP}})
    &\coloneqq(1+k_{\mr{OP}}^2)a_{\mr{OP}}^2
    -2m_{\mr{OP}}r_{\mr{OP}}
    +\left(1+k_{\mr{OP}}^2
    -\frac{e_{\mr{OP}}^2}{a_{\mr{OP}}^2}\right)r_{\mr{OP}}^2\,.
\end{align}
The functions in Eq.~\eqref{app:twisting-nonaccelerating-noNUT} obey
\begin{align}
    \Delta(\sqrt{h}\,r_{\mr{OP}})
    &=\Delta_{\mr{OP}}(r_{\mr{OP}})\,,\nn
    \frac{1+B^2(\sqrt{h}\,r_{\mr{OP}}-r_0)^2}{h}
    &=\mathcal{I}_{\mr{OP}}(r_{\mr{OP}})\,.
\end{align}
Consequently,
\begin{align}
    \mc{Q}_{\mr{OP}}(r_{\mr{OP}})
    &=\frac{\mc{Q}'(\sqrt{h}\,r_{\mr{OP}})}{h}
    =\mathcal{I}_{\mr{OP}}\Delta_{\mr{OP}}\,,\nn
    \Omega_{\mr{OP}}^2(r_{\mr{OP}},\theta)
    &=\frac{(\Omega')^2(\sqrt{h}\,r_{\mr{OP}},\theta)}{h}
    =\mathcal{I}_{\mr{OP}}
    -B_{\mr{OP}}^2\Delta_{\mr{OP}}\cos^2\theta\,,\nn
    \mc{P}(\theta)
    &=1+B_{\mr{OP}}^2\mu_{\mr{OP}}^2\cos^2\theta\,,\qquad
    \mu_{\mr{OP}}^2=h(M^2-Ia^2)\,.
    \label{app:KNBR-function-map}
\end{align}
These are the metric functions in Eq.~(2.2) of Ref.~\cite{Ovcharenko:2026tos}.
Together with Eq.~\eqref{app:KNBR-parameter-map}, they give the same line element.
The electromagnetic potential also agrees up to a gauge transformation and a constant electromagnetic duality rotation.
With compatible square-root branches for $\Omega'$ and $\Omega_{\mr{OP}}$, the choice $\gamma=\pi$ in Eq.~\eqref{app:twisting-genuine-potential} reproduces the magnetic-frame
potential in Eq.~(2.6) of that reference.

For a closed two-surface $\mathcal{S}$ surrounding the black hole, we define the electric and magnetic flux charges by
\begin{align}
    q_{\mr{e}}\coloneqq\frac{1}{4\pi}\int_{\mathcal{S}}{\ast F}\,,
    \qquad
    q_{\mr{m}}\coloneqq\frac{1}{4\pi}\int_{\mathcal{S}}F\,,
    \label{app:KNBR-flux-definitions}
\end{align}
with the orientation and charge conventions of Ref.~\cite{Ovcharenko:2026tos}.
Their Eq.~(4.13) gives $q_{\mr{e}}=\mc{C}e_{\mr{OP}}$ and $q_{\mr{m}}=0$ in the magnetic frame.
For the phase $\gamma$ used in our potential, the same result becomes
\begin{align}
    q_{\mr{e}}&=-\mc{C}e_{\mr{OP}}\cos\gamma\,,\qquad
    q_{\mr{m}}=\mc{C}e_{\mr{OP}}\sin\gamma\,.
    \label{app:KNBR-physical-charges}
\end{align}
In particular,
\begin{align}
    q_{\mr{e}}^2+q_{\mr{m}}^2
    &=\mc{C}^2a^2(1-I)
    =\mc{C}^2\frac{a^2B^2m^2}
    {1-B^2(a^2+\mr{q}^2)}\,.
    \label{app:KNBR-charge-magnitude}
\end{align}
The azimuthal period is unchanged by the coordinate transformation.
For regular axes, its value is fixed by
\begin{align}
    \mc{C}
    =\frac{1}{1+B^2(M^2-Ia^2)}
    =\frac{1}{1+B_{\mr{OP}}^2\mu_{\mr{OP}}^2}\,.
\end{align}
Thus, in this subclass $\mr{Q}^2=0$ even when the physical flux charges are nonzero.

The weak-field limit of this subclass clarifies its relation to Wald's family of test electromagnetic fields~\cite{Wald:1974np}.
Take $B\to0$ at fixed $(a,m,\mr{q}^2)$ along a smooth real branch with positive limiting mass:
\begin{align}
    r_0&\to c_0\coloneqq-m+M_0\,,
    \qquad
    M\to M_0\coloneqq\sqrt{m^2-\mr{q}^2}>0\,.
    \label{app:weak-field-Kerr-mass}
\end{align}
The limiting shift satisfies
\begin{align}
    \mr{q}^2+2mc_0+c_0^2=0\,.
\end{align}
Since $I\to1$ as $B\to0$, the metric in the shifted radial coordinate $r'$ tends to Kerr with mass $M_0$, even when $\mr{q}\neq0$.
This differs from the standard charged aligned limit taken at fixed nonzero acceleration in Sec.~\ref{twisting charged PD limit}.
In the magnetic frame $\gamma=\pi$, corresponding to a generally electrically charged black hole in an external magnetic field, the electromagnetic potential has the weak-field expansion, up to a local gauge transformation,
\begin{align}
    \bm{A}
    =\frac{B}{2}\bm{\psi}_{\mr{K}}^{\flat}
    +Ba\left(1-\frac{m}{2M_0}\right)
    \bm{\xi}_{\mr{K}}^{\flat}
    +O(B^3)\,,
    \label{app:Wald-test-field-limit}
\end{align}
where $\bm{\xi}_{\mr{K}}^{\flat}$ and $\bm{\psi}_{\mr{K}}^{\flat}$ are the metric duals of $\partial_t$ and $\partial_{\phiv}$ in the limiting Kerr geometry.
The leading-order field is contained in Wald's test-field family with net electric charge $Q=Bam$.
It is consistent with the exact flux formula, which gives $q_{\mr{e}}=Bam+O(B^3)$ when the axes are regularized.

\paragraph{Relation to the original Kerr--Bertotti--Robinson solution}

For $\mr{q}=0$ and $r_0=0$, one has $M=m$, $h=1$, and
\begin{align}
    I=1-\frac{B^2m^2}{1-B^2a^2}\,,\qquad
    e_{\mr{OP}}^2=\frac{a^2B^2m^2}{1-B^2a^2}\,.
\end{align}
This is the previously known charged Kerr--Bertotti--Robinson solution, called Kerr--$\mr{BR}_{s}$ in Ref.~\cite{Ovcharenko:2026tos}.
In contrast to the non-twisting sector, vanishing radial shift therefore does not imply vanishing physical charge.

\paragraph{Relation to the genuine Kerr--Bertotti--Robinson solution}

The uncharged genuine Kerr--$\mr{BR}_{0}$ solution is also contained.
For $aB\neq0$ and $1-B^2(a^2+\mr{q}^2)\neq0$, the choice
\begin{align}
    m=0\,,\qquad \mr{q}^2=-r_0^2
    \label{app:genuine-KBR-signed-parameter}
\end{align}
satisfies the shift constraint and gives
\begin{align}
    I=1\,,\qquad M=r_0\,,\qquad
    \mr{Q}^2=0\,,\qquad e_{\mr{OP}}=0\,,\qquad
    k_{\mr{OP}}=-m_{\mr{OP}}B_{\mr{OP}}\,.
\end{align}
The functions in Eq.~\eqref{app:twisting-nonaccelerating-noNUT} then match the genuine Kerr--$\mr{BR}_{0}$ solution in Eq.~(3.10) of Ref.~\cite{Ovcharenko:2026tos}, and $q_{\mr{e}}=q_{\mr{m}}=0$.
For example, $r_0>\abs{a}$ gives two distinct roots of $\Delta=(r')^2-2r_0r'+a^2$ and a positive angular function.
This provides an explicit example in which a negative value of $\mr{q}^2$ is compatible with a real magnetized black hole solution.
Moreover, its $B\to0$ limit at fixed $(a,r_0)$ is the Kerr solution with mass parameter $r_0$.
This is distinct from the standard aligned limit with vanishing radial shift considered in Sec.~\ref{Sec. OP-PD}.

\setstretch{0.65} 
\bibliography{reff} 
\bibliographystyle{utphys.bst}

\end{document}